\documentclass[prx,reprint,notitlepage,noeprint,superscriptaddress]{revtex4-2}

\usepackage[english]{babel}
\usepackage{amsmath,amssymb,amsfonts,bbm,xcolor,graphicx,comment,txfonts,mathdots,titlesec}
\definecolor{darkblue}{rgb}{0.,0.,0.5}
\usepackage[colorlinks,linkcolor=darkblue,citecolor=darkblue,urlcolor=darkblue]{hyperref}

\renewcommand\theparagraph{(\Roman{paragraph})}
\titleformat{\paragraph}[runin]{\normalfont\normalsize\itshape}{\theparagraph}{1em}{}

\newcommand{\id}{1}

\newcommand{\R}{\mathbb{R}}

\newcommand{\peq}{\mathrel{\hphantom{=}}}
\newcommand{\e}{\mathrm{e}}
\newcommand{\imag}{\mathrm{i}}
\newcommand{\diff}{\mathrm{d}}
\newcommand{\Diff}{\mathrm{D}}
\newcommand{\ket}[1]{\lvert #1 \rangle}
\newcommand{\bra}[1]{\langle #1 \rvert}
\newcommand{\braket}[1]{\left\langle #1 \right\rangle}

\newcommand{\svn}[1]{S_{#1}}

\newcommand{\dQG}{\delta Q_{\mathcal{G}}}
\newcommand{\dQS}{\delta Q_{\Sigma}}
\newcommand{\lind}{-}
\newcommand{\gind}{+}

\makeatletter
\newcommand*{\transpose}{%
  {\mathpalette\@transpose{}}%
}
\newcommand*{\@transpose}[2]{%
  \raisebox{\depth}{$\m@th#1\intercal$}%
}
\makeatother

\newcommand{\abs}[1]{\left\lvert #1 \right\rvert}

\newcommand{\tr}{\mathop{\mathrm{tr}}}
\newcommand{\Tr}{\mathop{\mathrm{Tr}}}
\newcommand{\Det}{\mathop{\mathrm{Det}}}
\newcommand{\trK}{\mathop{\mathrm{tr}_K}}
\newcommand{\trR}{\mathop{\mathrm{tr}_R}}
\newcommand{\trKR}{\mathop{\mathrm{tr}}}
\newcommand{\trCR}{\mathop{\mathrm{tr}_{CR}}}
\newcommand{\detK}{\mathop{\mathrm{det}_K}}
\newcommand{\detR}{\mathop{\mathrm{det}_R}}
\newcommand{\detKR}{\mathop{\mathrm{det}}}
\newcommand{\sgn}{\mathop{\mathrm{sgn}}}

\renewcommand{\Re}{\mathop{\mathrm{Re}}}
\renewcommand{\Im}{\mathop{\mathrm{Im}}}

\makeatletter
\DeclareFontFamily{OMX}{MnSymbolE}{}
\DeclareSymbolFont{MnLargeSymbols}{OMX}{MnSymbolE}{m}{n}
\SetSymbolFont{MnLargeSymbols}{bold}{OMX}{MnSymbolE}{b}{n}
\DeclareFontShape{OMX}{MnSymbolE}{m}{n}{
  <-6>  MnSymbolE5
  <6-7>  MnSymbolE6
  <7-8>  MnSymbolE7
  <8-9>  MnSymbolE8
  <9-10> MnSymbolE9
  <10-12> MnSymbolE10
  <12->   MnSymbolE12
}{}
\DeclareFontShape{OMX}{MnSymbolE}{b}{n}{
  <-6>  MnSymbolE-Bold5
  <6-7>  MnSymbolE-Bold6
  <7-8>  MnSymbolE-Bold7
  <8-9>  MnSymbolE-Bold8
  <9-10> MnSymbolE-Bold9
  <10-12> MnSymbolE-Bold10
  <12->   MnSymbolE-Bold12
}{}

\let\llangle\@undefined
\let\rrangle\@undefined
\DeclareMathDelimiter{\llangle}{\mathopen}%
{MnLargeSymbols}{'164}{MnLargeSymbols}{'164}
\DeclareMathDelimiter{\rrangle}{\mathclose}%
{MnLargeSymbols}{'171}{MnLargeSymbols}{'171}
\makeatother

\begin{document}

\title{Generalized Zeno effect and entanglement dynamics induced by fermion counting}

\author{Elias Starchl}

\affiliation{Institute for Theoretical Physics, University of Innsbruck, 6020
  Innsbruck, Austria}

\author{Mark H. Fischer}

\affiliation{Department of Physics, University of Zurich, Winterthurerstrasse
  190, CH-8057 Z\"urich, Switzerland}

\author{Lukas M. Sieberer}

\email{lukas.sieberer@uibk.ac.at}

\affiliation{Institute for Theoretical Physics, University of Innsbruck, 6020
  Innsbruck, Austria}

\date{\today}

\begin{abstract}
  We study a one-dimensional lattice system of free fermions subjected to a
  generalized measurement process: the system exchanges particles with its
  environment, but each fermion leaving or entering the system is counted. In
  contrast to the freezing of dynamics due to frequent measurements of lattice
  site occupation numbers, a high rate of fermion counts induces fast
  fluctuations in the state of the system. Still, through numerical simulations
  of quantum trajectories and an analytical approach based on replica Keldysh
  field theory, we find that instantaneous correlations and entanglement
  properties of free fermions subjected to fermion counting and local occupation
  measurements are strikingly similar. We explain this similarity through a
  generalized Zeno effect induced by fermion counting and a universal
  long-wavelength description in terms of a nonlinear sigma model. The physical
  requirements underlying this universal emergent behavior are conservation of
  the total number of particles in the system and its environment, and
  conservation of the purity of the state of the system by keeping a full record
  of all measurement outcomes. For both types of measurement processes, we
  present strong evidence against the existence of a critical phase with
  logarithmic entanglement and conformal invariance. Instead, we identify a
  finite critical range of length scales on which signatures of conformal
  invariance are observable. While area-law entanglement is established beyond a
  scale that is exponentially large in the measurement rate, the upper boundary
  of the critical range is only algebraically large and thus numerically
  accessible. Our finding that these properties do not rely on particle number
  conservation has far reaching implications for measurement-induced phenomena
  beyond noninteracting fermions, such as charge sharpening in random quantum
  circuits or generic interacting systems.
\end{abstract}

\maketitle

\section{Introduction}
\label{sec:introduction}

Frequent projective measurements induce the quantum Zeno effect, the freezing of
the evolution of a quantum system in an eigenstate of the measured
observable~\cite{Misra1977}. For measurements of local observables in a
spatially extended many-body system, these eigenstates exhibit area-law scaling
of the entanglement entropy. The quantum Zeno effect, thus, stabilizes area-law
entanglement even in systems that would in the absence of measurements unitarily
evolve toward volume-law entanglement. Reducing the rate at which measurements
are performed can lead to a novel type of dynamical phase transition between
area-law and volume-law scaling of the entanglement entropy. Such
measurement-induced phase transitions have first been described in quantum
circuits~\cite{Li2018, Skinner2019, Szyniszewski2019, Li2019, Bao2020,
  Gullans2020, Turkeshi2020, Zabalo2020, Bao2021, Ippoliti2021, Nahum2021,
  Lavasani2021, Sierant2022, Zabalo2022, Yu2022, Agrawal2022, Barratt2022,
  Jian2022, Jian2023, Oshima2023, Kelly2023, Weinstein2023a, Kelly2025} but
occur also in the continuous-time dynamics of fermionic~\cite{Cao2019, Chen2020,
  Nahum2020, Alberton2021, Tang2021, Buchhold2021, VanRegemortel2021,
  Coppola2022, Carollo2022, Ladewig2022, Buchhold2022, Turkeshi2022b, Yang2023a,
  Szyniszewski2023, Loio2023, LeGal2023, Poboiko2023, Poboiko2024, Kells2023,
  Fava2023, Merritt2023, Klocke2023, Chahine2024, Eissler2024, Lumia2024,
  Tsitsishvili2024, Fava2024, Guo2024, Poboiko2025}, bosonic~\cite{Fuji2020,
  Goto2020, Tang2020, Doggen2022, Minoguchi2022}, and spin
systems~\cite{Lang2020, Rossini2020, Botzung2021, Turkeshi2021a, Weinstein2023,
  Yang2023, Tirrito2023}. Experimental studies of measurement-induced phase
transitions have been performed with trapped ions~\cite{Noel2022, Agrawal2024}
and superconducting qubits~\cite{Hoke2023, Koh2023, Kamakari2024}. While the
freezing of dynamics provides an intuitive explanation for area-law entanglement
through repeated measurements, is it also a requirement?  As we detail in the
following, this question is of fundamental relevance for systems that are
subjected to generalized measurements.

A generalized measurement is described by a collection of measurement operators
$\hat{M}_n$, which obey the completeness relation
$\sum_n \hat{M}_n^{\dagger} \hat{M}_n^{\vphantom{\dagger}} = 1$, where the sum
is over possible measurement outcomes labeled by the index
$n$~\cite{Nielsen2010}. If the state of a quantum system immediately before a
generalized measurement performed at time $t$ is $\ket{\psi(t)}$, then the
outcome $n$ occurs with probability
$p_n(t) = \left\langle \psi(t) \middle| \hat{M}_n^{\dagger}
  \hat{M}_n^{\vphantom{\dagger}} \middle| \psi(t) \right\rangle$,
and the state of the system after the measurement is
\begin{equation}
  \label{eq:GM-state-update}
  \ket{\psi(t + 0^+)} = \frac{1}{\sqrt{p_n(t)}} \hat{M}_n \ket{\psi(t)}.
\end{equation}
Crucially, the measurement operators $\hat{M}_n$ need not be projectors. If they
are not, performing a generalized measurement immediately after a measurement
with outcome $n$ does generally not yield the same result $n$. As such, repeated
generalized measurements do not lead to a freezing of the dynamics.

\begin{figure}
  \centering
  \includegraphics[width = \linewidth]{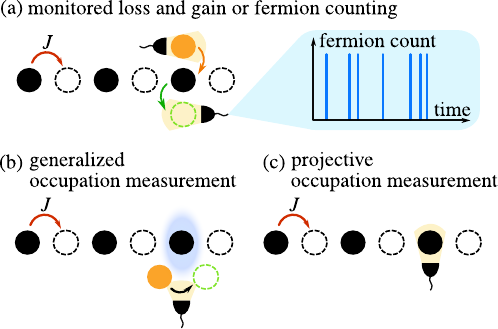}
  \caption{The two generalized measurement processes we consider are
    (a)~monitored loss and gain or fermion counting and (b)~generalized
    occupation measurements. (a)~Each site of the lattice (filled and empty
    circles) with hopping amplitude $J$ is coupled to two reservoirs acting as
    drain (empty green circle) and source (filled orange circle) of fermions,
    respectively. The occupations of the reservoirs are monitored continuously,
    such that each fermion leaving or entering the system is counted. (b)~Each
    site is coupled to a source and drain in a way that tunneling of reservoir
    fermions from the source to the drain is possible only if the system site is
    occupied. The detection of tunneling events thus indicates the presence of
    fermions in the system. Details are provided in Sec.~\ref{sec:models} and
    Appendices~\ref{sec:app-physical-models}
    and~\ref{sec:impl-gener-meas}. Generalized measurements in (a) and (b),
    which are implemented by coupling the system to auxiliary systems on which
    projective measurements are performed, should be contrasted with
    (c)~projective measurements of occupation numbers performed directly on the
    system.}
  \label{fig:1}
\end{figure}

Here, we describe a novel mechanism that stabilizes area-law entanglement
through measurements but does not require the freezing of the dynamics
associated with the conventional Zeno effect. We consider a one-dimensional (1D)
lattice model of free fermions subjected to monitored loss and gain as
illustrated in Fig.~\ref{fig:1}(a). That is, each lattice site is coupled to two
reservoirs, acting as drain and source of particles. The occupation of the
reservoirs is monitored continuously, such that each fermion that leaves or
enters the system is registered. This setup is analogous to photon counting in
quantum optics~\cite{Gardiner2014, Gardiner2015}. There, a physical system such
as an atom excited by a laser, or a leaky cavity, emits photons into the
surrounding electromagnetic field. The electromagnetic field thus acts as an
empty reservoir, and a photodetector continuously checks whether there is a
photon present. Even though each site of the fermionic lattice system we
consider is coupled to two reservoirs and can both ``emit'' and ``absorb''
fermions, the analogy to photon counting suggests to refer to monitored loss and
gain of fermions as \emph{fermion counting.}

The process of fermion counting can be theoretically modeled as a continuous
generalized measurement, where each measurement operator removes or adds a
fermion at a particular lattice site. A high rate of fermion counts causes the
number of fermions and thus the state of the system to fluctuate rapidly. At the
same time, the local removal and addition of particles efficiently disentangles
the many-body wave function similarly to generalized~\cite{Alberton2021} or
projective~\cite{Poboiko2023} measurements of local occupation numbers. These
latter two types of measurements, which are illustrated in Figs.~\ref{fig:1}(b)
and~(c), conserve the total number of particles in the system, and induce a Zeno
effect in the frequent-measurement limit.

In this work, we perform a comparative study of fermion counting and generalized
measurements of local occupation numbers to obtain a comprehensive understanding
of the similarities and differences between these scenarios. Despite their
starkly different dynamics, both models exhibit almost identical steady-state
correlations and entanglement properties. We trace this striking similarity back
to a generalized Zeno effect induced by fermion counting. The key feature the
conventional and generalized Zeno effects have in common is the suppression of
coherent dynamics, leading in turn to a suppression of the growth of
entanglement.

More formally, we explain the near indistinguishability of the correlations and
entanglement properties of the models through a common long-wavelength effective
field theory. In particular, both models are described by a nonlinear sigma
model (NLSM), which can be derived in the framework of replica Keldysh field
theory~\cite{Fava2023, Jian2023, Poboiko2023, Poboiko2024}. The predictions from
this analytical description are in good agreement with our numerical results and
provide strong evidence that both types of measurements lead to area-law
entanglement for any nonzero measurement rate $\gamma$.

Recent studies have led to contradictory results concerning the existence of a
measurement-induced entanglement transition in 1D free fermions with conserved
number of particles. Entanglement dynamics under continuous measurements of
occupation numbers, described by either quantum state
diffusion~\cite{Jacobs2006} or quantum jump trajectories~\cite{Wiseman2010} as
considered in this work, were studied in Ref.~\cite{Alberton2021}. For both
types of dynamics, there is numerical evidence for a measurement-induced
Kosterlitz-Thouless (KT) transition at a finite critical measurement rate
$\gamma_c$, separating a critical phase with logarithmic growth of the
entanglement entropy as characteristic for a one-dimensional conformal field
theory (CFT)~\cite{DiFrancesco1997} from an area-law phase. These findings
suggest that the precise way in which measurements are implemented might affect
nonuniversal properties such as the value of the critical measurement rate, but
not whether there is a transition or not. Therefore, one is led to assume that
this question is decided by the emergent long-wavelength behavior, which is
universal in the sense that it is determined solely by spatial dimensionality
and symmetries. In turn, symmetries are reflected in conservation laws and in
both quantum state diffusion and quantum jump trajectories, the number of
particles is the only conserved quantity. The existence of a measurement-induced
phase transition in one-dimensional free fermions with particle number
conservation is corroborated by replica Keldysh theory for Dirac fermions with a
linear dispersion relation~\cite{Buchhold2021}.

However, based on the above assumption of universality, we should compare the
results of Refs.~\cite{Alberton2021, Buchhold2021} to the findings of
Ref.~\cite{Poboiko2023}, which studied free fermions under random projective
measurements of occupation numbers. Analytical and numerical results obtained in
Ref.~\cite{Poboiko2023} indicate that there is no measurement-induced phase
transition, in agreement with numerical findings of Refs.~\cite{Coppola2022,
  Kells2023}. Instead, for any finite measurement rate $\gamma$, the logarithmic
growth of the entanglement entropy transitions into area-law behavior above an
exponentially large scale $\ln(l_{*}) \sim \gamma^{-1}$~\cite{Poboiko2023,
  Fava2024, Poboiko2025, Guo2024}. For small values of the measurement rate, for
which the theory is expected to become quantitatively accurate, this scale is
beyond numerically accessible system sizes.

Here, we show analytically and numerically that clear signatures of the
crossover to area-law entanglement can be observed on much smaller scales, well
within the reach of finite-size numerics. Approximately logarithmic growth of
the entanglement entropy is restricted to a critical range of length scales,
bounded from below by $l_0 \sim \gamma^{-1}$ and from above by
$l_c \sim \gamma^{-2}$. Only within this critical range, we observe clear
signatures of conformal invariance~\cite{Alberton2021}.

These results apply both to fermion counting and to generalized measurements of
local occupation numbers, and thus show that particle number conservation is, in
fact, not a necessary precondition for the observed phenomenology of stationary
correlations and entanglement. Instead, we show that the two necessary
requirements for the $\mathrm{SU}(R)$ symmetry underlying the NLSM to occur in
free fermionic systems are (i)~the conservation of the total number of particles
in the system and auxiliary reservoirs as illustrated in Fig.~\ref{fig:1}, and
(ii)~the conservation of the purity of the state of the system. The purity is
ensured by keeping a full record of the measurement outcomes. In particular, the
$\mathrm{SU}(R)$ symmetry can be broken by inefficient detection, which can be
modeled theoretically by averaging over a fraction of the measurement
results~\cite{Wiseman2010}. Combining measurements that obey conditions (i)~and
(ii)~with dynamics generated by a generic hopping Hamiltonian leads to an NLSM
with target manifold $\mathrm{SU}(R)$. Note that if the hopping matrix has
particle-hole symmetry---which is the case for the nearest-neighbor hopping with
real hopping amplitudes we consider in this work---the target manifold is
modified to $\mathrm{SU}(2 R) / \mathrm{Sp}(R)$~\cite{Jian2022, Fava2024,
  Poboiko2025}. However, both types of NLSMs have the same qualitative
properties.

Our work thus clarifies the impact of particle-number conservation on the
measurement-induced dynamics of pure states of 1D free fermionic systems: There
is no entanglement transition if the Hamiltonian does conserve the number of
particles, and measurements conserve the total number of particles in the system
and auxiliary reservoirs that are required to implement the measurements. This
includes the stronger condition that measurements conserve the number of
particles in the system alone. In contrast, there can be a transition if these
conditions are violated as in the Majorana model of Ref.~\cite{Fava2023}, where
the Hamiltonian and, depending on the choice of parameters, also the measurement
operators break particle-number conservation.

While we focus here on free fermionic systems, our finding that particle-number
conservation is not fundamental for the observed entanglement properties has far
reaching implications beyond this class of models. For example, random quantum
circuits with a conserved charge have been shown to exhibit a
measurement-induced charge-sharpening transition that separates phases in which
measurements can and cannot efficiently reveal the total charge of the
system~\cite{Agrawal2022, Barratt2022, Oshima2023, Agrawal2024}. Very recently,
charge sharpening has also been discussed for interacting
fermions~\cite{Guo2024, Poboiko2025}. Our findings show that charge sharpening
can occur even in the absence of charge conservation.

The rest of this paper is organized as follows. In Sec.~\ref{sec:key-results}, we summarize
our key results. The models we study are introduced in
Sec.~\ref{sec:models}. Then, in Sec.~\ref{sec:generalized-Zeno-effect}, we
describe signatures of the conventional and generalized Zeno effects for
occupation measurements and fermion counting, respectively. An analytical
description of our models in terms of a replica Keldysh field theory is
introduced in Sec.~\ref{sec:replica-keldysh-field-theory}. We analyze the
Gaussian field theory that applies in the weak-measurement limit in
Sec.~\ref{sec:gaussian-theory}, and discuss the effect of fluctuations beyond
the Gaussian theory in Sec.~\ref{sec:effective-field-theory}. A detailed
comparison between our analytical predictions and numerical results is provided
in Sec.~\ref{sec:correlations-and-entanglement}, where we consider spatial
correlations, measures of entanglement, and signatures of conformal invariance
in the steady state. Section~\ref{sec:conclusions-outlook} contains our
conclusions and an outlook on future research questions. Details of our
analytical and numerical studies are described in several appendices.

\section{Key results}
\label{sec:key-results}

\begin{figure}
  \centering
  \includegraphics{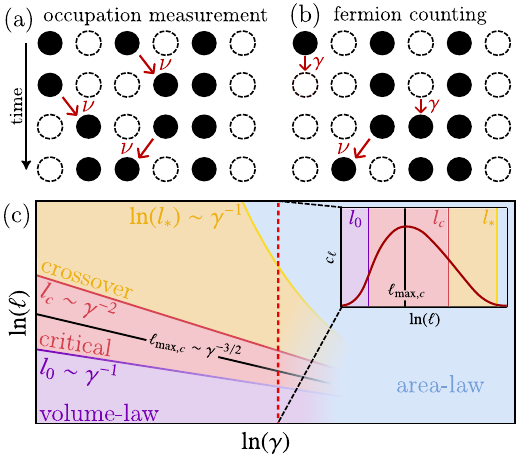}
  \caption{(a) Zeno effect for fermions on a one-dimensional lattice subjected
    to measurements of local occupation numbers. Frequent measurements freeze
    the system in a classical configuration with each lattice site either
    occupied or empty (filled and empty circles). Incoherent hopping occurs at
    the slow rate $\nu = J^2/\gamma$. (b) Generalized Zeno effect induced by
    fermion counting. A high rate $\gamma$ of fermion counts corresponds to fast
    transitions between classical configurations. As in the conventional Zeno
    effect, additional slow processes occur at the rate $\nu$. (c) Schematic
    finite-size phase diagram for fermion counting and occupation
    measurements. For low measurement rates, the entanglement exhibits
    volume-law scaling below $l_0 \sim \gamma^{-1}$ and grows logarithmically in
    the critical range between $l_0$ and $l_c \sim \gamma^{-2}$, before it
    slowly crosses over to area-low scaling, which is established beyond the
    exponentially large scale $\ln(l_{*}) \sim \gamma^{-1}$. Inset: the
    effective central charge $c_{\ell}$ has a maximum at
    $\ell_{\mathrm{max}, c} \sim \gamma^{-3/2}$.}
  \label{fig:2}
\end{figure}

We have obtained the following results through analytical and numerical studies
of free fermions on a 1D lattice subjected to fermion counting or generalized
measurements of local lattice site occupation numbers:

\paragraph{Frequent particle loss and gain suppress coherent dynamics and
  stabilize area-law entanglement via a generalized Zeno effect.}

Before we explain the generalized Zeno effect, let us briefly recall the
conventional Zeno effect for a single particle on a one-dimensional lattice,
subjected to repeated generalized measurements of lattice site occupation
numbers, $\hat{M}_l = \hat{n}_l$. The first measurement collapses the initial
wave function $\ket{\psi}$ of the particle to a state $\ket{l_0}$ that is
localized on a single lattice site $l_0$, chosen from $l \in \{ 1, \dotsc, L \}$
according to Born's rule with probability
$p_l = \braket{\psi | \hat{n}_l | \psi}$. Subsequent measurements, performed at
a rate $\gamma$, yield the same result $l_0$, since the probability distribution
after the first collapse is $p_l = \delta_{l, l_0}$. Coherent hopping with
amplitude $J \ll \gamma$ can induce transitions $\ket{l_0} \to \ket{l_0 \pm 1}$.
However, these processes occur at second order in perturbation theory, in other words
with a small rate $\nu = J^2/\gamma$. This suppression of coherent dynamics
through frequent measurements is the essence of the conventional Zeno
effect. The resultant slow dynamics are illustrated for a many-body system in
Fig.~\ref{fig:2}(a).

To illustrate the generalized Zeno effect, we first consider a single fermionic
lattice site with annihilation and creation operators $\hat{\psi}$ and
$\hat{\psi}^{\dagger}$, respectively. Monitored loss and gain or fermion
counting is described by generalized measurements performed at a rate $\gamma$
and with measurement operators $\hat{M}_- = \hat{\psi}$ and
$\hat{M}_+ = \hat{\psi}^{\dagger}$. The completeness relation
$\hat{M}_-^{\dagger} \hat{M}_- + \hat{M}_+^{\dagger} \hat{M}_+ = 1$ follows from
canonical anticommutation relations. Repeated measurements cause the occupation
of the lattice site to fluctuate between 0 and 1, realizing a random telegraph
process~\cite{Gardiner2009}. The autocorrelation function of the occupation
number $\hat{n} = \hat{\psi}^{\dagger} \hat{\psi}$ decays exponentially with a
rate $\sim \gamma$. That is, the dynamics become faster for increasing
measurement rate $\gamma$---the exact opposite of the freezing of the dynamics
in the conventional Zeno effect.

In an extended lattice system with monitored loss and gain on each lattice site,
the dynamics induced by measurements allow the system to ergodically explore all
classical configurations of $N \in \{ 0, \dotsc, L \}$ particles distributed on
$L$ lattice sites. Coherent hopping enables additional transitions between these
configurations. Crucially, in analogy to the conventional Zeno effect, these
transitions are suppressed for $J \ll \gamma$ and occur with a slow rate
$\nu = J^2/\gamma \ll \gamma$.  Therefore, we refer to this phenomenon, which is
illustrated in Fig.~\ref{fig:2}(b), as a generalized Zeno effect. The facts that
coherent dynamics are suppressed by frequent fermion counts and that the
classical configurations, which dominate the dynamics for $J \ll \gamma$, have
vanishing entanglement entropy rationalize that fermion counting stabilizes
area-law entanglement.

\paragraph{The generalized and conventional Zeno effects induce almost identical
  correlations and entanglement properties.}

Even though the dynamics are strikingly different for fermion counting and
occupation measurements, both lead to equivalent behavior in various measures of
correlations and entanglement in the steady state. This includes density-density
correlations, the entanglement entropy, and the bipartite mutual information. We
observe a qualitative difference only in higher-order correlations as quantified
through the tripartite mutual information, indicating that the tripartite mutual
information probes properties that are not captured by the long-wavelength
effective field theory, which is the same for both models as described next.

\paragraph{The similarity of correlations and entanglement properties can be
  explained in terms of a universal long-wavelength effective field theory.}

Starting from a microscopic model of the fermionic lattice system shown in
Fig.~\ref{fig:1}(a), with each site coherently coupled to particle reservoirs
acting as drain and source, a natural description of the dynamics of the system
alone can be given in terms of a stochastic Schr\"odinger
equation~\cite{Wiseman2010}. This equation describes the evolution of the state
$\ket{\psi(t)}$ of the system conditioned on the counting of fermions that are
transferred between the system and reservoirs. For equal rates of fermion loss
and gain, the same quantum trajectories $\ket{\psi(t)}$ are obtained from a
description of fermion counting as generalized measurements that are performed
at random times and with outcomes determined by Born's rule. This equivalence
enables an analytical description of the system dynamics in terms of a replica
Keldysh field theory~\cite{Fava2023, Jian2023, Poboiko2023, Poboiko2024,
  Fava2024, Guo2024, Chahine2024, Poboiko2025}, which we adopt to the case of
generalized measurements.

The field theory approach gives access to two types of observables:
(i)~Observables that are linear in the projector on the conditional state
$\hat{\rho}(t) = \ket{\psi(t)} \bra{\psi(t)}$, such as expectation values
$\overline{\left\langle \hat{A}(t) \right\rangle}$ or temporal correlation
functions $\overline{\left\langle \hat{A}(t) \hat{B}(t') \right\rangle}$, where
the overbar denotes the average over trajectories; (ii)~instantaneous
observables that are nonlinear in $\hat{\rho}(t)$, such as connected equal-time
correlation functions
$\overline{\left\langle \hat{A}(t) \hat{B}(t) \right\rangle - \left\langle
    \hat{A}(t) \right\rangle \left\langle \hat{B}(t) \right\rangle}$
or the entanglement entropy. Observables of type (i)~are encoded in the
replica-symmetric sector of the theory, and observables of type (ii)~in the
replica-asymmetric or replicon sector.

Fermion counting leads to exponential decay of the density autocorrelation
function, which is an observable of type (i). Instead, particle-number
conservation in the model with occupation measurements leads to diffusive
decay. Formally, this difference between exponential and algebraic decay is
reflected in the replica-symmetric sector being massive and massless for fermion
counting and occupation measurements, respectively. In contrast, the replicon
sector is described in both cases by the same massless long-wavelength effective
field theory that takes the form of an NLSM. While the Hamiltonian we consider
obeys particle-hole symmetry, leading to an NLSM with target manifold
$\mathrm{SU}(2 R) / \mathrm{Sp}(R)$, the target manifold for the generic case is
$\mathrm{SU}(R)$. Crucially, the qualitative behavior remains unchanged. This
common long-wavelength effective field theory description analytically explains
the numerically observed equivalence of correlations and entanglement
properties. Importantly, the theory predicts that there is no
measurement-induced entanglement transition in these models. Instead, area-law
entanglement is established for any nonzero measurement rate.

\paragraph{Particle-number conservation is not necessary for the occurrence of
  the $\mathrm{SU}(R)$ symmetry underlying the NLSM, and thus the absence of an
  entanglement transition.}

A peculiar feature of measurement-induced phenomena is that both the dynamics
and the observables of main interest are nonlinear in the quantum state of the
system. This nonlinearity of the dynamics is due to the necessity to normalize
the state after each measurement; the key nonlinear observable is the
entanglement entropy. Nevertheless, it is possible to obtain a theoretical
description that is linear in the state by introducing $R$ replicas of the
system~\cite{Jian2023, Bao2020}. Consequently, it is not sufficient to study the
symmetries of the unitary time evolution operator and measurement operators of a
single realization of the system to fully understand relevant symmetries of the
dynamics. Instead, one has to study symmetries of the dynamics of $R$ replicas
to obtain a comprehensive picture.

If the unitary dynamics and measurement operators of a single realization of the
system of interest have the symmetry group $\mathrm{G}$, then the symmetry group
of $R$ replicas is $\mathrm{G}^R \rtimes \mathrm{S}_R$, where $\mathrm{S}_R$ is
the group of permutations of $R$ replicas. Conservation of the number of
particles corresponds to $\mathrm{G} = \mathrm{U}(1)$---for the moment, we
disregard a possible additional particle-hole symmetry of the
Hamiltonian---leading to the symmetry group
$\mathrm{U}(1)^R \rtimes \mathrm{S}_R$ for generic interacting fermions. This is
enhanced to $\mathrm{U}(R)$ for free fermions, where it is possible to not only
perform discrete permutations of replicas but even continuous rotations between
them~\cite{Jian2022, Buchhold2021, Bao2021, Fava2023, Poboiko2023}. For example,
the dynamics of $R$ replicas of free fermions subjected to random projective
measurements of lattice site occupation numbers studied in
Ref.~\cite{Poboiko2023} are invariant under $\mathrm{U}(R)$ rotations. In a
replica Keldysh field theory description of this model, the decomposition
$\mathrm{U}(R) = \mathrm{U}(1) \rtimes \mathrm{SU}(R)$ has been found to emerge
naturally: Goldstone modes associated with $\mathrm{U}(1)$ and $\mathrm{SU}(R)$
determine the long-wavelength behavior of the replica-symmetric and replicon
sectors of the theory, respectively, and these sectors decouple for sufficiently
rare measurements, $\gamma \ll J$.  However it has not been recognized before,
that in the above decomposition only the first factor, $\mathrm{U}(1)$, requires
particle-number conservation; as we show, $\mathrm{SU(R)}$ symmetry is an exact
property, realized on the microscopic level of the theory, even for fermion
counting, where the number of particles is not conserved. Note, finally, that an
additional particle-hole symmetry of the hopping Hamiltonian can be incorporated
in the field theory description by introducing a combined $2 R$-dimensional
particle-hole and replica space~\cite{Poboiko2025}. Then, the microscopic
$\mathrm{SU}(R)$ symmetry is enhanced to $\mathrm{SU}(2 R)$, which becomes
$\mathrm{U}(2 R) = \mathrm{U}(1) \rtimes \mathrm{SU}(2 R)$ for measurements that
conserve the number of particles.

Our finding, that the $\mathrm{SU}(R)$ symmetry underlying the emergence of an
NLSM at long wavelengths does not require particle-number conservation,
indicates that phenomena that have so far been discussed for systems with
particle-number conservation can, in fact, occur also when particle-number
conservation is broken. A case in point is the charge-sharpening transition in
random quantum circuits~\cite{Agrawal2022, Barratt2022, Oshima2023, Agrawal2024}
and interacting fermionic systems~\cite{Guo2024, Poboiko2025}.

\paragraph{The necessary conditions for the $\mathrm{SU}(R)$ symmetry to occur
  are (i)~conservation of the total number of particles in the system and
  auxiliary reservoirs, and (ii)~preservation of the purity of the state.}

The implementation of fermion counting illustrated in Fig.~\ref{fig:1}(a) relies
on coupling each site of the fermionic lattice system to two auxiliary
reservoirs, such that the total number of particles in the system and reservoirs
is conserved. We show that under this condition and the additional condition
that all measurement outcomes are recorded such that the state of the system
remains pure, the dynamics of $R$ replicas feature the $\mathrm{SU}(R)$ symmetry
underlying both the $\mathrm{SU}(2R)/\mathrm{Sp}(R)$ and $\mathrm{SU}(R)$ NLSMs
for hopping Hamiltonians with and without particle-hole symmetry,
respectively. In contrast, the symmetry is broken for inefficient detection,
meaning some of the fermions that are transferred between the system and
reservoirs remain undetected. These two conditions, therefore, provide a
physical explanation for the observed phenomenology in terms of conservation
laws, which complements the more heuristic reasoning in terms of the generalized
Zeno effect. Furthermore, by identifying these conditions, we settle the
question about the requirements to observe a measurement-induced entanglement
transition in one-dimensional free fermions. Let us reiterate that such a
transition does not occur for particle-number-conserving models of random
generalized~\cite{Alberton2021} or projective measurements~\cite{Poboiko2023},
or for fermion counting with a conserved total number of particles. Instead,
such a transition does occur in the Majorana model of Ref.~\cite{Fava2023},
where particle-number conservation is broken by the Hamiltonian that generates
the unitary dynamics and, depending on the choice of parameters, also by the
measurement operators. We find the long-wavelength effective replicon field
theory to be identical for fermion counting as well as particle-number
conserving measurements of occupation numbers~\cite{Poboiko2023, Poboiko2024,
  Chahine2024, Fava2024, Guo2024, Poboiko2025}, but different for the Majorana
model~\cite{Fava2023}.

\paragraph{Logarithmic growth of the entanglement entropy and signatures of
  conformal invariance are observable within a well-defined and finite critical
  range of length scales.}

The renormalization-group (RG) flows of the $\mathrm{SU}(2R)/\mathrm{Sp}(R)$ and
$\mathrm{SU}(R)$ NLSMs are qualitatively the same and indicate that area-law
entanglement is established beyond a scale $l_{*}$ that is exponentially large
in $\gamma^{-1}$. However, the RG-corrected Gaussian field theory predicts that
for small measurement rates $\gamma$, properties that are characteristic for a
critical phase can be observed within a finite critical range of length scales
between $l_0 \sim \gamma^{-1}$ and $l_c \sim \gamma^{-2}$ as illustrated in
Fig.~\ref{fig:2}(c). These characteristics include logarithmic growth of the
entanglement entropy and signatures of conformal
invariance~\cite{Alberton2021}. Beyond $l_c$, there is a wide crossover region
that separates the critical range from the onset of area-law scaling. Since
$l_c$ is only algebraically large in $\gamma$, this scale is numerically
accessible also for relatively small values of the measurement rate, even if
area-law scaling beyond $l_{*}$ is not observable.

To precisely characterize the behavior of the entanglement entropy, we introduce
a scale-dependent effective central charge $c_{\ell}$ for a subsystem of size
$\ell$. On short scales $\ell \lesssim l_0 \sim \gamma^{-1}$, the effective
central charge $c_{\ell}$ grows with $\ell$, indicating volume-law scaling of
the entanglement entropy; on an intermediate scale
$\ell = \ell_{\mathrm{max}, c} \sim \gamma^{- 3/2}$, the central charge exhibits
a maximum and thus becomes stationary, leading to logarithmic growth of the
entanglement entropy; on large scales $\ell \gtrsim l_c \sim \gamma^{-2}$, the
central charge decreases; and finally, beyond the exponentially large scale
$l_{*}$, $c_{\ell}$ is expected to vanish, corresponding to area-law
entanglement. The numerically observed behavior of the effective central charge,
illustrated schematically in Fig.~\ref{fig:2}(c), is in good agreement with our
analytical predictions.

Evidence for emergent conformal invariance is provided by the collapse of the
mutual information as a function of the cross ratio with the form predicted by
conformal field theory~\cite{Calabrese2009, Alberton2021}. We perform a
systematic numerical analysis of the mutual information for varying subsystem
sizes, which reveals that this collapse occurs only for subsystem sizes within
the critical range, corroborating that conformal invariance is violated at both
short and large scales.

The results described above apply equally to fermion counting and occupation
measurements. Differences between the two types of generalized measurements and
deviations from conformal invariance even within the critical range can be seen
in the tripartite mutual information.

Our detailed analytical and numerical analysis of the crossover from the
critical to the area-law phase corroborates and refines the conclusion of
Ref.~\cite{Poboiko2023}, that the evidence for an entanglement transition of
free fermions presented in Ref.~\cite{Alberton2021} describes, in fact, a
finite-size crossover phenomenon.

\section{Models and time evolution}
\label{sec:models}

We consider two models of noninteracting fermions on a 1D lattice
undergoing unitary time evolution interspersed with generalized measurements. In
both models, the unitary dynamics are generated by the Hamiltonian
\begin{equation}
  \label{eq:Hamiltonian}
  \hat{H} = - J \sum_{l = 1}^L \left( \hat{\psi}_l^{\dagger} \hat{\psi}_{l +
      1}^{\vphantom{\dagger}} + \hat{\psi}_{l + 1}^{\dagger} \hat{\psi}_l^{\vphantom{\dagger}} \right),
\end{equation}
where $\hat{\psi}_l$ and $\hat{\psi}_l^{\dagger}$ are fermionic annihilation and
creation operators, respectively, on lattice sites $l \in \{ 1, \dotsc, L \}$
with periodic boundary conditions, $\hat{\psi}_{L + 1} = \hat{\psi}_1$. The two
models differ by the types of measurement processes, illustrated in Figs.~1(a)
and~(b), respectively: in the first model, we consider monitored loss and gain
or fermion counting; in the second model, we consider measurements of local
occupation numbers,
$\hat{n}_l^{\vphantom{\dagger}} = \hat{\psi}_l^{\dagger}
\hat{\psi}_l^{\vphantom{\dagger}}$.
Our main focus lies on the first model, with the second serving as a
reference. Entanglement dynamics under continuous and random measurements of
local occupation numbers have been studied in Refs.~\cite{Cao2019,
  Alberton2021}, and under projective random measurements in
Ref.~\cite{Poboiko2023}.

In the following, we present two equivalent ways to describe quantum
trajectories, that is, the evolution of the state vector of the fermionic
many-body system, conditioned on a sequence of measurement outcomes. The first
description in terms of a stochastic Schr\"odinger equation results naturally
from microscopic physical models, such as implementing
monitored loss and gain by coupling the system of interest to reservoirs, as illustrated in
Fig.~\ref{fig:1}~\cite{Wiseman2010}. In the second more formal description,
generalized measurements are performed directly on the system at an externally
imposed constant rate. This description has the technical advantage of lending
itself naturally toward a reformulation as a replica Keldysh field theory,
generalizing the construction of Ref.~\cite{Poboiko2023} for projective
measurements. Crucially, the statistics of measurement times and outcomes, as well as
the resulting quantum trajectory dynamics, are identical in both descriptions.

\subsection{Stochastic Schr\"odinger equation}
\label{sec:stochastic-schrodinger-equation}

A minimal physical model for fermion counting consists of a quantum dot, meaning
a single fermionic lattice site, tunnel-coupled to two fermionic reservoirs.  We
further assume these reservoirs to be in thermodynamic equilibrium at a low
temperature. The chemical potentials of the reservoirs are chosen such that
fermions can tunnel from the quantum dot to the first reservoir, the drain, and
from the second reservoir, the source, to the quantum dot, but the reverse
processes are inhibited. We obtain a theoretical description of fermion counting
by integrating the Schr\"odinger equation for the quantum dot and reservoirs in
discrete time steps $\Delta t$ using the Born and Markov approximations, in
other words treating the coupling to the reservoirs perturbatively and assuming
the reservoirs to have short correlation times as detailed in
Appendix~\ref{sec:app-physical-models}. At each time step, the occupations of
drain and source are measured projectively. These measurements indirectly count
the number of fermions leaving or entering the quantum dot: finding states of
the drain and source to be occupied and empty, respectively, indicates that a
fermion must have left or entered the quantum dot. A description of the dynamics
of the quantum dot alone can be obtained by modeling the sequence of measurement
results through appropriate stochastic processes. In the continuous-time limit
$\Delta t \to \diff t$ and generalizing the setup to $L$ fermionic lattice
sites, each in contact with its own drain and source as sketched in
Fig.~\ref{fig:1}(a), we obtain a stochastic Schr\"odinger equation for the state
$\ket{\psi(t)}$ of the fermionic lattice system,
\begin{multline}
  \label{eq:stochastic-schroedinger-equation}
  \diff \ket{\psi(t)} = \left( \left\{ - \imag \hat{H} - \sum_{\alpha = \pm}
      \sum_{l = 1}^L \left[ \hat{L}_{\alpha, l}^{\dagger} \hat{L}_{\alpha,
          l}^{\vphantom{\dagger}} - \tilde{p}_{\alpha, l}(t) \right] \right\}
    \diff t \right. \\ \left. + \sum_{\alpha = \pm} \sum_{l = 1}^L \left[
      \frac{\hat{L}_{\alpha, l}}{\sqrt{\tilde{p}_{\alpha, l}(t)}} - 1 \right]
    \diff N_{\alpha, l}(t) \right) \ket{\psi(t)}.
\end{multline}
The first line describes deterministic dynamics, governed by the Hamiltonian
$\hat{H}$ in Eq.~\eqref{eq:Hamiltonian} and the jump operators
\begin{equation}
  \label{eq:jump-operators-FC}
  \hat{L}_{\lind, l}^{\vphantom{\dagger}} = \sqrt{\gamma_{\lind}/2} \, \hat{\psi}_l^{\vphantom{\dagger}}, \qquad
  \hat{L}_{\gind, l}^{\vphantom{\dagger}} = \sqrt{\gamma_{\gind}/2} \, \hat{\psi}_l^{\dagger},
\end{equation}
where the loss and gain rates $\gamma_-$ and $\gamma_+$, respectively, are
determined by microscopic parameters of the model as specified in
Appendix~\ref{sec:app-physical-models}. Nonnormalized probabilities are defined
as
\begin{equation}
  \label{eq:nonnormalized-probabilities-FC}
  \tilde{p}_{\alpha, l}(t) = \left\langle \psi(t) \middle| \hat{L}_{\alpha,
      l}^{\dagger} \hat{L}_{\alpha, l}^{\vphantom{\dagger}} \middle| \psi(t) \right\rangle.
\end{equation}
The second line of Eq.~\eqref{eq:stochastic-schroedinger-equation} incorporates
the effect of measurements, where $\diff N_{\alpha, l}(t) = 0, 1$ are stochastic
increments. When $\diff N_{\alpha, l}(t) = 0$, no jump occurs and the wave
function evolves continuously. In contrast, $\diff N_{\alpha, l}(t) = 1$ indicates
a jump at time $t$ and site $l$. For $\alpha = -$, this jump corresponds to
loss of a fermion; the jump describes gain of a fermion for $\alpha = +$. After
a jump, the state of the system is
\begin{equation}
  \label{eq:quantum-jump}
  \ket{\psi(t + 0^+)} = \frac{1}{\sqrt{\tilde{p}_{\alpha,l}(t)}} \hat{L}_{\alpha, l}
  \ket{\psi(t)}.
\end{equation}
The stochastic increments $\diff N_{\alpha, l}(t)$ obey a Poisson point process
with mean
$\overline{\diff N_{\alpha, l}(t)} = 2 \tilde{p}_{\alpha, l}(t) \diff
t$~\cite{Wiseman2010}.
For the case of equal loss and gain rates,
$\gamma = \gamma_{\lind} = \gamma_{\gind}$, the rate of jumps per lattice site
is constant in time and given by
\begin{equation}
  \label{eq:jump-rate-FC}
  \frac{1}{L} \sum_{\alpha = \pm} \sum_{l = 1}^L \frac{\overline{\diff
      N_{\alpha, l}}}{\diff t} = \frac{2}{L} \sum_{\alpha = \pm} \sum_{l = 1}^L
  \tilde{p}_{\alpha, l}(t) = \gamma,
\end{equation}
leading to an exponential distribution of waiting times between jumps. To
integrate Eq.~\eqref{eq:stochastic-schroedinger-equation} numerically, we employ
a higher-order quantum jump algorithm~\cite{Daley2014}. In this algorithm,
waiting times are sampled from the exponential distribution, and the type of
jump that occurs is chosen randomly according to the normalized probabilities
given by
\begin{equation}
  \label{eq:probabilities-FC}
  p_{-, l}(t) = \left\langle \hat{n}_l(t) \right\rangle/L, \qquad p_{+, l}(t) =
  \left( 1 - \left\langle \hat{n}_l(t) \right\rangle \right) \! /L.
\end{equation}
Between jumps, the system undergoes unitary time evolution described by the
Hamiltonian $\hat{H}$.

The occupation of a quantum dot can be measured through a modification of the
above setup: instead of tunneling between the dot and reservoirs, we now
consider direct tunneling from the source to the drain, with the tunneling
amplitude proportional to the occupation number of the quantum
dot~\cite{Wiseman2010, Gustavsson2009}. Then, as described in
Appendix~\ref{sec:app-physical-models}, the detection of fermions in the drain
indirectly signals that also the quantum dot is occupied. In the resulting
stochastic Schr\"odinger equation for the extended lattice system shown in
Fig.~\ref{fig:1}(b), there is only a single type of jump operator at each site,
\begin{equation}
  \label{eq:jump-operators-OM}
  \hat{L}_l = \sqrt{\gamma} \hat{n}_l.
\end{equation}
The rate of jumps per lattice site depends on the number of particles $N$,
\begin{equation}
  \label{eq:jump-rate-OM}
  \frac{1}{L} \sum_{l = 1}^L \frac{\overline{\diff N_l}}{\diff t} = \frac{2 \gamma
    N}{L},
\end{equation}
and, given a jump occurs at time $t$, the lattice site $l$ at which the jump
operator Eq.~\eqref{eq:jump-operators-OM} is applied is
chosen with probability
\begin{equation}  
  \label{eq:probabilities-OM}
  p_l(t) = \langle \hat{n}_l(t) \rangle/N.
\end{equation}

The key difference between fermion counting and occupation measurements is the
conservation of the number of particles in the latter case. For fermion
counting, each quantum jump decreases or increases the number of particles by
one. Nevertheless, a meaningful quantitative comparison between the two models
is enabled by choosing parameters such that the mean number of particles in the
steady state and the rate of quantum jumps are the same. This is achieved by
setting $\gamma = \gamma_{\lind} = \gamma_{\gind}$ and choosing the initial
state as
\begin{equation}
  \label{eq:charge-density-wave}
  \ket{\psi_0} = \prod_{l = 1}^{L/2} \hat{\psi}_{2 l - 1}^{\dagger} \ket{0},
\end{equation}
which contains $N = L/2$ particles, corresponding to a fermion density of
$n = N/L = 1/2$. Then, according to Eqs.~\eqref{eq:jump-rate-FC}
and~\eqref{eq:jump-rate-OM}, the rate of quantum jumps per
lattice site is for both models given by $\gamma$. While we have obtained all of
our numerical results for $n = 1/2$, we will keep the explicit dependence on $n$
in analytical expressions.

In solving Eq.~\eqref{eq:stochastic-schroedinger-equation} numerically, a great
simplification results from the initial state,
Eq.~\eqref{eq:charge-density-wave}, being Gaussian and the dynamics generated by the
quadratic Hamiltonian, Eq.~\eqref{eq:Hamiltonian}, and the jump operators in
Eqs.~\eqref{eq:jump-operators-FC}
and~\eqref{eq:jump-operators-OM} preserving this property.
Gaussian states are fully determined by the $L \times L$ single-particle density
matrix
\begin{equation}
  \label{eq:single-particle-density-matrix}
  D_{l, l'}^{\vphantom{\dagger}}(t) = \left\langle \psi(t) \middle| \hat{\psi}_{l'}^{\dagger}
    \hat{\psi}_l^{\vphantom{\dagger}} \middle| \psi(t) \right\rangle.
\end{equation}
A quantum jump at lattice site $m$ as described by Eq.~\eqref{eq:quantum-jump}
modifies the single-particle density matrix as
\begin{equation}
  D_{l, l'}(t + 0^+) = \frac{\left\langle \psi(t) \middle| \hat{L}_{\alpha,
        m}^{\dagger} \hat{\psi}_{l'}^{\dagger} \hat{\psi}_l^{\vphantom{\dagger}}
      \hat{L}_{\alpha, m}^{\vphantom{\dagger}} \middle| \psi(t)
    \right\rangle}{\left\langle \psi(t) \middle| \hat{L}_{\alpha,
        m}^{\dagger} \hat{L}_{\alpha, m}^{\vphantom{\dagger}} \middle| \psi(t) \right\rangle}.
\end{equation}
By using Wick's theorem, the expectation values on the right-hand side can be
expressed in terms of $D_{l, l'}(t)$, which leads to a closed algorithm for
the stochastic dynamics of the single-particle density matrix.

The stochastic Schr\"odinger
equation~\eqref{eq:stochastic-schroedinger-equation} describes the evolution of
the state vector $\ket{\psi(t)}$, conditioned on a sequence of measurement
outcomes. Averaging over the stochastic increments $\diff N_{\pm, l}(t)$ yields a
quantum master equation in Lindblad form that describes the unconditional
evolution of the density matrix
$\overline{\hat{\rho}}(t) = \overline{\ket{\psi(t)}\bra{\psi(t)}}$. In the
unconditional dynamics, occupation measurements lead to dephasing between
lattice sites, which in turn induces heating to infinite temperature in the
steady state, $\overline{\hat{\rho}}_{\mathrm{ss}} = \hat{P}_N/\tr(\hat{P}_N)$,
where $\hat{P}_N$ is the projector on the subspace with $N$ particles. For
fermion counting the steady state is
$\overline{\hat{\rho}}_{\mathrm{ss}} = \prod_{l = 1}^L \left( 1 +
  \frac{\gamma_{\gind} - \gamma_{\lind}}{\gamma_{\gind} + \gamma_{\lind}}
  \hat{n}_l \right)$,
which reduces to $\overline{\hat{\rho}}_{\mathrm{ss}} = 1/2^L$ for our choice
$\gamma_{\lind} = \gamma_{\gind}$~\cite{Starchl2022, Starchl2024}. As a consequence,
the unconditional steady state is completely featureless for
both models,
irrespective of the value of the jump rate $\gamma$. Consequently, observables
that are linear in the state, such as expectation values
$\overline{\left\langle \hat{A}(t) \right\rangle} = \tr \! \left( \hat{A}
  \overline{\hat{\rho}}(t) \right)$,
do not exhibit any nontrivial effects of continuous monitoring. To see such
effects, it is necessary to consider observables that are nonlinear in the
projector on the conditional state
$\hat{\rho}(t) = \ket{\psi(t)} \bra{\psi(t)}$.

\subsection{Interpretation as random generalized measurements}
\label{sec:random-generalized-measurements}

An alternative description of the dynamics under continuous monitoring that is
equivalent to the stochastic Schr\"odinger
equation~\eqref{eq:stochastic-schroedinger-equation} can be given by using the
framework of generalized measurements introduced in Sec.~\ref{sec:introduction}.
Specifically, we consider unitary dynamics generated by the Hamiltonian
Eq.~\eqref{eq:Hamiltonian}, interspersed with generalized measurements. The
times at which measurements are performed and the measurement operators have to
be chosen such that the effect of measurements on the quantum state described by
Eq.~\eqref{eq:GM-state-update} and the statistics of measurement times and
outcomes reproduce the corresponding properties of quantum jumps in quantum
trajectories described by the stochastic Schr\"odinger
equation~\eqref{eq:stochastic-schroedinger-equation}.

As explained above, jumps occur at the rate $\gamma$ per lattice
site. Therefore, the number of jumps during the evolution from time $t_0$ to $t$
obeys a Poisson distribution with mean $\gamma L T$ where $T = t - t_0$, and the
times of individual jumps are distributed uniformly within the interval
$[t_0, t]$. We choose the times at which measurements are performed accordingly.

Monitored loss and gain or fermion counting, as described by the jump operators
in Eq.~\eqref{eq:jump-operators-FC}, corresponds to generalized measurements
with the measurement operators
\begin{equation}
  \label{eq:measurement-ops-FC}
  \hat{M}_{-, l} = \frac{1}{\sqrt{L}} \hat{\psi}_l, \qquad \hat{M}_{+, l} =
  \hat{M}_{-, l}^{\dagger} = \frac{1}{\sqrt{L}} \hat{\psi}_l^{\dagger}.
\end{equation}
These measurement operators differ from the corresponding jump operators in
Eq.~\eqref{eq:jump-operators-FC} only through prefactors, which are chosen to
obey the completeness relation
$\sum_{\alpha = \pm} \sum_{l = 1}^L \hat{M}_{\alpha, l}^{\dagger}
\hat{M}_{\alpha, l}^{\vphantom{\dagger}} = 1$.
This implies that the effect of generalized measurements and quantum jumps on
the state, described by Eqs.~\eqref{eq:GM-state-update}
and~\eqref{eq:quantum-jump}, respectively, are the same. Furthermore, the
probabilities
$p_{\alpha, l}^{\vphantom{\dagger}} = \left\langle \hat{M}_{\alpha, l}^{\dagger}
  \hat{M}_{\alpha, l}^{\vphantom{\dagger}} \right\rangle$
of different measurement outcomes agree with the corresponding probabilities of
the quantum jumps in Eq.~\eqref{eq:probabilities-FC}.

The reformulation of quantum trajectory dynamics with the jump operators in
Eq.~\eqref{eq:jump-operators-OM} as generalized measurements
of occupation numbers relies on the fact that the dynamics are restricted to a
subspace with a fixed number of particles $N$, which allows us to choose the
measurement operators as
\begin{equation}
  \label{eq:measurement-ops-OM}
  \hat{M}_l = \frac{1}{\sqrt{N}} \hat{n}_l.
\end{equation}
These operators obey the completeness relation
$\sum_{l = 1}^L \hat{M}_l^{\dagger} \hat{M}_l^{\vphantom{\dagger}} = \hat{N}/N = 1$, where the
last equality holds in the restriction to the relevant subspace in which the
particle number operator $\hat{N} = \sum_{l = 1}^L \hat{n}_l$ reduces to the
number $N$. Again, the measurement operators are proportional to the
corresponding jump operators, implying that they induce the same change of the
state, and the probabilities
$p_l^{\vphantom{\dagger}} = \left\langle \hat{M}_l^{\dagger} \hat{M}_l^{\vphantom{\dagger}} \right\rangle$ of
different measurement outcomes agree with the corresponding probabilities of
the quantum jumps in
Eq.~\eqref{eq:probabilities-OM}.

Note that in the reformulation of quantum jumps as generalized measurements, the
lattice site index $l$ is the \emph{outcome} of a generalized measurement and,
therefore, determined by the quantum state, and only the measurement times are
imposed externally. In contrast, in the model considered in
Ref.~\cite{Poboiko2023}, both the measurement times and the lattice sites at
which projective measurements of occupation numbers are performed are chosen
independently from the state of the system. The measurement operators for such
local projective measurements as illustrated in Fig.~\ref{fig:1}(c) are
$\hat{M}_{0, l} = 1 - \hat{n}_l$ and $\hat{M}_{1, l} = \hat{n}_l$, and obey the
local completeness relation $\hat{M}_{0, l} + \hat{M}_{1, l} = 1$, reflecting
that the measurement operators are projectors.

The physical difference between the generalized measurements of occupation
numbers described by Eq.~\eqref{eq:measurement-ops-OM} and projective
measurements is best understood by considering a single particle on a lattice in
a state $\ket{\psi}$. Generalized occupation measurements amount to asking:
``Where is the particle?'' The answer is a particular lattice site $l$, with
probability $p_l = \braket{\psi | \hat{n}_l | \psi}$. In contrast, performing a
projective measurement of $\hat{n}_l$ at a given lattice site $l$ means asking
the question: ``Is the particle at this site?'' The answer is yes or know, with
probabilities $p_1 = \braket{\psi | \hat{n}_l | \psi}$ and $p_0 = 1 - p_1$,
respectively.

The fact that the lattice site $l$ is the outcome of the measurement is crucial
for generalized measurements of occupation numbers, for which the distribution
of lattice sites in Eq.~\eqref{eq:probabilities-OM} is not uniform in space. In
contrast, for fermion counting with probabilities given in
Eq.~\eqref{eq:probabilities-FC}, the marginal distribution
$p_l(t) = p_{-, l}(t) + p_{+, l}(t) = 1/L$ is uniform.

We finish this discussion by noting that any generalized measurement on a given system can be implemented by coupling the
system to auxiliary systems through unitary operations and performing a
projective measurement on the auxiliary systems~\cite{Nielsen2010}. We describe
how this is achieved for the measurement operators in
Eqs.~\eqref{eq:measurement-ops-FC} and~\eqref{eq:measurement-ops-OM} in
Appendix~\ref{sec:impl-gener-meas}. The proposed implementations utilize
operations that are available on programmable quantum simulators of fermionic
circuits~\cite{Abrams1997, Ortiz2001, Bravyi2002, Ball2005, Verstraete2005,
  Whitfield2011, Barends2015, Whitfield2016, Gonzalez-Cuadra2023}.

\section{Generalized Zeno effect in the frequent-measurement limit}
\label{sec:generalized-Zeno-effect}

\begin{figure}
  \centering
  \includegraphics{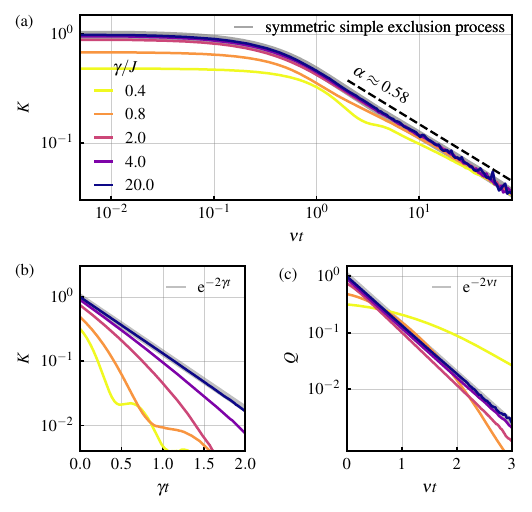}
  \caption{(a)~Density autocorrelation function,
    Eq.~\eqref{eq:density-autocorrelation-function}, for occupation
    measurements. For $\gamma \gg J$, the dynamics agree with a symmetric simple
    exclusion process. The Zeno effect is manifest in the collapse of the data
    after rescaling time with $\nu = J^2/\gamma$. Particle-number conservation
    causes algebraic decay $K(t) \sim t^{-\alpha}$ with $\alpha \approx 0.58$.
    (b)~For fermion counting, the density autocorrelation function exhibits fast
    exponential decay, described by a random telegraph process. (c)~The
    generalized Zeno effect is revealed by factoring out the telegraph process
    in Eq.~\eqref{eq:z-string-z-correlation}. We observe a collapse of the
    telegraph-reduced autocorrelation function after rescaling time with
    $\nu$. For the results shown in this figure, $L = 100$; in (a), the number
    of trajectories ranges from $N_{\mathrm{traj}} = 10^4$ to
    $N_{\mathrm{traj}} = 10^3$ for $\gamma = 0.4 J$ to $\gamma = 20 J$, and
    $N_{\mathrm{traj}} = 10^6$ for the symmetric simple exclusion process; in
    (b) and (c), $N_{\mathrm{traj}} = 5 \times 10^5$ to
    $N_{\mathrm{traj}} = 5 \times 10^4$ for $\gamma = 0.4 J$ to
    $\gamma = 20 J$.}
  \label{fig:3}
\end{figure}

While frequent generalized measurements of occupation numbers induce the
conventional quantum Zeno effect, fermion counting at high rates leads to a
generalized Zeno effect. Importantly, both types of Zeno
effects can be observed in the density autocorrelation function.

The simplest manifestation of the conventional quantum Zeno effect occurs in
projective measurements: a single projective measurement causes the state of a
quantum system to collapse to an eigenstate of the measured observable. Due to
this initial collapse, repeated measurements yield the same result. If
measurements are repeated at rate $\gamma$ and in between measurements there is
coherent evolution with a characteristic frequency $J$, then transitions to
other eigenstates of the measured observable occur at the perturbatively small
rate $\nu = J^2/\gamma$---this suppression of coherently induced transitions is
the essence of the Zeno effect.

Generalized measurements of local occupation numbers are not projective
measurements. However, the measurement operators
Eq.~\eqref{eq:measurement-ops-OM} are proportional to the projection operators
$\hat{n}_l$ and therefore, repeated measurements still induce a quantum Zeno
effect. Suppose the system is prepared in a superposition of pointer states, in
other words eigenstates of the measurement operators, which are classical
configurations of $N$ particles on $L$ lattice sites. Then, repeated
measurements gradually collapse this superposition to a single pointer state,
randomly chosen according to Born's rule. The pointer states are thus fixed
points of the measurement-only dynamics. Coherent hopping destabilizes these
fixed points. However, they remain metastable on a time scale
$\sim 1/\nu$~\cite{Bernard2018}. Since the pointer states are classical
configurations, signatures of the Zeno effect induced by generalized
measurements of local occupation numbers are both emergent classicality and the
suppression of coherent dynamics. This is formalized in the mapping to the
symmetric simple exclusion process~\cite{Bernard2018, Chou2011, Mallick2015}.

The symmetric simple exclusion process is a classical stochastic process of $N$
particles on a 1D lattice of size $L$. On each lattice site, there
is at most one particle---in the present context, this is a consequence of the
Pauli exclusion principle. Each particle has its own internal timer and waits
for a random duration, drawn from an exponential distribution with a mean
waiting time that is identical for all particles. After the waiting time has
elapsed for a given particle, the particle attempts a jump. The process being
symmetric means that the probabilities to jump to the left and to the right are
the same; for a simple process, jumps are only to neighboring sites. If the
target site is empty, the particle jumps there; otherwise, the particle stays at
its current location. In both cases, the timer of the particle is restarted, and
the particle waits for a random time before it attempts the next jump.

To observe emergent classicality and the suppression of coherent dynamics as
described above, we consider the average over trajectories of the conditional
density autocorrelation function in the steady state, which we define as
\begin{equation}
  \label{eq:density-autocorrelation-function}
  K(t - t') = 4 \left[ \overline{\left\langle \hat{n}_l(t) \right\rangle
      \left\langle \hat{n}_l(t') \right\rangle} - \overline{\left\langle
        \hat{n}_l(t) \right\rangle} \; \overline{\left\langle \hat{n}_l(t')
      \right\rangle} \right] = \overline{z_l(t) z_l(t')}.
\end{equation}
In preparation for our discussion of the generalized Zeno effect below, we have
introduced an alternative representation of the density autocorrelation function
in terms of the variables $z_l(t) = 2 \langle \hat{n}_l(t) \rangle - 1$, such
that an occupied or empty site corresponds to $z_l(t) = + 1$ or $z_l(t) = - 1$,
respectively, and we consider a system with average density
$\overline{\left\langle \hat{n}_l(t) \right\rangle} = 1/2$.

Figure~\ref{fig:3}(a) shows the density autocorrelation function for generalized
measurements of occupation numbers for a system of size $L = 100$. For large
values of $\gamma$, the data agree well with numerical simulations of the
symmetric simple exclusion process~\cite{Gillespie2007} with a jump rate of
$2 \nu$ where $\nu = J^2/\gamma$. In particular, the initial value $K(0)$
approaches 1, which reflects that individual trajectories become dominated by
pointer states, in other words classical configurations with $z_l(t) = \pm
1$.
The suppression of coherent dynamics is demonstrated by the collapse of the data
after rescaling time with $\nu = J^2/\gamma$. Finally, particle-number
conservation results in slow algebraic decay, $K(t) \sim t^{- \alpha}$. A fit to
the data for $\gamma = 20 J$ yields $\alpha \approx 0.58$, in reasonable
agreement with the diffusive scaling with $\alpha = 1/2$ expected in larger
systems~\cite{Daquila2011}. To observe this slow decay, it is necessary to
sample rare configurations with a strongly inhomogeneous distribution of
particles, which require long times $t \sim O(L)$ to relax and contribute
significantly to the algebraic tail of $K(t)$. Therefore, to obtain the data
shown in Fig.~\ref{fig:3}(a), we have initialized each trajectory in a randomly
chosen classical configuration with $N = L/2$ particles.

Monitored loss and gain or fermion counting can induce a generalized Zeno
effect. As for the case of generalized measurements of occupation numbers,
repeated application of the measurement operators in
Eq.~\eqref{eq:measurement-ops-FC} causes an initial superposition to collapse to
a pointer state. However, the measurement operators in
Eq.~\eqref{eq:measurement-ops-FC} are not proportional to projection
operators. Therefore, these operators do not stabilize pointer states, but
rather induce transitions between pointer states. In spite of this qualitative
difference, the dynamics in the frequent-measurement limit are again
characterized by emergent classicality and the suppression of coherent
evolution.

Figure~\ref{fig:3}(b) shows the density autocorrelation function for fermion
counting. Again, $K(0) \to 1$ for increasing $\gamma$ signals that the dynamics
are dominated by classical configurations. Indeed, for $\gamma = 20 J$, we
observe good agreement with exponential decay with a rate of $2 \gamma$,
corresponding to a classical random telegraph process~\cite{Gardiner2009}. That
is, the occupation of each lattice site fluctuates between zero and one at a
rate $\sim \gamma$, akin to the signal produced by a telegraph. In stark
contrast to the case of occupation measurements, here the measurement-only
dynamics do not cause the evolution to freeze but rather to accelerate. To
reveal the concomitant suppression of coherent dynamics, we consider the decay
of density autocorrelations after factoring out the fast local telegraph
process. This factorization can be achieved by noting that for a single
realization of the telegraph process on a lattice site $l$, the density
autocorrelation function can be written as
$z_l(t) z_l(t') = \left( -1 \right)^{N_l(t, t')}$, where $N_l(t, t')$ is the
fermion count at site $l$ integrated from time $t'$ to $t$. Put differently, each
fermion count increases $N_l(t, t')$ by one and switches the sign of
$z_l(t)$. Factoring out this classical contribution, we define the
telegraph-reduced autocorrelation function as
\begin{equation}
  \label{eq:z-string-z-correlation}
  Q(t - t') = \overline{z_l(t) \left( -1 \right)^{N_l(t, t')} z_l(t')}.
\end{equation}
By construction, $Q(t) = 1$ stays constant for measurement-only dynamics with
$J = 0$. In contrast, a nonzero hopping amplitude causes $Q(t)$ to decay, with
the decay rate providing a direct measure for the suppression of coherent
dynamics for large values of $\gamma$. Indeed, a perturbative calculation
presented in Appendix~\ref{sec:calc-telegr-reduc} and valid for $J \ll \gamma$
yields exponential decay, $Q(t) = \e^{- 2 \nu t}$, with a decay rate
$\nu = J^2/\gamma$ that vanishes for $\gamma \to \infty$. We interpret this
slowdown of coherent dynamics for high fermion count rates as a generalized
quantum Zeno effect. The exponential rather than algebraic decay is due to the
number of particles not being conserved.

Figure~\ref{fig:3}(c) shows numerical results for the telegraph-reduced
autocorrelation function $Q(t)$. To demonstrate the suppression of coherent
dynamics, we rescale time with $\nu = J^2/\gamma$. For increasing $\gamma$, the
data agree with the analytically predicted behavior.

To reveal the generalized Zeno effect, in Eq.~\eqref{eq:z-string-z-correlation},
we had to factor out the local telegraph process for each single trajectory
before taking the average over trajectories. This indicates that the generalized
Zeno effect is unique to conditional dynamics under continuous monitoring, and
that the suppression of coherent dynamics with an emergent slow decay rate
$\nu = J^2/\gamma$ cannot be seen in the unconditional dynamics. To provide
supporting evidence for this conjecture, we consider the density autocorrelation
function defined by
\begin{equation}
  \label{eq:C-0-autocorrelation}
  C_0(t - t') = \frac{1}{2} \overline{\left\langle \left\{
        \hat{n}_l(t), \hat{n}_l(t') \right\} \right\rangle} - \overline{\left\langle
      \hat{n}_l(t) \right\rangle} \; \overline{\left\langle \hat{n}_l(t')
    \right\rangle}.
\end{equation}
Here, unconditional two-time averages are determined by the quantum regression
theorem~\cite{Gardiner2014}. For example, for $t > t' > t_0$,
\begin{equation}
  \label{eq:quantum-regression}
  \overline{\left\langle \hat{n}_l(t) \hat{n}_l(t') \right\rangle} = \tr \!
  \left[ \hat{n}_l \e^{\mathcal{L} \left( t - t' \right)} \hat{n}_l
    \e^{\mathcal{L} \left( t' - t_0 \right)} \hat{\rho}_0 \right],
\end{equation}
where $\hat{\rho}_0 = \ket{\psi_0} \bra{\psi_0}$ is the projector on the initial
state, and the Liouvillian $\mathcal{L}$ is defined by
\begin{equation}
  \label{eq:Liouvillian}
  \mathcal{L} \hat{\rho} = - \imag \left[ \hat{H}, \hat{\rho} \right] +
  \sum_{\alpha = \pm} \sum_{l = 1}^L \left( 2 \hat{L}_{\alpha, l}^{\vphantom{\dagger}} \hat{\rho}
    \hat{L}_{\alpha, l}^{\dagger} - \left\{ \hat{L}_{\alpha, l}^{\dagger}
      \hat{L}_{\alpha, l}^{\vphantom{\dagger}}, \hat{\rho} \right\} \right).
\end{equation}
The sum over $\alpha$ is absent for occupation measurements, which are described
by a single type of jump operators. Let us anticipate that within the framework
of replica Keldysh field theory introduced next, we find the unconditional
density autocorrelation function for occupation measurements to be given by
\begin{equation}
  \label{eq:C-0-OM}
  C_0(t) \sim \frac{1}{8 \sqrt{\pi \nu \abs{t}}}.
\end{equation}
That is, we find slow, diffusive decay with a diffusion constant $\sim \nu$,
also in the unconditional dynamics. For fermion counting, the Liouvillian,
Eq.~\eqref{eq:Liouvillian}, is quadratic and the unconditional autocorrelation
function can be calculated exactly by elementary means. The exact result is
reproduced by the replica Keldysh field theory approach introduced below, which
yields
\begin{equation}
  \label{eq:C-0-FC}
  C_0(t) = \frac{1}{4} \e^{- 2 \gamma \abs{t}} J_0(2 J \abs{t})^2 \sim
  \frac{\e^{-2 \gamma \abs{t}}}{4 \pi J \abs{t}} \sin(2 J \abs{t} + \pi/4)^2,
\end{equation}
where $J_0$ is the Bessel function of the first kind and the asymptotic form
applies to $\abs{t} \to \infty$. As for the conditional
density correlation function Eq.~\eqref{eq:density-autocorrelation-function}, we
obtain exponential decay with a rate of $2 \gamma$, and there is no indication
of a suppression of coherent dynamics for $\gamma \gg J$.

Having established the suppression of coherent dynamics and emergent
classicality in the conditional time evolution as common traits of the
conventional and generalized Zeno effects, we now turn to the question of how
these properties are reflected in correlations and entanglement in the steady
state.

\section{Replica Keldysh field theory}
\label{sec:replica-keldysh-field-theory}

In Sec.~\ref{sec:models}, we have introduced two equivalent ways to describe the
dynamics generated by continuous monitoring: in terms of a stochastic
Schr\"odinger equation, and as random generalized measurements. We now focus on
the latter formulation, which is a suitable starting point for making analytical
progress using replica Keldysh field theory. To that end, we generalize the
formalism introduced for projective measurements in Ref.~\cite{Poboiko2023} to
the generalized measurements. Our presentation closely follows
Ref.~\cite{Poboiko2023}, but we highlight new aspects that are specific to our
models.

Recently, it has been pointed out that the long-wavelength effective field
theory derived in Ref.~\cite{Poboiko2023} is modified if the Hamiltonian has
particle-hole symmetry (PHS)~\cite{Fava2024}. This is the case, in particular,
for the 1D lattice Hamiltonian, Eq.~\eqref{eq:Hamiltonian}, considered in
Ref.~\cite{Poboiko2023} and our work: Through a gauge transformation,
$\hat{\psi}_l \mapsto \imag^l \hat{\psi}_l$, the hopping amplitudes can be made
purely imaginary; after this transformation, the hopping matrix,
Eq.~\eqref{eq:Hamiltonian-matrix} below, obeys the PHS $H = - H^{\transpose}$.
The PHS can be broken, for example, by adding next-nearest-neighbor hopping with
real amplitudes. Crucially, the modifications of the field theory due to PHS do
not lead to qualitatively different behavior. Therefore, we will present the
theory first for the technically simpler case of broken PHS, and we will
summarize changes due to PHS in the end~\cite{Poboiko2025}.

We anticipate that considering generalized measurements described by
Eqs.~\eqref{eq:measurement-ops-FC} and~\eqref{eq:measurement-ops-OM} instead of
projective measurements of occupation numbers as in Ref.~\cite{Poboiko2023}
results in a number of important conceptual, technical, and, ultimately,
physical differences: (i)~We have to account for the lattice site $l$ being an
outcome of a generalized measurement rather than selected randomly as for the
projective measurements in Ref.~\cite{Poboiko2023}. (ii)~The modified forms of
interaction vertices in the measurement action necessitate a novel perturbative
approach to preserve the causality structure of Keldysh field theory. (iii)~The
breaking of particle-number conservation due to monitored loss and gain or
fermion counting is reflected in explicitly broken symmetries of the Keldysh
action. We have already discussed the physical consequences of these broken
symmetries in Sec.~\ref{sec:key-results}. Below, we describe how they affect the
long-wavelength effective field theory.

\subsection{Observables}

As explained in Sec.~\ref{sec:stochastic-schrodinger-equation}, the unconditional steady state
is completely featureless for both fermion
counting and generalized occupation measurements for any value of the measurement rate
$\gamma$. Therefore, to observe nontrivial effects of continuous monitoring, we
have to consider observables that are nonlinear in the state, such that the
averaging over trajectories does not simply amount to replacing the conditional
by the unconditional state. As an important example, we will study here the von
Neumann entanglement entropy. For a pure state
$\hat{\rho} = \ket{\psi} \bra{\psi}$ and given a bipartition of the system into a
subsystem $A$ and its compliment $B$, the von Neumann entanglement entropy of
subsystem $A$ reads
\begin{equation}
  \label{eq:entanglement-entropy}
  \svn{A} = - \overline{\tr \! \left[ \hat{\rho}_A
      \ln\!\left(\hat{\rho}_A\right) \right]},
\end{equation}
where $\hat{\rho}_A = \mathop{\mathrm{tr}_B}(\hat{\rho})$ is the reduced density
matrix of subsystem $A$. For Gaussian states, the entanglement entropy is
related to the full counting statistics of the number of particles in subsystem
$A$~\cite{Klich2009, Song2011, Song2012, Thomas2015, Burmistrov2017},
\begin{equation}
  \label{eq:entropy-cumulant}
  \svn{A} = 2 \sum_{k = 1}^{\infty}  \zeta(2k) C_A^{(2 k)} = \frac{\pi^2}{3}
  C_A^{(2)} +  \frac{\pi^4}{45} C_A^{(4)} + \dotsb,
\end{equation}
where $\zeta(k)$ denotes the Riemann zeta function and $C_A^{(k)}$ is the $k$-th
cumulant of the subsystem particle number
$\hat{N}_A = \sum_{l \in A} \hat{n}_l$. In particular, the second cumulant is
given by
\begin{equation}
  \label{eq:cumulant}
  C_A^{(2)} = \overline{\left\langle \left( \hat{N}_A - \left\langle \hat{N}_A
        \right\rangle \right)^2 \right\rangle}.
\end{equation}
The task at hand is, therefore, to calculate the average over trajectories of
polynomials of quantum expectation values, which are taken in the pure
conditional state. Since we are considering a quantum many-body system, we wish
to perform this calculation within the framework of nonequilibrium quantum field
theory. This will allow us to employ standard techniques to obtain an
approximate solution in the long-wavelength limit.

\subsection{Replica trick}
\label{sec:replica-trick}

Rewriting nonlinear observables such as the second cumulant,
Eq.~\eqref{eq:cumulant}, as functional integrals in the framework of Keldysh
field theory can be achieved by introducing copies or \emph{replicas} of the system in two
steps. First, we note that the average over trajectories of a product of
$k$ expectation values can be reformulated as a single expectation value that
contains the average of $k$ replicas of the density matrix. For example, for two
operators $\hat{A}$ and $\hat{B}$ we can write
\begin{equation}
  \label{eq:A-B-replicas}
  \begin{split}
    \overline{\left\langle \hat{A}(t) \right\rangle \left\langle \hat{B}(t)
      \right\rangle} & = \overline{\tr \! \left[ \hat{A} \hat{\rho}(t) \right]
      \tr \! \left[ \hat{B} \hat{\rho}(t) \right]} \\ & = \tr \! \left\{ \left(
        \hat{A}_1 \otimes \hat{B}_2 \right) \overline{\left[ \hat{\rho}_1(t)
          \otimes \hat{\rho}_2(t) \right]} \right\}.
  \end{split}
\end{equation}
In the last equality, we have introduced the replica index $r \in \{ 1, 2 \}$ to
indicate on which copy of the system a particular operator is
acting~\footnote{Strictly speaking, the tensor product of fermionic Fock spaces
  is not well defined. However, one can give meaning to such a tensor product by
  first mapping the fermionic Fock spaces to bosonic ones through a
  Jordan-Wigner transformation, then forming the tensor product of the bosonic
  spaces, and finally reintroducing fermionic operators in the product
  space~\cite{Fava2023}.}. Generalizing Eq.~\eqref{eq:A-B-replicas}, a product
of $k$ expectation values can be expressed in terms of the $k$-replica density
matrix
\begin{equation}
  \label{eq:rho-k}
  \hat{\mu}_k(t) = \overline{\bigotimes_{r=1}^k \hat{\rho}_r(t)}.
\end{equation}
In the formulation of our models in terms of random generalized measurements,
the average over trajectories comprises three components: First, an average over
the total number of measurements $M$. As explained in
Sec.~\ref{sec:random-generalized-measurements}, the number of measurements $M$
during the total evolution time $T = t - t_0$ obeys a Poisson distribution with
mean $\gamma L T$,
\begin{equation}
  p_M(T) = \frac{1}{M!} \left( \gamma L T \right)^M \e^{- \gamma L T}.
\end{equation}
Second, for each measurement labeled by $m \in \{ 1, \dotsc, M \}$, the average
in Eq.~\eqref{eq:rho-k} contains an average over the measurement time $t_m$. The
latter is uniformly distributed in the interval $t_m \in [t_0, t]$ of length
$T = t - t_0$. Third, for each measurement, there is an average over measurement
outcomes $\alpha_m \in \{ +, - \}$ and $l_m \in \{ 1, \dotsc, L \}$. The index
$\alpha_m$, which distinguishes the measurement operators in
Eq.~\eqref{eq:measurement-ops-FC} for fermion counting, is absent for
generalized occupation measurements with a single type of measurement operator
given in Eq.~\eqref{eq:measurement-ops-OM}. For measurements at times
$\{ t_m \}$, we denote the probability to obtain a sequence of measurement
outcomes $\{ \alpha_m, l_m \}$ by $p_{\{ \alpha_m, l_m \}}(\{ t_m \})$. The
average in Eq.~\eqref{eq:rho-k} is thus
\begin{equation}
  \label{eq:mu-k-symbolic-sum}
  \hat{\mu}_k(t) = \sum_{\{ \alpha_m, l_m, t_m \}} p_{\{ \alpha_m, l_m \}}(\{
  t_m \}) \bigotimes_{r=1}^k \hat{\rho}_r(t),
\end{equation}
where the symbolic sum includes the average over both the Poisson distribution
of the number of measurements $M$ and the uniform distributions of measurement
times $\{ t_m \}$, as well as a summation over measurement outcomes $\{
\alpha_m, l_m \}$,
\begin{equation}
  \label{eq:alpha-l-t-average}
  \sum_{\{ \alpha_m, l_m, t_m \}} = \sum_{M = 0}^{\infty} p_M(T)
  \prod_{m = 1}^M \sum_{\alpha_m = \pm} \sum_{l_m = 1}^L \int_{t_0}^t
  \frac{\diff t_m}{T}.
\end{equation}

The probability $p_{\{ \alpha_m, l_m \}}(\{ t_m \})$ of a sequence of
measurement results $\{ \alpha_m, l_m \}$ can be expressed in terms of the
nonnormalized density matrix for a single copy of the system. If we omit the
normalization factor in Eq.~\eqref{eq:GM-state-update}, the state of the system
after a time $t$ can be written as
\begin{equation}
  \label{eq:nonnormalized-density-matrix}
  \hat{D}(t) = \hat{V}(t) \hat{\rho}_0 \hat{V}(t)^{\dagger},
\end{equation}
with the initial state $\hat{\rho}_0 = \ket{\psi_0} \bra{\psi_0}$ and
\begin{equation}
  \label{eq:V}
  \hat{V}(t) = \hat{U}(t - t_M) \hat{M}_{\alpha_M, l_M} \hat{U}(t_M - t_{M - 1}) \dotsb
  \hat{M}_{\alpha_1, l_1} \hat{U}(t_1 - t_0).
\end{equation}
In $\hat{V}(t)$, unitary evolution $\hat{U}(t) = \e^{- \imag \hat{H} t}$ is
interspersed with quantum jumps $\hat{M}_{\alpha_m,l_m}$ at times $t_m$. We
leave the dependence of $\hat{V}(t)$ on the sequence of measurement outcomes and
times $\{ \alpha_m, l_m, t_m \}$ implicit. The normalized density matrix is
obtained by reinstating the factors $p_{\alpha_m, l_m}(t)$ in
Eq.~\eqref{eq:GM-state-update},
\begin{equation}
  \label{eq:rho-D}
  \hat{\rho}(t) = \frac{1}{\prod_{m = 1}^M p_{\alpha_m, l_m}(t_m)} \hat{D}(t).
\end{equation}
Finally, using that $\tr(\hat{\rho}(t)) = 1$, we obtain the probability of the
entire sequence of measurement outcomes, $p_{\{ \alpha_m, l_m \}}(\{ t_m \})$,
given by the product of probabilities of individual measurement outcomes,
$p_{\alpha_m, l_m}(t_m)$, as
\begin{equation}
  \label{eq:born-probabilities}
  p_{\{ \alpha_m, l_m \}}(\{ t_m \}) = \prod_{m = 1}^M
  p_{\alpha_m, l_m}(t_m) = \tr \! \left[ \hat{D}(t) \right].
\end{equation}
Recall the discussion in Sec.~\ref{sec:random-generalized-measurements} that
treating generalized occupation measurements requires us to consider the site
indices $l_m$ as measurement outcomes. Their distribution is contained in the
factors $p_{\alpha_m, l_m}(t_m)$. In contrast, the projective measurements
studied in Ref.~\cite{Poboiko2023} are assumed to be distributed uniformly in
space, an the uniform probability $1/L$ is combined with the summation over
$l_m$ in Eq.~\eqref{eq:alpha-l-t-average}. Inserting the above expression for
$p_{\{ \alpha_m, l_m \}}(\{ t_m \})$ in Eq.~\eqref{eq:mu-k-symbolic-sum}, we
obtain
\begin{equation}
  \label{eq:rho-k-before-replica-trick}
  \hat{\mu}_k(t) = \sum_{\{ \alpha_m, l_m, t_m \}} \left. \left[ \bigotimes_{r=1}^k
      \hat{D}_r(t) \right] \middle/ \tr \! \left[ \hat{D}(t) \right]^{k-1} \right..
\end{equation}
The denominator in Eq.~\eqref{eq:rho-k-before-replica-trick} obstructs the
application of the usual Keldysh construction to obtain a functional integral
representation of the time evolution of the $k$-replica density matrix. We can
get rid of this denominator by using the replica trick, which amounts here to
first introducing additional replicas with indices
$r \in \{ k + 1, \dotsc, R \}$ and then taking the replica limit
$R \to 1$~\cite{Jian2023, Bao2020, Fava2023, Poboiko2023}:
\begin{equation}
  \hat{\mu}_k(t) = \lim_{R \to 1} \sum_{\{ \alpha_m, l_m, t_m \}}
  \mathop{\mathrm{tr}_{r = k + 1, \dotsc, R}} \! \left[ \bigotimes_{r=1}^R
    \hat{D}_r(t) \right].
\end{equation}
We note that the replica limit $R \to 1$, which is required to reproduce the
probability according to Born's rule given in Eq.~\eqref{eq:born-probabilities},
is different from the usual replica limit $R \to 0$ in the theory of disordered
systems.

We can now obtain averages over
trajectories of nonlinear observables such as the cumulant,
Eq.~\eqref{eq:cumulant}, by calculating expectation values that are linear in
the nonnormalized density matrix of $R$ replicas of the original system and
taking the replica limit $R \to 1$ in the end. Note that the way in which we
have introduced replicas in Eq.~\eqref{eq:A-B-replicas} allows us to obtain
\emph{equal-time} correlations of conditional expectation values
$\left\langle \hat{A}(t) \right\rangle$ and
$\left\langle \hat{B}(t) \right\rangle$. However, unconditional averages, such
as the autocorrelation function, Eq.~\eqref{eq:C-0-autocorrelation}, can also be
obtained for operators $\hat{A}(t)$ and $\hat{B}(t')$ at different times
$t \neq t'$. A convenient way to generate different forms of
expectation values is by introducing source terms in the $R$-replica Keldysh
partition function,
\begin{equation}
  \label{eq:keldysh-partition}
  Z_R(t) = \sum_{\{ \alpha_m, l_m, t_m \}} \tr \! \left[ \bigotimes_{r = 1}^R
    \hat{D}_r(t) \right].
\end{equation}
By construction, the $R$-replica Keldysh partition function is normalized such
that $Z_R(t) \to 1$ for $R \to 1$. We proceed to derive a functional integral
representation of $Z_R(t)$, where source fields can be introduced as required at
a later stage.

\subsection{Replica Keldysh action}
\label{sec:repl-keldysh-action}

In Keldysh field theory~\cite{Altland2010a, Kamenev2023, Sieberer2016a,
  Sieberer2023}, the $R$-replica Keldysh partition function,
Eq.~\eqref{eq:keldysh-partition}, is expressed as a functional integral over two
independent sets of Grassmann fields $\psi_{k, r, l}(t)$ and
$\psi^*_{k, r, l}(t)$ with Keldysh index $k$ and replica index $r$, which we
collect into $2 R$-component vectors denoted by $\psi_l(t)$ and
$\psi^*_l(t)$. Time evolution of the nonnormalized density matrix,
Eq.~\eqref{eq:nonnormalized-density-matrix}, is visualized as proceeding along
two branches, the forward branch described by $\hat{V}(t)$ acting on
$\hat{\rho}_0$ from the left, and the backward branch described by
$\hat{V}(t)^{\dagger}$ acting on $\hat{\rho}_0$ from the right. Replacing
operators on the forward and backward branches by fields with Keldysh indices
$k = +$ and $k = -$, respectively, we obtain
\begin{multline}
  \label{eq:Z-R-M-M-dagger}
  Z_R(t) = \int \Diff[\psi^*, \psi] \, \e^{\imag \psi^{\dagger} G_0^{-1} \psi} \\ \times
  \sum_{\{ \alpha_m, l_m, t_m \}} \prod_{r = 1}^R M_{+, r, \alpha_m, l_m}(t_m)
  M_{-, r, \alpha_m, l_m}^{\dagger}(t_m).
\end{multline}
We omit the matrix element of the initial state $\hat{\rho}_0$, which fixes the
number of particles for occupation measurements but is otherwise irrelevant in
the steady state. The latter can be described by taking the limit
$t_0 \to - \infty$.  Unitary evolution contained in $\hat{V}(t)$,
Eq.~\eqref{eq:V}, gives rise to the quadratic Keldysh action, where we use a
shorthand notation leaving the integration over time and summation over lattice
sites implicit,
\begin{equation}
  \label{eq:G-0}
  \psi^{\dagger} G_0^{-1} \psi = \int_{t_0}^t \diff t'
  \sum_{l, l' = 1}^L \psi^*_l(t') G_{0, l, l'}^{-1} \psi_{l'}(t').
\end{equation}
The inverse Green's function reads
$G_0^{-1} = \left( \imag \partial_t - H \right) \sigma_z$, where $\sigma_z$ is a
Pauli matrix in Keldysh space and the matrix $H$ describes hopping according to
Eq.~\eqref{eq:Hamiltonian},
\begin{equation}
  \label{eq:Hamiltonian-matrix}
  H_{l, l'} = - J \left( \delta_{l + 1, l'} + \delta_{l, l' + 1} \right).
\end{equation}
To further simplify the notation, we have omitted the identity $1_R$ in replica
space in the definition of the inverse Green's function. In the limit
$t_0 \to - \infty$, the Green's function should be augmented by an infinitesimal
regularization term, fixing the correct causality properties and particle
number~\cite{Poboiko2023}. We omit this term, anticipating that the Green's
function will be dressed due to measurements.

According to the rule formulated above for replacing operators by fields, the
measurement operators for monitored loss and gain or fermion counting,
Eq.~\eqref{eq:measurement-ops-FC}, are represented by
\begin{equation}  
  \begin{split}
    M_{\pm, r, -, l}(t) & = M_{\pm, r, +, l}^{\dagger}(t) = \pm
    \frac{1}{\sqrt{L}} \psi_{\pm, r, l}(t), \\ M_{\pm, r, +, l}(t) & = M_{\pm,
      r, -, l}^{\dagger}(t) = \pm \frac{1}{\sqrt{L}} \psi^*_{\pm, r, l}(t),
  \end{split}
\end{equation}
where we include the minus sign acquired by Grassmann fields on the backward
branch~\cite{Sieberer2023}. For the measurement operators describing generalized
measurements of occupation numbers, Eq.~\eqref{eq:measurement-ops-OM}, we obtain
\begin{equation}
  \label{eq:OM-K-K-dagger-fields}
  M_{\pm, r, l}(t) = M_{\pm, r, l}^{\dagger}(t) = \frac{1}{\sqrt{N}} \psi^*_{\pm,
    r, l}(t) \psi_{\pm, r, l}(t).
\end{equation}

Inserting the explicit expression, Eq.~\eqref{eq:alpha-l-t-average}, for the
average over the number of measurements as well as the measurement times and
outcomes in the functional integral representation of the $R$-replica Keldysh
partition function in Eq.~\eqref{eq:Z-R-M-M-dagger}, we find that each factor in
the product over measurements labeled by $m$ is identical. We may thus omit the
label $m$ to obtain
\begin{multline}
  \label{eq:Z-R-M-M-dagger-average-explicit}
  Z_R(t) = \int \Diff[\psi^*, \psi] \, \e^{\imag \psi^{\dagger} G_0^{-1} \psi}
  \sum_{M = 0}^{\infty} \frac{\e^{- \gamma L T}}{M!} \\
  \times \left[ \gamma L \sum_{\alpha = \pm} \sum_{l = 1}^L \int_{t_0}^t \diff t
    \prod_{r = 1}^R M_{+, r, \alpha, l}(t) M_{-, r, \alpha, l}^{\dagger}(t)
  \right]^M.
\end{multline}
When we now perform the sum over the number of measurements $M$, the expression
in the second line of Eq.~\eqref{eq:Z-R-M-M-dagger-average-explicit} is
exponentiated. This leads to
$Z_R(t) = \int \Diff[\psi^*, \psi] \e^{\imag S[\psi^*, \psi]}$, where
the action reads
\begin{equation}
  \label{eq:S-fermionic}
  S[\psi^*, \psi] = \int_{t_0}^t \diff t' \sum_{l, l' = 1}^L
  \psi^*_l G_{0, l, l'}^{-1} \psi_{l'} + \gamma \int \diff^2
  \mathbf{x} \, \mathcal{L}_M[\psi^*_l, \psi_l].
\end{equation}
Here and occasionally in the following, we abbreviate expressions by omitting
field indices and time arguments. Furthermore, we abbreviate integration over
space and time as
$\int \diff^2 \mathbf{x} = \sum_{l=1}^L \int_{t_0}^t \diff t'$. The Lagrangian
density due to random generalized measurements reads
\begin{equation}
  \label{eq:measurement-Lagrangian}
  \imag \mathcal{L}_M[\psi^*, \psi] = \frac{1}{L^{R - 1}} \sum_{\alpha =
    \pm} \prod_{r=1}^R V_{\alpha}[\psi^*_r, \psi_r] - 1,
\end{equation}
where the prefactor $L^{1 - R}$ vanishes in the replica limit and will be
omitted in the following. The last term in the measurement Lagrangian stems from
the factor $\e^{- \gamma L T}$ in
Eq.~\eqref{eq:Z-R-M-M-dagger-average-explicit}, where we have used
$L T = \int \diff^2 \mathbf{x} \, 1$. For fermion counting, the vertices in the
measurement Lagrangian are given by
\begin{equation}
  \label{eq:FC-vertices}
  V_-[\psi^*, \psi] = - \psi_+ \psi^*_-, \qquad V_+[\psi^*, \psi] =
  - \psi^*_+ \psi_-.
\end{equation}
The product over replicas in Eq.~\eqref{eq:measurement-Lagrangian} thus yields
an interaction vertex containing $2 R$ fermionic fields. For occupation
measurements, there is a single vertex that depends explicitly on the fermion
density $n = N/L$,
\begin{equation}
  \label{eq:OM-vertex}
  V[\psi^*, \psi] = \frac{1}{n} \psi^*_+ \psi_+ \psi^*_- \psi_-.
\end{equation}
In this case, the product over replicas in Eq.~\eqref{eq:measurement-Lagrangian}
yields an interaction vertex containing $4 R$ fermionic fields.

The replica Keldysh construction we have presented above provides an exact
reformulation of the dynamics under continuous monitoring as a field theory. To
make further progress, we now have to introduce suitable approximations. The
first step is to identify the relevant degrees of freedom, which dominate
correlations and---via Eq.~\eqref{eq:entropy-cumulant}---the dynamics of
entanglement at long wavelengths. For generalized measurements of occupation
numbers, we expect that the relevant slow modes are the same as for the case of
projective measurements, that is, fermionic bilinears which we collect in the
matrix
$\mathcal{G}_l(t) = - \imag \psi_l(t) \psi^*_l(t)$~\cite{Poboiko2023}. As we
will see comparing with our numerical results, this turns out to be the correct
choice also for fermion counting.

Specifically, we want $\left\langle \mathcal{G}_l(t) \right\rangle$ to represent
the fermionic Green's function at equal positions and in the symmetrized limit
of equal time arguments of the fields $\psi_l(t)$ and $\psi^*_l(t)$,
\begin{equation}
  \label{eq:G-symmetric-limit}
  \left\langle \mathcal{G}_l(t) \right\rangle = - \frac{\imag}{2} \left(
    \left\langle \psi_l(t) \psi^*_l(t + 0^+) \right\rangle + \left\langle
      \psi_l(t) \psi^*_l(t - 0^+) \right\rangle \right).
\end{equation}
Two technical issues arise: (i)~in the construction of the Keldysh field
integral, which is based on a discretization of time, products of field
operators $\hat{n}_l = \hat{\psi}_l^{\dagger} \hat{\psi}_l^{}$ as occur in the
measurement operators for occupation measurements in
Eq.~\eqref{eq:measurement-ops-OM} are replaced by fields that are
a discrete time step apart; (ii)~the discrete-time Green's function at equal
discrete times differs from the symmetrized limit of equal time arguments in the
continuous-time formulation~\cite{Kamenev2023}, and we want
$\left\langle \mathcal{G}_l(t) \right\rangle$ to represent the latter. These
discrepancies can be resolved by using the regularization procedure detailed in
Appendix~\ref{sec:caus-struct-regul}.

The measurement operators for fermion counting are linear in fermionic field
operators. Therefore, in this case the issues described above do not arise, and
the vertices in Eq.~\eqref{eq:FC-vertices} are not affected by the
regularization. After a Larkin-Ovchinnikov rotation~\cite{Altland2010a,
  Kamenev2023},
\begin{equation}
  \label{eq:Larkin-Ovchinnikov-rotation}
  \psi_{1,2} = \frac{1}{\sqrt{2}} \left(\psi_+ \pm \psi_-\right), \qquad
  \psi^*_{1,2} = \frac{1}{\sqrt{2}} \left(\psi^*_+ \mp \psi^*_-\right),
\end{equation}
these vertices take the form
\begin{equation}
  \label{eq:FC-vertices-rotated}
  V_{\pm}[\psi^*, \psi] = -\frac{1}{2} \left[ \psi^*_1 \psi_1 -
    \psi^*_2 \psi_2 \mp \left( \psi^*_1 \psi_2 - \psi^*_2\psi_1
    \right) \right].
\end{equation}
In contrast, we obtain a modified vertex for
generalized measurements of occupation numbers through the regularization,
\begin{equation}
  \label{eq:OM-vertex-rotated}
  V[\psi^*, \psi] = \frac{1}{n} \left[ \frac{1}{4} + \frac{1}{2} \psi^{\dagger}
    \sigma_x \psi - \psi^*_1 \psi_1
    \psi^*_2 \psi_2 \right] = \frac{1}{4 n} \e^{2 \psi^{\dagger} \sigma_x
    \psi}.
\end{equation}
Up to the prefactor $1/n$, the regularized vertex is identical to one of the
regularized vertices for projective occupation measurements of
Ref.~\cite{Poboiko2023}.

The vertices in Eqs.~\eqref{eq:FC-vertices-rotated}
and~\eqref{eq:OM-vertex-rotated} contain the term $\psi^*_2 \psi_1$, which
would correspond to an anti-Keldysh component of the self-energy, and thus
violates the familiar causality structure of fermionic Keldysh field
theory~\cite{Kamenev2023}. For fermion counting and in the replica limit
$R \to 1$, this term cancels in the sum over $\alpha$ in
Eq.~\eqref{eq:measurement-Lagrangian}. A similar cancellation occurs for
projective measurements of occupation numbers~\cite{Poboiko2023}. However, this
is not the case for generalized occupation measurements. In
Appendix~\ref{sec:caus-struct-regul}, we discuss how the appearance of an
anti-Keldysh component can be reconciled with the normalization of the Keldysh
partition function in the replica limit, $Z_R(t) \to 1$ for $R \to 1$. The
violation of the usual causality structure is only due to the part of the
Lagrangian for occupation measurements that is quadratic in fermionic fields,
\begin{equation}
  \label{eq:L-M-Q}
  \imag \mathcal{L}_{M, Q}[\psi^*, \psi] = \frac{1}{2^{2 R - 1} n^R} \sum_{r
    = 1}^R \left( \psi^*_{1, r} \psi_{2, r} + \psi^*_{2, r} \psi_{1, r}
  \right),
\end{equation}
which we therefore separate from the nonquadratic part,
$\mathcal{L}_{M, \mathit{NQ}} = \mathcal{L}_M - \mathcal{L}_{M, Q}$. To preserve
the usual structure of the Keldysh formalism, we will treat $\mathcal{L}_{M, Q}$
perturbatively, so that the self-energy, which nonperturbatively dresses the
Green's function, is fully determined by $\mathcal{L}_{M, \mathit{NQ}}$. For
brevity, we will drop the subscript $\mathit{NQ}$ in the following.

With these precautions, we introduce Hermitian $2R \times 2R$ matrix fields
$\mathcal{G}$ and $\Sigma$, corresponding to the equal-time fermionic Green's
function and the self-energy, by means of a generalized Hubbard-Stratonovich
transformation~\cite{Poboiko2023}. Specifically, we include the factor
\begin{equation}
  \label{eq:HS-identity}
  1 = \int \Diff[\mathcal{G}, \Sigma] \, \e^{- \frac{\epsilon}{2} \Tr 
    \left( \Sigma^2 \right) - \imag \Tr \left( \mathcal{G} \Sigma
    \right) - \psi^{\dagger} \Sigma \psi}
\end{equation}
in the functional integral over $\psi$ and $\psi^*$. The trace
$\Tr\!\left( \, \cdot \, \right)$ acts in Keldysh, replica, lattice, and time
spaces, with the matrices $\mathcal{G}$ and $\Sigma$ being diagonal in lattice
and time spaces. Convergence of the integration over $\mathcal{G}$ and $\Sigma$
is ensured by the term proportional to $\epsilon$. We omit this term in the
following, with the understanding that the limit $\epsilon \to 0$ has to be
taken at the end of the calculation. In this limit, the integral over $\Sigma$
reduces to a delta functional fixing
$\mathcal{G} = - \imag \psi \psi^{\dagger}$~\cite{Poboiko2023}. Using this relation,
decoupling the measurement Lagrangian simultaneously in all possible slow
channels is achieved by taking the average of $\mathcal{L}_M$
with respect to the Gaussian action
$\psi^{\dagger} \mathcal{G}^{-1} \psi$~\cite{Poboiko2023}. For fermion counting,
this is done most conveniently with the form of the vertices given in
Eq.~\eqref{eq:FC-vertices}, and the result can be expressed as a trace $\trK$ in
Keldysh space and a determinant $\detR$ in replica space,
\begin{equation}
  \label{eq:lagrangian-FC}
  \imag \mathcal{L}_M[\mathcal{G}] = \sum_{\alpha = \pm} \detR \! \left[ \trK \!
    \left( \tau_{\alpha} \mathcal{G} \right) \right] - 1,
\end{equation}
where $\tau_{\pm} = (\imag \sigma_z \pm \sigma_y)/2$ and $\sigma_i$ are the
Pauli matrices. The Lagrangian for occupation measurements contains a
determinant $\det( \, \cdot \, )$ and a trace $\tr( \, \cdot \, )$ in both
Keldysh and replica spaces,
\begin{equation}
  \label{eq:lagrangian-occupation}
  \imag \mathcal{L}_M[\mathcal{G}] = \frac{1}{n^R} \detKR \! \left( \frac{1}{2} -
    \imag \sigma_x \mathcal{G} \right) + \frac{\imag}{2^{2 R - 1} n^R} \trKR \!
  \left( \sigma_x \mathcal{G} \right) - 1.
\end{equation}
This form of the measurement Lagrangian is similar to the one for projective
measurements~\cite{Poboiko2023}: the latter does not contain the prefactor
$1/n^R$ and the two types of projection operators, $\hat{n}_l$ and
$1 - \hat{n}_l$, result in two determinant contributions with opposite signs of
the term containing $\mathcal{G}$, in contrast to our model with only one
determinant term. The trace in Eq.~\eqref{eq:lagrangian-occupation} stems from
the subtraction of $\mathcal{L}_{M, Q}$, Eq.~\eqref{eq:L-M-Q}, and ensures that
$\mathcal{L}_M$ does not contain terms that are linear in $\mathcal{G}$.

After decoupling the measurement Lagrangian, the action is quadratic in the
fermionic fields $\psi$ and $\psi^*$, which can thus be integrated out, leading to
\begin{equation}
  \label{eq:action-G-Sigma}
  S[\mathcal{G}, \Sigma] = S_0[\mathcal{G}, \Sigma] + \gamma \int \diff^2
  \mathbf{x} \, \mathcal{L}_M[\mathcal{G}_l],
\end{equation}
where
\begin{equation}
  \label{eq:S-0-G-Sigma-nonperturbative-A}
  \imag S_0[\mathcal{G}, \Sigma] = \Tr \! \left\{ \ln \! \left[ G_0^{-1} +
      \imag \left( \Sigma + A \right) \right] -\imag \mathcal{G} \Sigma \right\}.
\end{equation}
In the basis of fields introduced in the Larkin-Ovchinnikov rotation in
Eq.~\eqref{eq:Larkin-Ovchinnikov-rotation}, the bare Green's function is
proportional to the identity in both Keldysh and replica spaces,
$G_0^{-1} = \imag \partial_t - H$. The matrix $A$ results from expressing the
quadratic part of the measurement action, Eq.~\eqref{eq:L-M-Q}, as
\begin{equation}  
  \imag \gamma \int \diff^2 \mathbf{x} \, \mathcal{L}_{M, Q}[\psi^*_l,
  \psi_l] = - \psi^{\dagger} A \psi,
\end{equation}
where $A = 0$ for fermion counting and, for occupation measurements,
\begin{equation}
  \label{eq:anti-Keldysh-self-energy}
  A = - \frac{\gamma}{2^{2 R - 1} n^R} \sigma_x.
\end{equation}
As anticipated, we treat $A \sim \gamma$ perturbatively. To first order in $A$,
the action in Eq.~\eqref{eq:S-0-G-Sigma-nonperturbative-A} reads
\begin{equation}
  \label{eq:S-0-G-Sigma}
  \imag S_0[\mathcal{G}, \Sigma] = \Tr \! \left[ \ln\!\left( G^{-1} \right) +
    \imag \left( G A - \mathcal{G} \Sigma \right) \right],
\end{equation}
where the dressed Green's function is given by
\begin{equation}
  \label{eq:dressed-GF}
  G^{-1} = G_0^{-1} + \imag \Sigma.
\end{equation}

\subsection{Symmetries of the Keldysh action}
\label{sec:symmetry-keldysh-action}

The long-wavelength behavior of our models is dominated by strong fluctuations
of soft modes, which are related to symmetries of the Keldysh action. Therefore,
a prerequisite for deriving a long-wavelength effective field theory is to
identify the relevant symmetries. This analysis provides important insights into
the consequences of particle-number conservation for generalized occupation
measurements. As explained at the beginning of
Sec.~\ref{sec:replica-keldysh-field-theory}, we consider here the case of broken
PHS. Modifications due to PHS will be discussed further below.

The fermionic replica Keldysh action in Eq.~\eqref{eq:S-fermionic} is defined in
terms of $2R$-component vectors of Grassmann fields $\psi$ and $\psi^*$. In
the Larkin-Ovchinnikov basis, Eq.~\eqref{eq:Larkin-Ovchinnikov-rotation}, the
bare Green's function $G_0$ is diagonal in Keldysh and replica spaces, and,
therefore, rotations of the fields described by $\psi \mapsto \mathcal{R} \psi$
and $\psi^{\dagger} \mapsto \psi^{\dagger} \mathcal{R}^{-1}$ with
$\mathcal{R} \in \mathrm{U}(2R)$ leave the first term in the action of
Eq.~\eqref{eq:S-fermionic}, which encodes free evolution in the absence of
measurements, invariant. Which rotations are symmetries of the full action
including the measurement Lagrangian is most conveniently analyzed after
performing the generalized Hubbard-Stratonovich transformation, meaning for the
action in Eq.~\eqref{eq:S-0-G-Sigma} and the measurement Lagrangians given in
Eqs.~\eqref{eq:lagrangian-FC} and~\eqref{eq:lagrangian-occupation} in
terms of the matrix fields $\mathcal{G}$ and $\Sigma$. Rotations of the
fermionic fields act on the matrix fields as
$\mathcal{G} \mapsto \mathcal{R} \mathcal{G} \mathcal{R}^{-1}$ and
$\Sigma \mapsto \mathcal{R} \Sigma \mathcal{R}^{-1}$. Interestingly, the
measurement Lagrangians have different symmetries for $R = 1$ and $R > 1$,
corresponding to the unconditional evolution of observables that are linear in
the system state and to the conditional evolution of nonlinear observables,
respectively.

We consider first the case $R = 1$. Then, the measurement Lagrangian for fermion
counting, Eq.~\eqref{eq:lagrangian-FC}, reduces to
\begin{equation}
  \imag \mathcal{L}_M[\mathcal{G}] = \imag \trK \!
  \left( \sigma_z \mathcal{G} \right) - 1.
\end{equation}
This Lagrangian and the action in Eq.~\eqref{eq:S-0-G-Sigma}, where $A = 0$ for
fermion counting, are invariant under phase rotations
$\mathcal{R}_{\eta} = \e^{\imag \eta/2}$ and
$\mathcal{R}_{\zeta} = \e^{\imag \zeta \sigma_z / 2}$, with
$\eta, \zeta \in \R$. For occupation measurements, the Lagrangian
Eq.~\eqref{eq:lagrangian-occupation} simplifies for $R = 1$ to
\begin{equation}
  \label{eq:measurement-Lagrangian-OM-R=1}
  \imag \mathcal{L}_M[\mathcal{G}] = \frac{1}{n} \left[ \frac{1}{4} +
    \detK(\mathcal{G}) \right] - 1,
\end{equation}
which is invariant under arbitrary rotations $\mathcal{R} \in \mathrm{U}(2)$.
However, the matrix $A$, Eq.~\eqref{eq:anti-Keldysh-self-energy}, appearing in
the action in Eq.~\eqref{eq:S-0-G-Sigma} is proportional to $\sigma_x$, thus
reducing the symmetries of the action to rotations of the form
$\mathcal{R}_{\eta}$ and $\mathcal{R}_{\phi} = \e^{\imag \phi \sigma_x / 2}$. An
inverse Larkin-Ovchinnikov rotation, Eq.~\eqref{eq:Larkin-Ovchinnikov-rotation},
shows that $\mathcal{R}_{\phi}$ corresponds to phase rotations with opposite
signs on the forward and backward branches of the Keldysh contour, which is the
quantum~\cite{Sieberer2016a, Sieberer2023} or strong symmetry~\cite{Buca2012,
  Albert2014} that is associated with particle-number conservation.

We now turn to the case $R > 1$. Interestingly, there is now an additional type
of rotations, which is a symmetry of the action, Eq.~\eqref{eq:S-0-G-Sigma}, and
the measurement Lagrangian for fermion counting, Eq.~\eqref{eq:lagrangian-FC},
given by $\mathcal{R}_{\Phi} = \e^{\imag \Phi \sigma_x / 2}$ where $\Phi$ is a
traceless $R \times R$ matrix, $\trR(\Phi) = 0$. Such rotations form the group
$\mathrm{SU}(R)$. We show that $\mathcal{R}_{\Phi}$ is a symmetry of
Eq.~\eqref{eq:S-0-G-Sigma} in Appendix~\ref{sec:app-replicon-symmetry}. For the
case of occupation measurements, it follows directly from inspection of
Eqs.~\eqref{eq:lagrangian-occupation} and~\eqref{eq:S-0-G-Sigma}
with~\eqref{eq:anti-Keldysh-self-energy} that rotations of the form
$\mathcal{R}_{\Phi}$ are a symmetry.

It is informative to trace the symmetry of the Keldysh action under
$\mathcal{R}_{\Phi}$ back to a symmetry of the time-evolved state of $R$
replicas of the original fermionic lattice system. As we discuss in detail in
Appendix~\ref{sec:app-replicon-symmetry}, this symmetry is specific to
noninteracting systems, for which coherent dynamics and measurements preserve
the Gaussianity of the state. Furthermore, as anticipated in
Secs.~\ref{sec:introduction} and~\ref{sec:key-results}, there are two
requirements for the symmetry to occur: (i)~The number of particles has to be
conserved by the coherent dynamics; measurements are allowed to change the
number of particles by integer values, but they have to keep the system in a
state with a well-defined number of particles. This is the case, in particular,
for the measurement operators in Eqs.~\eqref{eq:measurement-ops-FC}
and~\eqref{eq:measurement-ops-OM}, which change the number of particles by
$\Delta N = \pm 1$ and $\Delta N = 0$, respectively. For the Majorana model of
Ref.~\cite{Fava2023}, this condition is not met. (ii)~The outcomes of all
measurements have to be recorded. The symmetry is broken, in particular, for
inefficient detection, which is commonly modeled by averaging over a fraction of
the measurement outcomes~\cite{Wiseman2010}.

These two conditions, combined with the symmetry under permutations of replicas
that is built into the formalism, imply that the $R$-fold replicated state of
the system is at all times a Slater determinant with a well-defined number $N$
of particles in each replica,
\begin{equation}
  \label{eq:R-replica-N-particle-Slater-determinant}
  \ket{\psi_{R, N}} = \prod_{r = 1}^R \prod_{n = 1}^N \sum_{l = 1}^L \psi_{l, n}^{}
  \hat{\psi}_{r, l}^{\dagger} \ket{0}.
\end{equation}
Here, $\psi_n$ are $N$ orthonormal single-particle states with amplitude
$\psi_{l, n}$ on lattice site $l$. The fermionic field operators
$\hat{\psi}_{r, l}$ act in the Hilbert space of $R$ replicas of the original
lattice system. Furthermore, the $R$-replica vacuum state obeys
$\hat{\psi}_{r, l} \ket{0} = 0$.

Due to the symmetry under permutations of replicas, each single-particle state
$\psi_n$ is occupied in all replicas. Furthermore, fermionic statistics imply
that each state $\psi_n$ remains fully occupied after a single-particle
transformation that corresponds to a change of basis in replica space but not in
the Hilbert space of individual replicas. Therefore, each unitary transformation
that describes such a change of basis is a symmetry of the above Slater
determinant. Such transformations can be written as
\begin{equation}
  \label{eq:G-Phi}
  \hat{G}_{\Phi} = \e^{\imag \hat{\Phi}/2}, \qquad \hat{\Phi} = \sum_{r, r' =
    1}^R \sum_{l = 1}^L \hat{\psi}_{r, l}^{\dagger} \Phi_{r, r'}^{} \hat{\psi}_{r',
  l}^{},
\end{equation}
where $\Phi$ is a Hermitian $R \times R$ matrix. The detailed derivation in
Appendix~\ref{sec:app-replicon-symmetry} shows that $\hat{G}_{\Phi}$ is a
symmetry of the $R$-replica Slater determinant,
Eq.~\eqref{eq:R-replica-N-particle-Slater-determinant}, up to a phase factor
that reduces to unity for $\trR(\Phi) = 0$. We thus recover the symmetry group
$\mathrm{SU}(R)$. As we explain in Appendix~\ref{sec:app-replicon-symmetry},
this symmetry of the $R$-replica state translates to a strong symmetry of the
superoperator that describes the time evolution of the $R$-replica nonnormalized
density matrix~\cite{Buca2012, Albert2014}, which in turn is reflected in the
symmetry of the Keldysh action under $\mathcal{R}_{\Phi}$.

Having clarified the technical requirements for the $\mathrm{SU}(R)$ symmetry to
arise, let us finally provide a physical explanation in terms of conservation of
the total number of particles. We first note that according to the discussion
below Eq.~\eqref{eq:measurement-Lagrangian-OM-R=1}, conservation of the number
of particles in each of $R$ replicas is associated with the symmetry of the
Keldysh action under rotations of the form
$\mathcal{R}_{\Phi} = \e^{\imag \Phi \sigma_x/2}$ with
$\Phi_{r, r'} = \delta_{r, r_0} \delta_{r', r_0}$ for
$r_0 \in \{ 1, \dotsc, R \}$. Evidently, these rotations are not contained in
the symmetry group $\mathrm{SU}(R)$ characterized by $\trR(\Phi) = 0$. This
observation reaffirms that this symmetry does not require particle-number
conservation within the system and its replicas. However, it is interesting to
note that the implementations of fermion counting described in
Appendices~\ref{sec:app-physical-models} and~\ref{sec:impl-gener-meas} require
coupling the fermionic lattice system to auxiliary reservoirs in such a way that
the total number of particles in the system and reservoirs is conserved. Indeed,
the conservation of matter in a closed system implies that the implementation of
any measurement that obeys condition (i)~above requires coupling the system to
reservoirs to compensate the change of the number of particles in the
system. Therefore, we can regard conservation of the total number of particles
as the physical reason for the occurrence of the $\mathrm{SU}(R)$
symmetry. However, an important caveat is formulated in condition (ii):
Averaging over measurement outcomes corresponds to a loss of information that is
reflected in the state of the system becoming mixed rather than pure as in
Eq.~\eqref{eq:R-replica-N-particle-Slater-determinant}. Then, as detailed in
Appendix~\ref{sec:app-replicon-symmetry}, the $\mathrm{SU}(R)$ symmetry is
broken, even though the total number of particles is still conserved.

\subsection{Saddle-point manifold}
\label{sec:saddle-point-manifold}

The field integral over $\mathcal{G}$ and $\Sigma$ is dominated by the spatially
homogeneous and time-independent saddle points of the Keldysh action. Due to the
symmetries of the Keldysh action discussed in the previous section, there is, in
fact, a manifold of saddle points. To establish the manifold, it is sufficient
to find one particular saddle point. The full manifold is then obtained by
applying all rotations that are symmetries of the action to the particular
saddle point.

We consider first the variation of the action with respect to $\Sigma$, which
yields the saddle-point equation
\begin{equation}
  \label{eq:G-saddle-point-equation}
  \mathcal{G} = \int_{-\pi}^{\pi} \frac{\diff q}{2 \pi} \int_{-\infty}^{\infty}
  \frac{\diff \omega}{2 \pi} \frac{1}{\omega - \xi_q + \imag
    \Sigma} ,
\end{equation}
where $\xi_q = - 2 J \cos(q)$ is the dispersion relation of the Hamiltonian in
Eq.~\eqref{eq:Hamiltonian}. We omit a contribution to the saddle point of
$\mathcal{G}$ that contains $A$. Since $\Sigma \sim \gamma$ at the saddle point
as shown below, such a contribution would lead to terms of second order in
$\gamma$ when we insert $\mathcal{G}$ in the action
Eq.~\eqref{eq:S-0-G-Sigma}. In accordance with Eq.~\eqref{eq:G-symmetric-limit},
the integration over frequencies in Eq.~\eqref{eq:G-saddle-point-equation} has
to be regularized by introducing a factor
$\bigl( \e^{\imag \omega t} + \e^{- \imag \omega t} \bigr)\big/2$, where the
limit $t \to 0^+$ is taken after the integration. By writing
$ \Sigma = \mathcal{V} \lambda \mathcal{V}^{-1}$, where $\lambda$ is a diagonal
matrix, we thus find~\cite{Poboiko2023}
\begin{equation}
  \label{eq:G-saddle-point}
  \mathcal{G} = - \imag Q/2, \quad Q = \sgn \! \left[ \Re(\Sigma) \right] =
  \mathcal{V} \sgn \! \left[\Re(\lambda)\right] \mathcal{V}^{-1}.
\end{equation}
Note that the matrix $Q$ obeys the nonlinear constraint $Q^2 = 1$.

To simplify the analysis of the saddle-point equation that follows from the
variation of the action with respect to $\mathcal{G}$, we use that---as
explained above---to establish the full manifold of saddle points, it is
sufficient to find one particular saddle point. We thus focus on
replica-symmetric saddle points, $Q = Q_0 \otimes 1_R$.
Equation~\eqref{eq:G-symmetric-limit} suggests that a particular solution of the
saddle-point equation for $\mathcal{G}$ is given by the symmetrized equal-time
limit of the Green's function, which, as discussed in
Appendix~\ref{sec:caus-struct-regul}, is fully determined by fermionic
anticommutation relations and the distribution function
$\left\langle \hat{n}_l \right\rangle_{\mathrm{ss}} = n$ in the steady
state. This particular solution is obtained from Eq.~\eqref{eq:G-saddle-point}
by setting $\Sigma \propto \Lambda$, leading to $Q_0 = \Lambda$, where
\begin{equation}
  \label{eq:lambda-OM}
  \Lambda =
  \begin{pmatrix}
    1 & 2 \left( 1 - 2 n \right) \\ 0 & - 1
  \end{pmatrix}.
\end{equation}
For $n = 1/2$, we thus find $\Lambda = \sigma_z$ for fermion counting. The
variation of the action with respect to $\mathcal{G}$ can be obtained
conveniently by inserting
$\mathcal{G} = - \imag \left( Q_0 + \dQG \right) \! \big/2$ in the measurement
Lagrangian and performing an expansion in $\dQG$ as detailed in
Appendix~\ref{sec:fluctuation-expansion}. Accounting for the symmetries of the
Keldysh action discussed in Sec.~\ref{sec:symmetry-keldysh-action}, we then
obtain the manifold of saddle points for fermion counting:
\begin{equation}
  \label{eq:saddle-points-FC}
  \Sigma = \gamma Q/2^{R - 1}, \qquad Q = \mathcal{R}_{\Phi} \sigma_z \mathcal{R}_{\Phi}^{-1}.
\end{equation}
The symmetries of the Keldysh action $\mathcal{R}_{\eta}$ and
$\mathcal{R}_{\zeta}$ do not rotate the saddle point, and, therefore, do not
further enlarge the saddle-point manifold. Note that the relation between
$\Sigma$ and $Q$ is consistent with the definition of $Q$ in
Eq.~\eqref{eq:G-saddle-point}. For occupation measurements, we find
\begin{equation}
  \label{eq:saddle-points-OM}
  \Sigma = \gamma Q/(2 n), \qquad Q = \mathcal{R}_{\Phi} \mathcal{R}_{\phi}
  \Lambda \mathcal{R}_{\phi}^{-1} \mathcal{R}_{\Phi}^{-1}.
\end{equation}
Here, we omit a shift of the saddle point that vanishes in the replica limit
$R \to 1$ and is given explicitly in Appendix~\ref{sec:fluctuation-expansion}.

Each particular saddle point in the manifold spontaneously breaks the symmetries
of the Keldysh action under $\mathcal{R}_{\Phi}$ and, for occupation
measurements, also $\mathcal{R}_{\phi}$. This gives rise to Goldstone modes. The
Goldstone mode associated with $\mathcal{R}_{\phi}$ leads to the slow algebraic
decay of the unconditional density autocorrelation function
for occupation measurements, Eq.~\eqref{eq:C-0-OM}. In
contrast, for fermion counting, there are no Goldstone modes in the replica
limit $R \to 1$ that describes the unconditional dynamics---by definition, the
traceless $R \times R$ matrix $\Phi$ reduces to $\Phi = 0$ in this case. The
absence of Goldstone modes is reflected in the exponential decay of the
autocorrelation function in Eq.~\eqref{eq:C-0-FC}. However, as we discuss
further below, the Goldstone mode associated with $\mathcal{R}_{\Phi}$ is
described by the same long-wavelength effective theory for both fermion counting
and occupation measurements, leading to almost identical correlations and
entanglement properties in the steady state.

\section{Gaussian theory}
\label{sec:gaussian-theory}

Properly treating strong massless fluctuations within the saddle-point manifold
requires an RG analysis of the corresponding nonlinear sigma model. However,
first important insights can be obtained from considering quadratic fluctuations
of $\mathcal{G}$ and $\Sigma$ around a particular saddle point within the
manifold. A convenient expansion point is given by $Q = \Lambda$, where
$n = 1/2$ such that $\Lambda = \sigma_z$ for fermion counting. The Gaussian
approximation is controlled for $\gamma \ll J$ and valid up to intermediate
length scales, where renormalization corrections are small.  We will go beyond
the Gaussian approximation in the next section,
Sec.~\ref{sec:effective-field-theory}.

\subsection{Expansion in fluctuations}

Arbitrary fluctuations around the saddle point $Q = \Lambda$ can be parametrized
as
\begin{equation}
  \label{eq:G-Sigma-fluctuations}
  \mathcal{G} = -\imag \left( \Lambda + \dQG{} \right) \! \big/2, \qquad \Sigma
  = \gamma \left( \Lambda + \dQS{} \right) \! \big/(2 n),
\end{equation} 
with Hermitian $2 R \times 2 R$ matrices $\dQG$ and $\dQS$. For simplicity, we
set $R = 1$ in all numerical factors. Then, the expansion point for $\Sigma$
lies within the manifold for both fermion counting,
Eq.~\eqref{eq:saddle-points-FC}, and occupation measurements,
Eq.~\eqref{eq:saddle-points-OM}, when we set $n = 1/2$ for fermion counting. As
detailed in Appendix~\ref{sec:fluctuation-expansion}, the Keldysh action
vanishes to zeroth order in fluctuations, and there are no contributions to
first order for an expansion around a saddle point. At second order, we find
\begin{equation}
  S^{(2)} = S_0^{(2)} + \gamma \int \diff^2
  \mathbf{x} \, \mathcal{L}_M^{(2)},
\end{equation} 
where~\cite{Poboiko2023}
\begin{multline}
  \label{eq:S-0-2-B}
  \imag S_0^{(2)} = \frac{1}{\left(4 \tau_0\right)^2} \int \diff^2 \mathbf{x} \,
  \diff^2 \mathbf{x}' \, \mathcal{B}_{l - l'}(t - t') \\ \times \Tr \!  \left[
    \delta Q_{\Sigma, l}(t) \left( 1 + \Lambda \right) \delta Q_{\Sigma, l'}(t')
    \left( 1 - \Lambda \right) \right] \\ - \frac{1}{4 \tau_0} \int \diff^2
  \mathbf{x} \Tr \! \left[ \delta Q_{\Sigma, l}(t) \delta Q_{\mathcal{G}, l}(t)
  \right],
\end{multline}
with the mean free time $\tau_0 = n/\gamma$ and the block of the diffusion
ladder $\mathcal{B}_{l - l'}(t - t')$, which, in momentum and frequency space, reads
\begin{equation}
  \label{eq:diffusion-ladder-block}
  \mathcal{B}_q(\omega) = \left\{ \left( \imag \omega - 1/\tau_0 \right)^2 +
    \left[ 4 J \sin(q/2) \right]^2 \right\}^{-1/2}.
\end{equation}
The expansion of the measurement Lagrangian is carried out in
Appendix~\ref{sec:fluctuation-expansion}. There are in general two contributions,
\begin{equation}
  \mathcal{L}_M^{(2)} = \mathcal{L}_M^{(2, 1)} + \mathcal{L}_M^{(2, 2)}.
\end{equation}
For fermion counting, the first contribution contains the trace of a square and
the second contribution the square of a trace,
\begin{equation}
  \label{eq:lagrangian-gain-loss-1-2}
  \begin{split}    
    \imag \gamma \mathcal{L}_M^{(2,1)} & = -\frac{1}{16 \tau_0}
    \Tr\!\left[ \Bigl( \sigma_z \dQG \Bigr)^2 - \left( \sigma_y \dQG \right)^2
    \right], \\
    \imag \gamma \mathcal{L}_M^{(2,2)} & = \frac{1}{16 \tau_0}
    \left[ \Tr \Bigl( \sigma_z \dQG \Bigr)^2 - \Tr\!\left( \sigma_y \dQG
      \right)^2 \right].
  \end{split}
\end{equation}
Similarly, for generalized occupation measurements we find
\begin{equation}
  \label{eq:lagrangian-occupation-1-2}
  \begin{split}
    \imag \gamma \mathcal{L}_M^{(2, 1)} & = - \frac{1}{32 n
      \tau_0} \Tr \!
    \left\{ \left[ \left( \Lambda - \sigma_x \right) \dQG \right]^2 \right\}, \\
    \imag \gamma \mathcal{L}_M^{(2, 2)} & = \frac{1}{32 n \tau_0}
    \Tr \! \left[ \left( \Lambda - \sigma_x \right) \dQG \right]^2.
  \end{split}
\end{equation}
At this point, it is useful to decompose the fluctuation matrices into
longitudinal or replica-symmetric and transversal or replicon modes according to
\begin{equation}
  \delta Q = \delta Q^{\parallel} \otimes \id_R + \delta Q^{\perp},
\end{equation}
where, by construction, $\delta Q^{\parallel} = \left( 1/R \right) \trR(\delta Q)$ and $\delta
Q^{\perp} = \delta Q - \delta Q^{\parallel} \otimes \id_R$. These modes are orthogonal,
\begin{equation}
  \trR \! \left[ \left( \delta Q^{\parallel} \otimes 1_R \right) \delta Q^{\perp} \right] = 0,
\end{equation} 
and, therefore, the theory splits into two sectors,
$Z_R = Z_R^{\parallel} Z_R^{\perp}$. The replica-symmetric and replicon sectors
describe, respectively, unconditional and conditional correlation functions.

\subsection{Density correlations}
\label{sec:dens-corr-funct}

We consider two types of connected density correlation functions: the
unconditional dynamics are described by
\begin{equation}
  \label{eq:C-0-definition}
  C_{0, l, l'}(t, t') = \frac{1}{2} \overline{\left\langle \left\{
        \hat{n}_l(t), \hat{n}_{l'}(t') \right\} \right\rangle} - \overline{\left\langle
      \hat{n}_l(t) \right\rangle} \; \overline{\left\langle \hat{n}_{l'}(t')
    \right\rangle},
\end{equation}
which reduces to Eq.~\eqref{eq:C-0-autocorrelation} for $l = l'$, and equal-time
correlations under conditional dynamics are captured by
\begin{equation}
  \label{eq:C-l-l-prime-t}
  C_{l, l'}(t) = \frac{1}{2} \overline{\left\langle \left\{ \hat{n}_l(t),
        \hat{n}_{l'}(t) \right\} \right\rangle} - \overline{\left\langle
      \hat{n}_l(t) \right\rangle \left\langle \hat{n}_{l'}(t) \right\rangle}.
\end{equation}
A unified description of both types of correlation functions can be obtained in
replica Keldysh field theory by considering
\begin{equation}
  \label{eq:C-r-r-prime}
  C_{r, r', l, l'}(t, t') = \left\langle \delta \rho_{r, l}(t) \delta \rho_{r',
      l'}(t') \right\rangle,
\end{equation}
where $r$ and $r'$ are replica indices and density fluctuations are related to
the fluctuation matrix $\dQG$ by~\cite{Poboiko2023}
\begin{equation}
  \delta \rho_{r, l}(t) = - \frac{1}{4} \trK \! \left[ \sigma_x \delta
    Q_{\mathcal{G}, r, r, l}(t) \right].
\end{equation}
In particular, replica-diagonal and replica-offdiagonal components determine the
unconditional and conditional density correlation functions as
\begin{align}
  \label{eq:replica-symmetric-density-correlation}
  C_{0, l, l'}(t, t') & = C_{r, r, l, l'}(t, t'), \\
  \label{eq:replicon-density-correlation}
  C_{l, l'}(t)  & = C_{r, r, l, l'}(t, t) - C_{r, r', l, l'}(t, t).
\end{align}
Due to the symmetry under permutations of replicas, the result does not depend
on the choice of $r \neq r'$. It is worthwhile to recall that the formalism is
constructed to enable the computation of conditional averages as in
Eqs.~\eqref{eq:A-B-replicas} and~\eqref{eq:C-l-l-prime-t} at equal times. While
Eq.~\eqref{eq:C-r-r-prime} can be evaluated at arbitrary times $t$ and $t'$,
only $t = t'$ has a clear physical meaning for $r \neq r'$.

The correlation function, Eq.~\eqref{eq:C-r-r-prime}, can be evaluated by
introducing sources that couple to density fluctuations, that is, by adding to
the action a contribution
\begin{equation}
  \label{eq:S-xi}
  \imag S_{\xi} = \imag \int \diff^2 \mathbf{x} \sum_{r = 1}^R
  \xi_{r, l}(t) \delta \rho_{r, l}(t),
\end{equation}
and taking derivatives with respect to the sources after performing the Gaussian
integral over $\dQG$ and $\dQS$. To that end, an explicit parametrization of the
Hermitian matrices $\dQG$ and $\dQS$ has to be chosen. It is convenient to
expand the fluctuation matrices in a basis of Pauli matrices in Keldysh
space. Replica-symmetric fluctuations $\delta Q^{\parallel}$ are fully
determined by the scalar coefficients in this expansion. For replicon modes
$\delta Q^{\perp}$, the coefficients themselves are traceless Hermitian
$R \times R$ matrices, which can be parametrized by decomposing them into their
real and imaginary parts, taking into account the restrictions imposed by
tracelessness and Hermiticity.

We consider correlations in the steady state, which are invariant under
translations in both space and time, such that
$C_{0, l, l'}(t, t') = C_{0, l - l'}(t - t')$, whereas
$C_{l, l'}(t) = C_{l - l'}$ becomes independent of $t$. For fermion counting, we
then find the unconditional correlation function
\begin{equation}
  \label{eq:C-0-l-FC}
  C_{0, l}(t) = \frac{1}{4} \e^{-\abs{t}/\tau_0} J_l(2 J \abs{t})^2,
\end{equation}
which reduces to Eq.~\eqref{eq:C-0-FC} for $l = 0$. By taking the
derivative with respect to time of Eq.~\eqref{eq:C-0-definition} and using the
explicit representation of two-time averages in
Eq.~\eqref{eq:quantum-regression}, it is straightforward to check that this is,
in fact, the exact result~\cite{Starchl2022, Starchl2024}.

For generalized occupation measurements, we obtain
\begin{equation}
  \label{eq:C-0-l-OM}
  C_{0, l}(t) \sim n \left( 1 - n \right) \frac{\e^{- l^2/(4 \nu
      \abs{t})}}{\sqrt{4 \pi \nu \abs{t}}},
\end{equation}
with $\nu = 2 n J^2/\gamma$. This result is valid in the limit of large
distances $l \gg l_0$, where the mean free path is given by
\begin{equation}
  \label{eq:l-0}
  l_0 = \sqrt{2} J n / \gamma,
\end{equation}
or at long times $t \gg \tau_0$. Equation~\eqref{eq:C-0-l-OM} agrees with the
corresponding result for random projective measurements obtained in
Ref.~\cite{Poboiko2023} when we set $n = 1/2$ in the expression for $\nu$ but
not in the prefactor of the correlation function. For $l = 0$ and $n = 1/2$, we
recover Eq.~\eqref{eq:C-0-OM}. From the structure of the measurement-induced
interaction vertices, it follows that the asymptotic behavior described by
Eq.~\eqref{eq:C-0-l-OM} obtained in Gaussian approximation is actually
exact~\cite{Poboiko2023}.

In contrast to the completely different unconditional correlation functions,
Eqs.~\eqref{eq:C-0-l-FC} and~\eqref{eq:C-0-l-OM}, we find the conditional
density correlation function in the steady state and for $n = 1/2$ to be
identical for fermion counting and both generalized and projective measurements
of occupation numbers~\cite{Poboiko2023}. This will be justified further in
Sec.~\ref{sec:effective-field-theory} where, instead of considering arbitrary
Gaussian fluctuations as in Eq.~\eqref{eq:G-Sigma-fluctuations}, we focus on
fluctuations of massless Goldstone modes, which obey the same long-wavelength
effective theory in all three models.

To calculate the conditional correlation function, one has to take the boundary
condition induced by stopping the measurement process at a finite time $t$ into
account~\cite{Poboiko2023}. For projective measurements of occupation numbers,
this leads to the correlation function in momentum space, $C_q$, being expressed
in terms of the solution of an integral equation that depends on a single
parameter, $u = 2 l_0 \sin(q/2)$. However, one obtains approximately the same
result for the correlation function by omitting the boundary condition and
instead rescaling $u$ by a factor of two, which yields~\cite{Poboiko2023}
\begin{equation}
  \label{eq:C-q-Gaussian}
  C_q \approx n \left( 1 - n \right) \tilde{c}(q l_0),
\end{equation}
where
\begin{equation}
  \label{eq:tilde-c-FC}
  \tilde{c}(u) = \frac{2}{\pi} \int_0^{\infty} \diff v \, \frac{\Re[b(2 u, v)] -
    \abs{b(2 u, v)}^2}{1 - \Re[b(2 u, v)]},
\end{equation}
and
\begin{equation}
  b(u, v) = \left[ \left( 1 - \imag v \right)^2 + 2 u^2 \right]^{-1/2}.
\end{equation}
For $n = 1/2$, we find the same expressions also for fermion counting and for
generalized occupation measurements; for generalized occupation measurements
with $n \neq 1/2$, Eq.~\eqref{eq:tilde-c-FC} is modified, with
$f = 1 - 2n$, as
\begin{equation}
  \label{eq:tilde-c-generalized}
  \tilde{c}(u) = \frac{4 n}{\pi} \int_0^{\infty} \diff v \, \frac{2 n \Re[b(2 u,
    v)] -
    \abs{b(2 u, v)}^2}{4 n^2 \left\{ 1 - \Re[b(2 u, v)] \right\} - f \abs{b(2 u,
      v)}^2}.
\end{equation}

According to Eq.~\eqref{eq:C-q-Gaussian}, $l_0$ is the only length scale that
determines the behavior of spatial correlations. Taking the inverse Fourier
transform of Eq.~\eqref{eq:C-q-Gaussian}, one finds algebraic decay on scales
$l \gg l_0$~\cite{Poboiko2023},
\begin{equation}
 \label{eq:connected-density-correlation-function-Gaussian}
 C_l \sim - n \left( 1 - n \right) \frac{2 l_0}{\pi l^2}.
\end{equation}
From the conditional density correlation function, the second cumulant,
Eq.~\eqref{eq:cumulant}, of the subsystem number of particles follows as
\begin{equation}
  \label{eq:cumulant-density-correlation}
  C_A^{(2)} = \sum_{l,l'\in A} C_{l - l'},
\end{equation}
which allows us to calculate the entanglement entropy to leading order in the
cumulant expansion, Eq.~\eqref{eq:entropy-cumulant}. The algebraic decay of $C_l$,
Eq.~\eqref{eq:connected-density-correlation-function-Gaussian}, leads to
logarithmic growth of the entanglement entropy of a subsystem of length
$\ell$~\cite{Poboiko2023},
\begin{equation}
 \label{eq:S-vN-Gaussian}
 \svn{\ell} \sim \frac{4 \pi}{3} n \left( 1 - n \right) l_0 \ln(\ell/l_0).
\end{equation}
Algebraic decay of spatial correlations and logarithmic growth of the
entanglement entropy are characteristic for a critical phase with conformal
invariance~\cite{Alberton2021}. However, strong fluctuations of Goldstone modes
that are not captured by the Gaussian approximation lead to a substantial
renormalization of the effective measurement rate, which invalidates the
Gaussian approximation at large scales. To treat this effect properly, we now
turn to an NLSM for fluctuations of Goldstone modes within the saddle-point
manifold.

\section{Effective field theory in the rare-measurement limit}
\label{sec:effective-field-theory}

Going beyond the Gaussian approximation, we generalize
Eq.~\eqref{eq:G-Sigma-fluctuations} to incorporate nonlinear fluctuations around
the saddle point $\Lambda$~\cite{Poboiko2023}:
\begin{equation}
  \label{eq:G-Sigma-nonlinear-fluctuations}
  \mathcal{G} = - \imag Q/2, \qquad \Sigma = \gamma Q/(2 n), \qquad Q =
  \mathcal{R} \Lambda \mathcal{R}^{-1},
\end{equation}
again with the understanding that $n = 1/2$ and thus $\Lambda = \sigma_z$ for
fermion counting. There is a subgroup $\mathrm{U}(R) \times \mathrm{U}(R)$ of
rotations $\mathcal{R}$ that leave the saddle point $\Lambda$
invariant. Therefore, nontrivial rotations of the saddle point form the
symmetric space $\mathrm{U}(2 R)/\mathrm{U}(R) \times \mathrm{U}(R)$. We
parametrize these rotations as
\begin{equation}
  \label{eq:Q-parametrized}
  Q = \mathcal{R}_{\Phi} \mathcal{R}_{\Theta} Q_0 \mathcal{R}_{\Theta}^{-1}
  \mathcal{R}_{\Phi}^{-1},
\end{equation}
with
\begin{equation}
  \label{eq:R-Phi-R-Theta}
  \mathcal{R}_{\Phi} = \e^{\imag \Phi \sigma_x/2}, \qquad \mathcal{R}_{\Theta} =
  \e^{\imag \Theta \sigma_y/2},
\end{equation}
where $\Phi$ and $\Theta$ are traceless Hermitian matrices in replica
space. Furthermore,
\begin{equation}
  \label{eq:Q0-parametrized}
  Q_0 = \mathcal{R}_{\phi} \mathcal{R}_{\theta} \Lambda
  \mathcal{R}_{\theta}^{-1} \mathcal{R}_{\phi}^{-1},
\end{equation}
where $\mathcal{R}_{\phi}$ and $\mathcal{R}_{\theta}$ describe replica-symmetric
rotations by the angles $\phi$ and $\theta$,
\begin{equation}
  \label{eq:R-phi-R-theta}
  \mathcal{R}_{\phi} = \e^{\imag \phi \sigma_x/2}, \qquad \mathcal{R}_{\theta} =
  \e^{\imag \theta \sigma_y/2}.
\end{equation}
This parametrization is chosen such that the symmetries of the action,
identified in Sec.~\ref{sec:symmetry-keldysh-action}, appear explicitly as
factors in the rotations $\mathcal{R}$. In particular, for fermion counting the
only symmetry is given by $\mathcal{R}_{\Phi}$, whereas for occupation
measurements also $\mathcal{R}_{\phi}$ is a symmetry. Considering now matrices
$\Phi$ and $\Theta$ and angles $\phi$ and $\theta$ that vary slowly in space and
time, an effective long-wavelength field theory, given by an NLSM for the
massless Goldstone modes associated with the broken symmetries of the action, is
obtained by integrating out massive modes in the Gaussian approximation. Technical details of
this procedure are presented in Appendix~\ref{sec:derivation-nlsm}.

\subsection{Replica-symmetric sector}

We first consider the replica-symmetric sector of the theory, where we can set
$R = 1$ and which describes fluctuations of the modes $\phi$ and $\theta$. For
fermion counting, both modes are massive and an expansion to second order
yields the Lagrangian
\begin{equation}
  \label{eq:L-replica-symmetric-FC}
  \imag \mathcal{L}[Q_0] = - \frac{\imag}{2} \theta
  \partial_t \phi - \frac{\nu}{4} \left[ \left( \partial_x \phi \right)^2 +
    \left( \partial_x \theta \right)^2 \right] - \frac{\gamma}{2} \left(
    \theta^2 + \phi^2 \right).
\end{equation}
To describe long-wavelength fluctuations, we have taken a spatial continuum
limit, in which the lattice site index $l$ is replaced by a continuous position
variable $x$, and we have performed an expansion in spatial and temporal
derivatives as detailed in Appendix~\ref{sec:derivation-nlsm}. As an artefact of
the gradient expansion, the diffusion coefficient $\nu = J^2/\gamma$, which does
not occur in the exact correlation function, Eq.~\eqref{eq:C-0-l-FC}, appears in
Eq.~\eqref{eq:L-replica-symmetric-FC}.

For occupation measurements, $\phi$ is massless but we can still expand the
Lagrangian in $\theta$. We obtain
\begin{multline}
  \label{eq:L-replica-symmetric-OM}
  \imag \mathcal{L}[Q_0] = - \frac{\imag}{4} \left( 2 + f \theta \right)
  \theta
  \partial_t \phi \\ - \frac{\nu}{4} \left\{ \left[ 1 - f^2 + 2 f \theta -
      \left( 1 - f^2 \right) \theta^2 \right] \left( \partial_x \phi \right)^2
  \right. \\ \left. - \imag 2 f \left( f - \theta \right) \left( \partial_x \phi
    \right) \left( \partial_x \theta \right) + \left( 1 + f^2 \right)
    \left( \partial_x \theta \right)^2 \right\},
\end{multline}
with $f = 1 - 2 n$. Recall that we have expanded the action to first order in
$\gamma$, both in Eq.~\eqref{eq:S-0-G-Sigma} and in the regularization described
in Appendix~\ref{sec:caus-struct-regul}. At this order of approximation, the
expected mass term $\sim \theta^2$ does not occur in
Eq.~\eqref{eq:L-replica-symmetric-OM}.

\subsection{Replicon sector}
\label{sec:replicon-sector}

We now turn to the replicon sector. For both fermion counting and occupation
measurements, the traceless Hermitian matrices $\Phi$ are Goldstone modes. These
modes parametrize the group $\mathrm{SU}(R)$ as $U = \e^{\imag \Phi}$.
Integration over the massive modes $\Theta$ yields an NLSM with Lagrangian given
by
\begin{equation}
  \label{eq:L-replicon}
  \imag \mathcal{L}[\Phi] = -\frac{g[Q_0]}{2} \trR\!\left(
    \frac{1}{v[Q_0]} \partial_t U^{\dagger} \partial_t U +
    v[Q_0] \partial_x U^{\dagger} \partial_x U \right).
\end{equation}
For fermion counting and occupation measurements, the velocity $v[Q_0]$ and
coupling constant $g[Q_0]$ are, respectively,
\begin{equation}
  \begin{aligned}
    v[Q_0] & = \left(2 l_0/\tau_0\right) \sqrt{\rho_0 \left( 1 - \rho_0
      \right)}, &
    g[Q_0] & = l_0 \sqrt{\rho_0 \left( 1 - \rho_0 \right)}, \\
    v[Q_0] & = \left(l_0/\tau_0\right) \sqrt{2 \left( 1 - \rho_0\right)}, &
    g[Q_0] & = l_0 \rho_0 \sqrt{2 \left( 1 - \rho_0 \right)}.
  \end{aligned}
\end{equation}
A coupling between the replica-symmetric and replicon sectors of the theory is
established through the dependence of these quantities on
\begin{equation}
  \label{eq:rho-0}
  \rho_0 = \frac{1}{4} \trK \! \left( 1 - Q_0 \sigma_x \right).
\end{equation}
However, to leading order in $\gamma \ll J$, we can neglect fluctuations in the
replica-symmetric sector and set $Q_0 = \Lambda$ such that $\rho_0 = n$. This
leads to
\begin{equation}
  v_0 = v[\Lambda] = \frac{l_0}{\tau_0} \sqrt{2 \left( 1 - n \right)},
  \quad g_0 = g[\Lambda] = l_0 n \sqrt{2 \left( 1 - n \right)},
\end{equation}
where $n = 1/2$ for fermion counting. Indeed, for $n = 1/2$, the velocity $v_0$
and the coupling constant $g_0$ and, therefore, the NLSM Lagrangian in
Eq.~\eqref{eq:L-replicon}, are identical to the corresponding expressions for
projective measurements of occupation numbers~\cite{Poboiko2023}. Therefore, we
can expect the same quantitative behavior of correlations and entanglement in
the conditional dynamics for all three types of measurements: fermion counting
and random generalized as well as projective measurements of occupation
numbers.

As indicated at the beginning of Sec.~\ref{sec:replica-keldysh-field-theory}, so
far, we have not taken the PHS of the Hamiltonian, Eq.~\eqref{eq:Hamiltonian},
into account. PHS does not affect the theory on the Gaussian level. However, PHS
does modify the target manifold of the NLSM. The target manifold can be
determined even without constructing the NLSM explicitly. Key steps of this
analysis are summarized in Appendix~\ref{sec:NLSM-with-PHS}. Focusing on the
replicon sector of the theory, we find that the target manifold of the NLSM for
a Hamiltonian with PHS is $\mathrm{SU}(2 R) / \mathrm{Sp}(R)$. This applies to
fermion counting as well as for generalized occupation measurements, and is in
agreement with other models with a conserved number of particles~\cite{Fava2024,
  Poboiko2025}. Crucially, the $\mathrm{SU}(2 R) / \mathrm{Sp}(R)$ and
$\mathrm{SU}(R)$ NLSMs for Hamiltonians with and without PHS, respectively, lead
to the same qualitative behavior of entanglement and correlations on large
scales. This behavior is determined by the RG flow of the coupling constant
$g_0$. To one-loop order, the RG flow is given by~\cite{Poboiko2025, Hikami1981,
  Wegner1989, Evers2008}
\begin{equation}
  \label{eq:g-renormalized}
  g = g_0 - \frac{1}{4 \pi \beta} \ln(l/l_0),
\end{equation}
where $\beta = 1/2$ and $\beta = 1$ for intact and broken PHS, respectively. In
both cases, and for any value of the bare coupling $g_0$, the flow reaches the
strong-coupling regime $g \lesssim 1$ at the exponentially large scale
\begin{equation}
  \label{eq:l-star}
  l_{*} = l_0 \e^{4 \pi \beta g_0},
\end{equation}
indicating that a frequent-measurement phase with area-law entanglement is
established in the thermodynamic limit.

Numerically observing emergent area-law behavior due to the renormalization of
$g$ on exponentially large scales is challenging. However, as we show in the
following, the renormalization of $g$ leaves clear fingerprints on much shorter
scales.

\section{Correlations and entanglement in the steady state}
\label{sec:correlations-and-entanglement}

We will now compare analytical predictions from replica Keldysh field theory to
numerical results obtained from the quantum jump algorithm outlined in
Sec.~\ref{sec:stochastic-schrodinger-equation}. In contrast to
Sec.~\ref{sec:generalized-Zeno-effect}, where we have studied time-dependent
correlations, here we focus on equal-time correlations and entanglement in the
steady state. Studying carefully the crossover to area-law entanglement on large
scales, we reconcile numerical observations of algebraic correlations,
logarithmic growth of the entanglement entropy, and conformal invariance
characteristic of a critical phase~\cite{Alberton2021}, with the absence of a
measurement-induced phase transition in the thermodynamic limit. Crucially,
clear numerical signatures of the absence of a critical phase in the
thermodynamic limit are visible on length scales much shorter than the
exponentially large scale $l_{*}$ in Eq.~\eqref{eq:l-star}, and these signatures
can readily be explained by the theory introduced above. We first investigate
algebraic scaling of the connected density correlation function, then focus on
the entanglement entropy and the effective central charge, and finally we study
the bipartite and tripartite mutual information.

\subsection{Connected density correlation function}
\label{sec:connected-density-correlation-function}

In the steady state, the conditional density correlation function,
Eq.~\eqref{eq:C-l-l-prime-t}, becomes time-independent and translationally
invariant, $C_{l, l'}(t) = C_{l - l'}$. Using Wick's theorem, the density
correlation function can be expressed in terms of the single-particle density
matrix $D$ defined in Eq.~\eqref{eq:single-particle-density-matrix}:
\begin{equation}
  \label{eq:connected-density-correlation-function}
  C_{l - l'} = \overline{\left\langle \hat{n}_l \hat{n}_{l'} \right\rangle} -
  \overline{\left\langle \hat{n}_l \rangle \langle \hat{n}_{l'}
    \right\rangle} = \frac{\delta_{l, l'}}{2}
  - \overline{\abs{D_{l, l'}}^2}.
\end{equation}
As explained in Sec.~\ref{sec:stochastic-schrodinger-equation}, the quantum jump
algorithm gives direct access to the single-particle density matrix. Here and in
the following, we set
$\overline{\left\langle \hat{n}_l \right\rangle} = n = 1/2$ as in our numerical
studies. Furthermore, for all numerical results, the overbar indicates averaging
over both trajectories and position.

\subsubsection{Analytical predictions}
\label{sec:connected-density-correlation-function-analytics}

As explained in Sec.~\ref{sec:dens-corr-funct}, on the Gaussian level the
connected density correlation is identical for fermion counting and random
generalized as well as projective measurements of occupation numbers. In
momentum space, it is given by Eq.~\eqref{eq:C-q-Gaussian}. To incorporate the
renormalization of $l_0 = 2 g_0$, we expand the Gaussian result for low momenta
and replace $g_0$ by its renormalized value $g$, Eq.~\eqref{eq:g-renormalized},
where we identify $l = 1/q$ with the external momentum of the correlation
function that cuts off the RG flow in the infrared~\cite{Poboiko2023}. We thus
obtain the rescaled correlation function
\begin{equation}
  \label{eq:C-q-renormalized}
  \frac{C_q}{g_0 q} \approx 1 - 2 q l_0 + \frac{1}{4 \pi \beta g_0} \ln(q l_0).
\end{equation}
For $q \to 0$, the logarithmic renormalization dominates, pushing $C_q/(g_0 q)$
to small values, while $C_q/(g_0 q) \to 1$ for $q \to 0$ in the Gaussian
theory. The momentum scale at which the logarithm starts to dominate marks the
beginning of the crossover from Gaussian behavior---with algebraic correlations,
Eq.~\eqref{eq:connected-density-correlation-function-Gaussian}, and logarithmic
growth of the entanglement entropy, Eq.~\eqref{eq:S-vN-Gaussian}---to the
emergent frequent-measurement regime. We identify this crossover scale with the
maximum of the rescaled correlation function at
$q_c = \bigl( 4 \pi \beta l_0^2 \bigr)^{-1} \sim \gamma^2$. Associated with this
momentum scale is a crossover length scale $l_c \sim \gamma^{-2}$. Importantly,
the crossover scale $l_c$ is only algebraically large in $\gamma$ and,
therefore, numerically accessible.

\subsubsection{Numerics}
\label{sec:connected-density-correlation-function-numerics}

\begin{figure}
  \centering   
  \includegraphics[width = \linewidth]{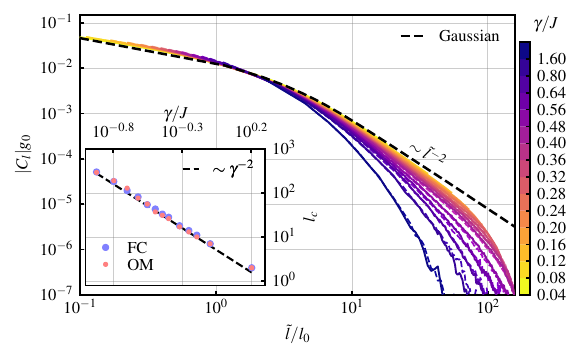}
  \caption{Rescaled density correlation function for fermion counting (solid
    lines) and generalized occupation measurements (colored dashed lines) for
    various monitoring rates $\gamma$. For $\gamma \ll J$, the data approaches
    the prediction from the Gaussian theory (black dashed line) with an
    algebraic scaling of $\sim \tilde{l}^{-2}$ for $\tilde{l}/l_0 \gg 1$. Inset:
    Upper boundary of the critical range $l_c$ for fermion counting (FC, blue
    dots) and occupation measurements (OM, red dots), agreeing well with the
    analytical prediction $l_c \sim \gamma^{-2}$ (black dashed line). The system
    size is $L = 1000$ and data are averaged over $N_{\mathrm{traj}} = 160$
    trajectories for this and all subsequent figures.}
  \label{fig:4}
\end{figure}

Figure~\ref{fig:4} shows the density correlation function,
Eq.~\eqref{eq:connected-density-correlation-function}, as a function of the
chord length $\tilde{\ell} = \left( L/\pi \right) \sin(\pi \ell/L)$, which
accounts for the finite system size and periodic boundary conditions, for both
fermion counting (solid lines) and generalized occupation measurements (dashed
lines). The quantitative agreement between both models despite their
fundamentally different dynamics is remarkable. For comparison, we show the
Gaussian result, obtained by numerically taking the inverse Fourier transform of
Eq.~\eqref{eq:C-q-Gaussian} (black dashed line). The Gaussian result transitions
from slow decay at short distances $\tilde{l} \ll l_0$ to algebraic scaling
$C_l \sim \tilde{l}^{-2}$ on large scales $\tilde{l} \gg l_0$. For small values
of $\gamma$, the numerical data follows the Gaussian result up to intermediate
scales. However, in accord with the renormalized correlation function,
Eq.~\eqref{eq:C-q-renormalized}, the decay of the correlation function visibly
becomes faster than algebraic beyond a crossover scale $l_c$. To quantify the
deviation from the Gaussian prediction, we define $l_c$ as the scale beyond
which the numerical data deviates from a line $\sim \tilde{l}^{-2}$, put
tangentially to the data, by more than $10 \, \%$. As shown in the inset of
Fig.~\ref{fig:4}, the crossover scale exhibits the scaling
$l_c \sim \gamma^{-2}$, confirming our expectation based on replica Keldysh
field theory. The critical range of algebraic behavior of spatial correlations
is thus bounded from below by $l_0$ and from above by $l_c$ as sketched in
Fig.~\ref{fig:2}(c).

\subsection{Entanglement entropy}
\label{sec:entanglement-entropy}

We now turn to the von Neumann entanglement entropy $S_{\ell}$,
Eq.~\eqref{eq:entanglement-entropy}, of a subsystem $A$ consisting of $\ell$
contiguous lattice sites. The RG flow of the NLSM indicates that area-law
scaling of the entanglement entropy is established beyond the exponentially
large scale $l_{*}$ in Eq.~\eqref{eq:l-star}. However, logarithmic growth of
$S_{\ell}$ as characteristic for a one-dimensional conformal field theory with
central charge $c$~\cite{Calabrese2004, Calabrese2009},
\begin{equation}
  \label{eq:entanglement-entropy-CFT}
  \svn{\ell} \sim \frac{c}{3} \ln \! \left( \ell \right),
\end{equation}
can be observed within the critical range $\ell \in [l_0, l_c]$.

\subsubsection{Analytical predictions}

The Gaussian theory predicts logarithmic growth of the entanglement entropy,
Eq.~\eqref{eq:S-vN-Gaussian}, with a central charge given by $2 \pi g_0$. To
quantify the agreement of our numerical results with this prediction, we find it
useful to introduce an effective scale-dependent central charge
$c_{\ell}$. Equation~\eqref{eq:entanglement-entropy-CFT} suggests to define
$c_{\ell}$ as the derivative of $S_\ell$ with respect to $\ln(\ell)$. However,
since $\ell$ is an integer, we define $c_{\ell}$ as the discrete difference,
\begin{equation}
  \label{eq:effective-central-charge}
  c_{\ell} = 3 \frac{\svn{\ell'} - \svn{\ell}}{\ln(\ell'/\ell)},
\end{equation}
where $\ell'$ should be chosen such that $\ell'/\ell = \e^{\epsilon}$ with
$\epsilon \ll 1$. For logarithmic growth of the entanglement entropy as in
Eq.~\eqref{eq:entanglement-entropy-CFT}, this definition yields $c_{\ell} = c$;
volume-law scaling results in $c_{\ell} \sim \ell$, while area-law behavior leads
to $c_{\ell} = 0$.

To obtain an analytical prediction for $c_{\ell}$, we extend the Gaussian
result, Eq.~\eqref{eq:S-vN-Gaussian}, by including higher orders in an
asymptotic expansion in $\ell \gg l_0 \gg 1$ and incorporating the
renormalization of $g_0$, Eq.~\eqref{eq:g-renormalized}. This can be achieved by
expressing the entanglement entropy $S_{\ell}$ using the cumulant expansion,
Eq.~\eqref{eq:cumulant-density-correlation}, and Eq.~\eqref{eq:entropy-cumulant}
as
\begin{equation}
  \label{eq:entanglement-entropy-C-q}
  S_{\ell} = \frac{2 \pi}{3} \int_0^{\infty} \frac{\diff q}{q^2} C_q \left[1 -
    \cos(q\ell)\right]
\end{equation}
and expanding the Gaussian result for $C_q$, Eq.~\eqref{eq:C-q-Gaussian}, to third
order in $q l_0$. We thus obtain
\begin{equation}
  \label{eq:entanglement-entropy-replica-Keldysh}
  \frac{\svn{\ell}}{2 \pi g_0/3} \sim \ln \! \left( \frac{\ell}{l_0} \right) +
  s_0 + \frac{7 l_0^2}{\ell^2} - \frac{1}{8 \pi \beta g_0} \ln \! \left(
    \frac{\ell}{l_0} \right)^2,
\end{equation}
where $s_0$ is a constant that does not depend on $\ell$. The first three terms
on the right-hand side describe the asymptotic behavior of $S_{\ell}$ predicted
by the Gaussian theory, with the leading contribution reproducing the CFT form,
Eq.~\eqref{eq:entanglement-entropy-CFT}, with a central charge $c = 2 \pi g_0$.
Renormalization effects are contained in the last term. The effective central
charge, Eq.~\eqref{eq:effective-central-charge}, is then
\begin{equation}
  \label{eq:effective-central-charge-replica-Keldysh}
  \frac{c_{\ell}}{2 \pi g_0} \sim 1 - \frac{14 l_0^2}{\ell^2} - \frac{1}{4 \pi
    \beta g_0} \ln \! \left( \frac{\ell}{l_0} \right).
\end{equation}
As explained above, $c_{\ell} \to 2 \pi g_0$ for $\ell \gg l_0$ would correspond
to logarithmic growth of the entanglement entropy. However, the last term in
Eq.~\eqref{eq:effective-central-charge-replica-Keldysh}, which is due to the
renormalization of $g_0$, causes the value of $c_{\ell}$ to decrease with
$\ell$. Consequently, $c_{\ell}$ is never constant but only takes a maximum at
\begin{equation}
  \label{eq:l-max-c}
  \ell_{\mathrm{max}, c} \sim 2 \sqrt{14 \pi \beta} l_0^{3/2}.
\end{equation}
The scaling $\ell_{\mathrm{max}, c} \sim l_0^{3/2} \sim \gamma^{-3/2}$ implies
that the maximum of the effective central charge $c_{\mathrm{max}}$ is within
the critical range, which is bounded from below by $l_0 \sim \gamma^{-1}$ and
from above by the crossover scale $l_c \sim \gamma^{-2}$. In the vicinity of the
maximum, the entanglement entropy again obeys the CFT form,
Eq.~\eqref{eq:entanglement-entropy-CFT}, but with
$c = c_{\mathrm{max}} < 2 \pi g_0$ approaching the Gaussian result
$c = 2 \pi g_0$ only for $\gamma \ll J$. It is worthwhile to reiterate that
while the Gaussian theory predicts unbounded logarithmic growth of the
entanglement entropy on scales $\ell \gtrsim l_0$ and with $c = 2 \pi g_0$, the
renormalization of $g_0$ results in the range of logarithmic growth being
restricted to $\ell \lesssim l_c$, and with a reduced value of the effective
central charge $c_{\mathrm{max}} < 2 \pi g_0$.

\subsubsection{Numerics}

\begin{figure}
  \centering
  \includegraphics[width = \linewidth]{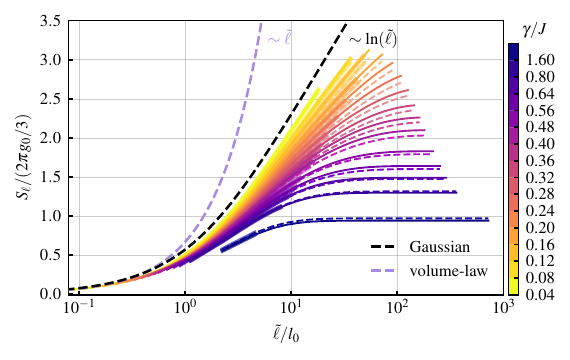}
  \caption{Rescaled entanglement entropy for fermion counting (solid lines) and
    generalized occupation measurements (colored dashed lines), for a subsystem
    of size $\ell$ of a system with periodic boundary conditions (see
    inset). Thicker lines indicate the extent of the critical range
    $\tilde{\ell} \in [l_0, l_c]$. The critical range is analogously indicated
    in the figures below. For $\gamma \ll J$, the numerical data approach
    the Gaussian result obtained by inserting Eq.~\eqref{eq:C-q-Gaussian} in
    Eq.~\eqref{eq:entanglement-entropy-C-q} (black dashed line), which grows
    logarithmically $\sim \ln(\tilde{\ell})$ for $\tilde{\ell} \gg l_0$.
    Volume-law scaling with $\svn{\ell} \sim \ell$ can be observed at
    $\tilde{\ell} \lesssim l_0$ (purple dashed line).}
  \label{fig:5}
\end{figure}

To confirm these predictions, we have determined the entanglement entropy and
effective central charge numerically. For a subsystem $A$ of size $\ell$, the
entanglement entropy, Eq.~\eqref{eq:entanglement-entropy}, can be obtained from
the eigenvalues $\lambda_l$ of the reduced single-particle density matrix
$D_A = \left( D \right)_{l, l' \in A}$ as~\cite{Peschel2009}
\begin{equation}
  \svn{\ell} = - \sum_{l = 1}^{\ell} \overline{\left[ \lambda_l \ln(\lambda_l) +
      \left( 1 - \lambda_l \right) \ln(1 - \lambda_l) \right]},
\end{equation}
where the overbar again denotes an average over both trajectories and position.

Figure~\ref{fig:5} shows the entanglement entropy for
$\ell \in \{1, \dotsc, L/2\}$ for fermion counting (solid lines) and generalized
occupation measurements (dashed lines). Qualitatively, these results appear to
support the existence of an entanglement transition between a critical phase
with logarithmic growth of the entanglement entropy and an area law phase for
larger values of $\gamma$. However, an analysis of the effective central charge
$c_\ell$ reveals clear evidence for an area-law phase for all values of
$\gamma$, as predicted analytically.

\begin{figure}
  \centering
  \includegraphics[width = \linewidth]{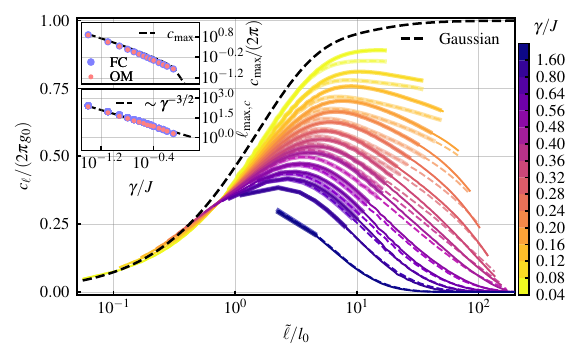}
  \caption{Rescaled effective central charge calculated by applying
    Eq.~\eqref{eq:effective-central-charge-finite-L} to the data shown in
    Fig.~\ref{fig:5} for fermion counting (solid lines) and generalized
    occupation measurements (colored dashed lines). For $\gamma \ll J$, the data
    approach the Gaussian result (black dashed line). Upper inset: scaling of
    the maximum of the central charge $c_{\mathrm{max}}$ for fermion counting
    (blue dots) and occupation measurements (red dots) compared to the
    prediction obtained by inserting Eq.~\eqref{eq:l-max-c} in
    Eq.~\eqref{eq:effective-central-charge-replica-Keldysh} (black dashed
    line). Lower inset: scaling of the position of the maximum
    $\ell_{\mathrm{max}, c}$ of the central charge. The data agree with the
    predicted scaling $\sim \gamma^{-3/2}$, Eq.~\eqref{eq:l-max-c}.}
  \label{fig:6}
\end{figure}

To account for the finite system size and periodic boundary conditions imposed
in our numerical studies, we rewrite the definition of the effective
scale-dependent central charge, Eq.~\eqref{eq:effective-central-charge}, in terms
of chord lengths,
\begin{equation}
  \label{eq:effective-central-charge-finite-L}
  c_{\ell} = 3 \frac{\svn{\ell'} - \svn{\ell}}{\ln(\tilde{\ell'}/\tilde{\ell})}.
\end{equation}
Ideally, the ratio $\tilde{\ell}'/\tilde{\ell}$ should be close to one and not
depend on $\ell$. In practice, due to $\ell$ being an integer, these conditions
cannot be met uniformly over the entire range of values from $\ell = 1$ to
$\ell = L/2$. To calculate $c_{\ell}$ numerically, we have chosen $N_{\ell}$
values $\ell_i$ with $\ell_1 = 1$ and $\ell_{N_{\ell}} = L/2$ such that
$\tilde{\ell}'/\tilde{\ell} = \tilde{\ell}_{i + 1}/\tilde{\ell}_i =
\e^{\epsilon_i}$
is approximately constant. For $N_{\ell} = 66$, we obtain $\epsilon_i < 0.1$ for
$i > 10$, ensuring that corrections to the value of $c_{\ell}$ obtained in the
limit $\epsilon_i \to 0$ are small.

Figure~\ref{fig:6} shows the effective central charge for fermion counting
(solid lines) and generalized occupation measurements (colored dashed lines). As
anticipated in Fig.~\ref{fig:2}(c), $c_{\ell}$ first grows with $\ell$, then
reaches a maximum $c_{\mathrm{max}}$ at a scale $\ell_{\mathrm{max}, c}$, and
eventually starts to decrease, indicating the crossover to the area-law
regime. The lower inset shows that the position of the maximum scales as
$\ell_{\mathrm{max}, c} \sim \gamma^{- 3/2}$, in good agreement with
Eq.~\eqref{eq:l-max-c} and within the critical range between
$l_0 \sim \gamma^{-1}$ and $l_c \sim \gamma^{- 2}$. This finding demonstrates
that the logarithmic growth of the entanglement entropy observed in
Fig.~\ref{fig:5} for small values of $\gamma$ and for $\ell \to L/2$ is due to
$l_c$ becoming larger than $L/2$, and that the apparent transition from a
critical phase to an area-law phase at larger $\gamma$ is, in fact, due to $l_c$
dropping below $L/2$.

Another signature of an apparent KT transition is a discontinuous jump of the
central charge at a finite critical value $\gamma_c$~\cite{Alberton2021}. As we
discuss in Sec.~\ref{sec:mutual-information-conformal-invariance} next, it is
the maximum value $c_{\mathrm{max}}$ of $c_{\ell}$ that yields best agreement
with CFT predictions within the critical range. The dependence of this maximum
value on $\gamma$ is shown in the upper inset in Fig.~\ref{fig:6}, where we
observe good agreement with the prediction obtained by inserting
Eq.~\eqref{eq:l-max-c} in
Eq.~\eqref{eq:effective-central-charge-replica-Keldysh}. In particular, the
maximum value $c_{\mathrm{max}}$ varies smoothly with $\gamma$, providing
further evidence for the absence of a KT transition at a finite $\gamma_c$.

\subsection{Mutual information and conformal invariance}
\label{sec:mutual-information-conformal-invariance}

The scaling of the entanglement entropy, Eq.~\eqref{eq:entanglement-entropy},
with subsystem size is the key property distinguishing different
measurement-induced phases. More refined measures of entanglement such as the
bi- and tripartite mutual information have proven useful for locating critical
points and characterizing critical phases. In particular, these quantities carry
signatures of conformal invariance.

The bipartite mutual information of two disjoint subsystems $A$ and $C$,
separated by a third subsystem $B$, is defined as
\begin{equation}
  \label{eq:mutual-information}
  I_2 = \svn{A} + \svn{C} - \svn{A \cup C},
\end{equation}
and provides a measure for the correlations between subsystem $A$ and
$C$~\cite{Wolf2008}. In studies of measurement-induced transitions in quantum
circuits, for subsystems of size $\ell_A = \ell_C = L/8$ and $\ell_B = 3 L/8$,
the bipartite mutual information has been found to decay exponentially with
system size $L$ in both area-law and volume-law phases~\cite{Skinner2019,
  Li2019}. At the critical point, $I_2$ exhibits a peak that remains finite in
the thermodynamic limit. Likewise, the tripartite mutual information, defined as
\begin{multline}
  \label{eq:tripartite-mutual-information}
  I_3 = \svn{A} + \svn{B} + \svn{C} + \svn{A \cup B \cup C} -
  \svn{A \cup B} - \svn{B \cup C} - \svn{A \cup C},
\end{multline}
remains finite at the critical point~\cite{Gullans2020, Zabalo2020,
  Ippoliti2021}. In the area-law phase, $I_3$ vanishes, and it has an
extensive negative value in the volume-law phase.

Apart from being indicators of measurement-induced phase transitions, the
interest in the bi- and tripartite mutual information is due to the existence of 
analytical predictions for these quantities for one-dimensional CFTs. These predictions
are expected to apply at critical points and in critical
phases~\cite{Alberton2021}. In CFTs, $I_2$ and $I_3$ depend
only on the cross ratio~\cite{Calabrese2009a, Maric2023, Maric2023a}
\begin{equation}
  \label{eq:cross-ratio}
  x = \frac{\ell_A \ell_C}{\ell_{A \cup B} \ell_{B \cup C}}
\end{equation}
and are given by
\begin{equation}
  \label{eq:I2-I3-CFT}  
  I_2 = \frac{c}{3} \ln \! \left( \frac{1}{1 - x} \right) +
  \mathcal{G}(x), \qquad
  I_3  = \mathcal{G}(x),
\end{equation}
where $c$ is the central charge of the CFT and $\mathcal{G}(x)$ is a universal
function that depends on the full operator content of the theory.
For finite systems with periodic boundary conditions, the subsystem sizes
$\ell_A$, $\ell_C$, $\ell_{A \cup B}$, and $\ell_{B \cup C}$ should be replaced
by the corresponding chord lengths.

\subsubsection{Analytical predictions}

Proceeding as described in Sec.~\ref{sec:entanglement-entropy} for the
entanglement entropy, we can obtain analytical predictions for the mutual
information, Eq.~\eqref{eq:mutual-information}. For subsystems of size
$\ell_A = \ell_C = \ell$ and $\ell_B = \ell'$ with $\ell, \ell' \gg l_0 \gg 1$,
we find
\begin{equation}
  \label{eq:bipartite-mutual-information-replica-Keldysh}
  \frac{I_2}{2 \pi g_0/3} \sim \ln\!\left(\frac{1}{1-x}\right)
  \left[ 1 - \frac{\ln(\ell/l_0)}{4 \pi \beta g_0} \right] - \frac{14 \left( 3 - x
    \right) x^2 l_0^2}{\left(1 - x \right)^2 \ell^2}.
\end{equation}
For fixed cross ratio $x$ and considered as a function of $\ell$, the bipartite
mutual information exhibits a maximum at
\begin{equation}
  \label{eq:l-max-I2}
  \ell_{\mathrm{max}, I_2} = 2 \sqrt{14 \pi \beta \left( 3 - x \right)}
  \ln\!\left(\frac{1}{1-x}\right)^{-1/2} \frac{x}{\abs{1 - x}} l_0^{3/2},
\end{equation}
which scales with the measurement rate as
$\ell_{\mathrm{max}, I_2} \sim \gamma^{-3/2}$, in analogy to the position of the
maximum of the effective central charge given in Eq.~\eqref{eq:l-max-c}.

On the same order of approximation that we have used to obtain predictions for
the entanglement entropy and the bipartite mutual information, the tripartite
mutual information vanishes. Indeed, generalizing
Eq.~\eqref{eq:cumulant-density-correlation} to
$C_{A, B}^{(2)} = \sum_{l \in A, l'\in B} C_{l - l'}$ such that
$C_A^{(2)} = C_{A, A}^{(2)}$, and using that the density correlation function,
Eq.~\eqref{eq:C-l-l-prime-t}, is symmetric under the exchange of $l$ and $l'$,
we obtain
$C_{A \cup B, A \cup B}^{(2)} = C_A^{(2)} + C_B^{(2)} + 2 C_{A, B}^{(2)}$ for
any two disjoint subsystems $A$ and $B$. Then, keeping only the contribution
from the second cumulant in Eq.~\eqref{eq:entropy-cumulant}, we find $I_3 = 0$,
indicating that the leading contribution to the tripartite mutual information
comes from the fourth cumulant. However, the fourth cumulant vanishes in a
Gaussian theory. In fact, the numerical results for $I_3$ presented below show
qualitatively different behavior for fermion counting and generalized occupation
measurements. Explaining these differences analytically would require going
beyond not only the Gaussian theory but even the long-wavelength effective field
theory developed in Sec.~\ref{sec:effective-field-theory}, which is identical
for both types of measurements.

\subsubsection{Numerics}

\begin{figure}
  \centering
  \includegraphics[width = \linewidth]{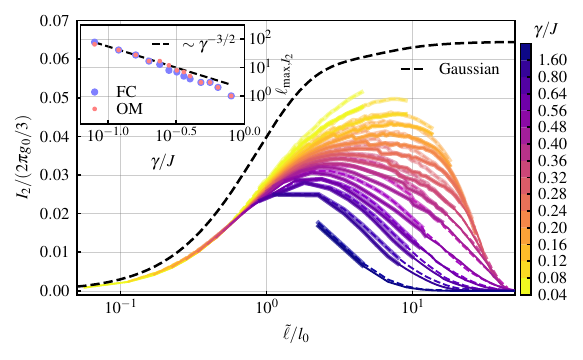}
  \caption{Rescaled bipartite mutual information for fermion counting (solid
    lines) and generalized occupation measurements (colored dashed lines) for
    subsystem sizes $\ell_A = \ell_C = \ell$ and $\ell_B \approx 3 \ell$ (see
    lower inset) chosen to approximately fix the cross ratio,
    Eq.~\eqref{eq:cross-ratio}, to $x \approx 1/16$, with deviations
    $< 1 \, \%$.  The Gaussian approximation is shown for comparison (black
    dashed line). Upper inset: scaling of the position of the maximum of the
    bipartite mutual information $\ell_{\mathrm{max},I_2}$ for fermion counting
    (blue dots) and occupation measurements (red dots), approaching the
    theoretical prediction, Eq.~\eqref{eq:l-max-I2}, for $\gamma \ll J$ (black
    dashed line). The considered subsystems are sketched in the lower left
    corner for periodic boundary conditions.}
  \label{fig:7}
\end{figure}

Figure~\ref{fig:7} shows the bipartite mutual information for
$\ell_A = \ell_C = \ell$ and $\ell_B \approx 3 \ell$ chosen such that the cross
ratio, Eq.~\eqref{eq:cross-ratio}, is $x \approx 1/16$ for all values of
$\ell \in \{ 1, \dotsc, L/8 \}$. The variations of $\ell_B/\ell \approx 3$
result in slight discontinuities in the data. Combining the finite-size phase
diagram sketched in Fig.~\ref{fig:2}(c) with the known behavior of $I_2$ across
measurement-induced phase transitions~\cite{Skinner2019, Li2019}, we expect the
bipartite mutual information to be finite only within the critical range, and to
decay toward zero for both $\tilde{\ell} \ll l_0$ and $\tilde{\ell} \gg l_c$.
This expectation is confirmed in Fig.~\ref{fig:7}, where we find the mutual
information to reach a maximum at $\ell_{\mathrm{max}, I_2} \sim \gamma^{-3/2}$,
within the critical range and in good agreement with the scaling predicted by
Eq.~\eqref{eq:l-max-I2}. We note that agreement between the numerical results
for $\gamma \ll J$ and the results obtained from the Gaussian theory is worse
for the bipartite mutual information than for the entanglement entropy shown in
Fig.~\ref{fig:5}. As we have confirmed numerically, this discrepancy can be
traced back to contributions from cumulants of higher than second order in
Eq.~\eqref{eq:entropy-cumulant}, which we neglect in the Gaussian results, and
which are more important for the mutual information.

\begin{figure}
  \centering  
  \includegraphics[width = \linewidth]{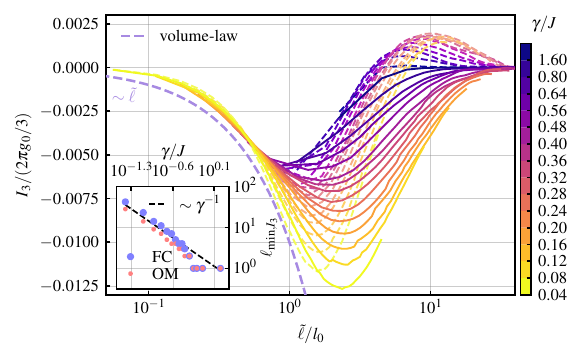}
  \caption{Rescaled tripartite mutual information for fermion counting (solid
    lines) and generalized occupation measurements (colored dashed lines) for
    the same choice of subsystems as in Fig.~\ref{fig:7}. Volume-law behavior
    $\sim \tilde{\ell}$ is shown for comparison (purple dashed line). Inset: the
    position of the minimum of the tripartite mutual information
    $\ell_{\mathrm{min},I_3}$ scales as $\sim \gamma^{-1}$ (black dashed line)
    both for fermion counting (blue dots) and occupation measurements (red
    dots).}
  \label{fig:8}
\end{figure}

The tripartite mutual information is shown in Fig.\ref{fig:8} for the same
choice of subsystems as in Fig.~\ref{fig:7}. On short scales
$\tilde{\ell} \lesssim l_0$, the tripartite mutual information behaves as
expected in a volume-law phase, growing extensively to negative values for both
fermion counting (solid lines) and occupation measurements (dashed lines). The
minimum value of $I_3$ is reached at
$\ell_{\mathrm{min}, I_3} \sim \gamma^{-1}$, close to the lower boundary of the
critical phase at $l_0 \sim \gamma^{-1}$. Notably, on scales
$\tilde{\ell} \gtrsim l_0$ beyond the volume-law regime, the tripartite mutual
information shows qualitatively different behavior for fermion counting and
generalized occupation measurements. For fermion counting, $I_3$ remains
negative, and approaches $I_3 \to 0$ from below for $\tilde{\ell} \to \infty$,
as expected in an area-law phase. In contrast, for generalized occupation
measurements, $I_3$ crosses to positive values and eventually approaches zero
from above. These findings confirm that the tripartite mutual information probes
behavior that is not captured by the long-wavelength effective field theory
developed in Sec.~\ref{sec:effective-field-theory}, as discussed above.

\begin{figure}
  \centering
  \includegraphics[width = \linewidth]{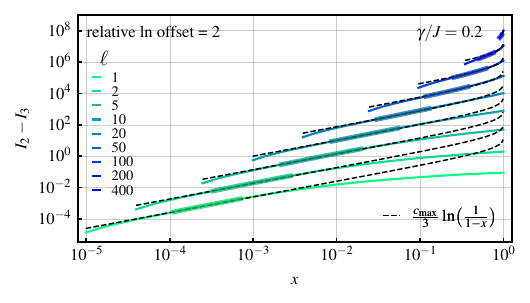}
  \caption{Difference between bi- and tripartite mutual information for fermion
    counting with $\gamma / J = 0.2$ (colored lines). The subsystem size
    $\ell_A = \ell_C = \ell$ is fixed for each curve according to the legend,
    while $\ell_B = \ell'$ is varied to sample different values of the cross
    ratio $x$. For better visibility, each curve with $\ell >1$ is shifted by a
    factor of $\e^2$ relative to the curve with the next smaller value of
    $\ell$, as indicated in the top center. The CFT prediction (black dashed
    line) coincides with the data when $\ell'$ is within the critical range
    (thicker colored lines). Violations of conformal invariance occur both on
    short scales $\tilde{\ell}' \lesssim a l_0$ (large values of $x$) and large
    scales $\tilde{\ell}' \gtrsim l_c$ (small values of $x$), where the factor
    $a = 8$ is introduced such that the lower bound agrees with the onset of
    algebraic scaling observed in Fig.~\ref{fig:4}.}
  \label{fig:9}
\end{figure}

So far, we have considered the bi- and tripartite mutual information for fixed
cross ratio $x$ and as a function of subsystem size $\ell$. We now turn to a
more stringent test of conformal invariance within the critical range. As we
have discussed in Sec.~\ref{sec:entanglement-entropy}, on scales
$\ell \gtrsim l_0$, the Gaussian theory predicts logarithmic growth of the
entanglement entropy with a central charge $c = 2 \pi g_0$. However, due the
renormalization of $g_0$ induced by strong fluctuations of massless replicon
modes, the range of approximately logarithmic growth is actually bounded by
$\ell \lesssim l_c$, and the value of the central charge, given by the maximum
value $c_{\mathrm{max}}$ of $c_{\ell}$, is reduced as compared to the Gaussian
result. These insights serve as a guideline in searching for signatures of
emergent conformal symmetry in $I_2$ and $I_3$.

For a CFT, it follows from Eq.~\eqref{eq:I2-I3-CFT} that the difference $I_2 -
I_3$ is fully determined by the central charge and the cross ratio,
\begin{equation}
  \label{eq:I2-I3-collapse}
  I_2 - I_3 = \frac{c}{3} \ln \! \left( \frac{1}{1 - x} \right).
\end{equation}
This difference is shown for fermion counting in Fig.~\ref{fig:9}. We have
obtained similar results for generalized occupation measurements, but do not
show them for the sake of brevity. A moderate value of $\gamma/J = 0.2$ is
chosen here to best illustrate deviations from conformal invariance at both
short and large scales. Each curve corresponds to a fixed value of
$\ell_A = \ell_C = \ell$, while $\ell_B = \ell'$ is varied to sample different
values of the cross ratio $x$, Eq.~\eqref{eq:cross-ratio}. We find good
agreement with the CFT prediction, Eq.~\eqref{eq:I2-I3-collapse}, for values
$\tilde{\ell}' \in [a l_0, l_c]$ within the critical range and when the central
charge $c$ in Eq.~\eqref{eq:I2-I3-collapse} is set to the maximum value
$c_{\mathrm{max}}$ of the effective central charge $c_{\ell}$ shown in
Fig.~\ref{fig:6}. The factor $a = 8$ is introduced such that the lower bound
agrees with the onset of algebraic scaling on scales $\tilde{l} \gg l_0$ in
Fig.~\ref{fig:4}. These results provide strong evidence for emergent conformal
invariance within the critical range.

It should be noted that for the data shown in Fig.~\ref{fig:9}, $I_3$ is
typically several orders of magnitude smaller than $I_2$. Therefore, the picture
does not change when $I_2$ is considered instead of $I_2 - I_3$. The collapse of
$I_2$ as a function of the cross ratio for continuous measurements of occupation
numbers described by a quantum state diffusion equation has been shown in
Ref.~\cite{Alberton2021} for a selection of subsystems. However, only the
systematic analysis of the dependence on subsystem size presented above, which
avoids an undersampling of subsystem sizes outside of the critical range, has
allowed us to demonstrate that there is agreement with the CFT expectation only
within the critical range and for $c = c_{\mathrm{max}}$.

\begin{figure}
  \centering  
  \includegraphics[width = \linewidth]{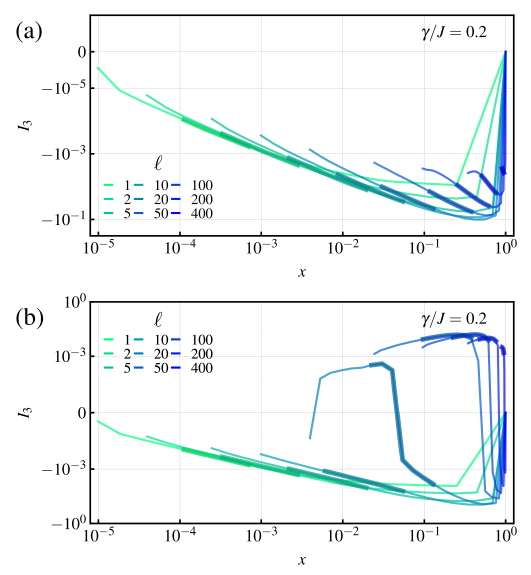}
  \caption{Tripartite mutual information as a
    function of the cross ratio $x$ for (a) fermion counting and (b) occupation
    measurements on a symmetric logarithmic scale and with $\gamma/J = 0.2$. In
    contrast to $I_2 - I_3$ shown in Fig.~\ref{fig:9}, the tripartite mutual
    information $I_3$ shows qualitatively different behavior for fermion
    counting and occupation measurements, and does not collapse to a single
    function of the cross ratio within the critical range.}
  \label{fig:20}
\end{figure}

Finally, we apply the same analysis to the tripartite mutual information. We
have argued in Sec.~\ref{sec:entanglement-entropy} that logarithmic growth of
the entanglement entropy within the critical range is described by the Gaussian
theory with perturbative corrections due the nonlinearity of the sigma model
manifold. This level of the theoretical description should, therefore, capture
all signatures of emergent conformal invariance. However, as shown above, the
tripartite mutual information vanishes in the Gaussian theory, and cannot be
obtained from the NLSM. These observations suggest that $\mathcal{G}(x) = 0$ in
Eq.~\eqref{eq:I2-I3-CFT}, and that nonzero values of $I_3$ can be attributed to
nonuniversal fluctuations on short scales. Indeed, we do not observe a collapse
of $I_3$ plotted as a function of the cross ratio in Fig.~\ref{fig:20}. As we
have already seen in Fig.~\ref{fig:8}, the tripartite mutual information
exhibits qualitatively different behavior for fermion counting and occupation
measurements, with $I_3$ assuming only negative values for fermion counting but
changing sign for occupation measurements.

\section{Conclusions and outlook}
\label{sec:conclusions-outlook}

We have presented an in-depth comparative study of two 1D models of free
fermions subjected to different types of generalized measurement processes:
monitored loss and gain or fermion counting on the one hand, and generalized
measurements of local occupation numbers on the other hand. A high rate of
fermion counts signals fast fluctuations of the quantum state of the fermionic
many-body system. In contrast, generalized or projective measurements of
occupation numbers lead to a freezing of the dynamics. Despite this striking
difference, the dynamics of both models have important similarities in the
frequent-measurement limit: emergent classicality and a suppression of coherent
dynamics, which are hallmarks of the quantum Zeno effect for occupation
measurements and which we attribute to a generalized Zeno effect for fermion
counting. These similarities explain heuristically why both models exhibit close
to identical results in various observables in the steady state, including
density correlations, the entanglement entropy, and the bipartite mutual
information. There are qualitative differences only in the tripartite mutual
information.

Combining analytical insights from replica Keldysh field theory with numerical
results from simulations of quantum jump trajectories, we have provided strong
evidence for the absence of a measurement-induced entanglement transition in
both models. In particular, we have shown that characteristic signatures of a
critical phase can be observed only within a well-defined and finite range of
length scales. Crucially, the upper boundary $l_c \sim \gamma^{-2}$ of this
critical range is only algebraically large in the measurement rate, and,
therefore, observable in numerics.

Formally, these properties can be understood to result from the universal
long-wavelength description in terms of an NLSM that applies to both models. We
have clarified the origin of the underlying $\mathrm{SU}(R)$ symmetry, and
provided a physical explanation for the occurrence of this symmetry in terms of
(i)~conservation of the total number of particles in the system and auxiliary
reservoirs and (ii)~preservation of the purity of the quantum state of the
system. Our work thus settles the question about the requirements to observe a
measurement-induced entanglement transition in 1D free fermions.

Furthermore, our work opens up several interesting directions for future
research. While there is no measurement-induced phase transition in the
considered models in 1D, such a transition does occur in higher
dimensions~\cite{Poboiko2024, Chahine2024}. It will be interesting to study the
impact of the breaking of particle-number conservation by fermion counting on
the dynamical critical behavior. A classification of \emph{dynamical}
criticality based on symmetries and conservation laws---in analogy to the
classification of dynamical criticality in thermodynamic
equilibrium~\cite{Hohenberg1977}---is an interesting prospect. However,
exploring this possibility analytically by studying, for example, two-time
correlation functions such as
$\overline{\left\langle \hat{A}(t) \right\rangle \left\langle \hat{B}(t')
  \right\rangle}$
will require a nontrivial extension of the replica Keldysh formalism.

From a more formal perspective, the stochastic Schr\"odinger
equation~\eqref{eq:stochastic-schroedinger-equation} can be regarded as an
unraveling of a quantum master equation that is recovered by taking the average
over the stochastic increments $\diff N_{\alpha, l}(t)$. While here we have
considered cases for which the stationary state of the unconditional dynamics
described by the master equation is trivial, this is generically not the case,
and jump operators can be chosen so as to prepare a state of interest. For
example, consider a 1D fermionic chain that is connected to particle reservoirs
only at its ends and in such a way that a finite current is driven through the
system. An unraveling of the corresponding master equation represents an
interesting setup to study the interplay of transport and entanglement dynamics
on the level of individual quantum trajectories. However, such an unraveling
can, in general, not be formulated as a random generalized measurement as we
have done here, and the derivation of a replica Keldysh field theory described
in Sec.~\ref{sec:replica-keldysh-field-theory} has to be modified. Based on such
an extension of the formalism, a further interesting perspective is to
investigate quantum jump unravelings of quadratic fermionic Liouvillians
belonging to different symmetry classes~\cite{Altland2021, Lieu2020, Sa2023,
  Kawabata2023}. The resulting replica Keldysh field theories and NLSMs will
shed further light on the connections between the dynamics of continuously
monitored quantum systems in $d$ spatial dimensions and the physics of Anderson
localization in $d + 1$ dimensions~\cite{Evers2008}.

Finally, let us discuss implications of our work beyond noninteracting
fermions. As we have already mentioned, our results show that charge sharpening,
which has recently been discussed in random quantum circuits~\cite{Agrawal2022,
  Barratt2022, Oshima2023, Agrawal2024} and interacting fermionic
systems~\cite{Guo2024, Poboiko2025}, can occur even in the absence of charge
conservation: For free fermionic systems, Eq.~\eqref{eq:entropy-cumulant}
relates the entanglement entropy to the variance of the subsystem charge, that
is, the second cumulant of the subsystem particle number; in particular,
area-law scaling of the entanglement entropy, which we have found to be induced
by fermion counting, implies that the system is in a charge-sharp
phase~\cite{Oshima2023}. It is an interesting prospect for future studies to
explore charge sharpening in interacting fermionic systems and random quantum
circuits with broken charge conservation.

\begin{acknowledgements}
  E.S.\ and L.S.\ acknowledge support from the Austrian Science Fund (FWF)
  through the projects 10.55776/P33741 and 10.55776/COE1, and from the European
  Union - NextGenerationEU. For open access purposes, the authors have applied a
  CC BY public copyright license to any author accepted manuscript version
  arising from this submission.
\end{acknowledgements}

\subsection*{Data availability}

The data that support the findings of this article are openly
available~\cite{Starchl2025}.

\appendix

\section{Minimal physical models and their description through a stochastic
  Schr\"odinger equation}
\label{sec:app-physical-models}

In this appendix, we outline the derivation of the stochastic Schr\"odinger
equation~\eqref{eq:stochastic-schroedinger-equation} for minimal physical models
of fermion counting and generalized occupation measurements. These models are
inspired by theoretical descriptions of continuous measurements in mesoscopic
electronics~\cite{Wiseman2010}.

\subsection{Fermion counting}
\label{sec:app-physical-models-FC}

We consider the setup of a quantum dot coupled to two reservoirs described in
Sec.~\ref{sec:stochastic-schrodinger-equation}~\cite{Wiseman2010}. The quantum
dot corresponds to a single fermionic lattice site, with annihilation and
creation operators $\hat{\psi}$ and $\hat{\psi}^{\dagger}$, respectively, and
Hamiltonian
\begin{equation}
  \label{eq:H-quantum-dot}
  \hat{H}_0 = \omega_0 \hat{\psi}^{\dagger} \hat{\psi}.
\end{equation}
We denote the reservoir annihilation and creation operators by
$\hat{b}_{\alpha,k}$ and $\hat{b}_{\alpha,k}^{\dagger}$, where $\alpha = -$ for
the reservoir that acts as the drain, $\alpha = +$ for the reservoir that acts
as the source, and $k$ is a discrete index that labels the reservoir states. The
Hamiltonians of the reservoirs states with energies $\omega_{\alpha, k}$ read
\begin{equation}
  \label{eq:H-source-drain}
  \hat{H}_{\alpha}^{} = \sum_k \omega_{\alpha, k}^{} \hat{b}_{\alpha, k}^{\dagger} \hat{b}_{\alpha,k}^{},
\end{equation}
and the reservoirs are coupled to the quantum dot by
\begin{equation}
  \label{eq:H-coup-FC}
  \hat{H}_{\mathrm{coup}}^{} = \imag \sum_{\alpha, k} \left(T_{\alpha,k}^{}
    \hat{\psi}^{\dagger} \hat{b}_{\alpha,k}^{} - T_{\alpha,k}^{\ast}
    \hat{b}_{\alpha,k}^{\dagger} \hat{\psi} \right),
\end{equation}
with tunneling coefficients $T_{\alpha,k}$. Both reservoirs are assumed to be in
thermodynamic equilibrium at a low temperature, with chemical potentials chosen
such that the reservoir states at the energy $\omega_0$ are empty in the drain
and occupied in the source. We model this scenario by assuming the dot and
reservoirs to be in the pure state $\ket{\Psi(0)} = \ket{\psi(0), 0_-, 1_+}$ at
time $t = 0$, with $0_-$ and $1_+$ denoting empty and occupied states of the
drain and source, respectively.

Our goal is to derive a stochastic Schr\"odinger equation describing the
dynamics of the quantum dot. As a first step, we move to an interaction frame
with respect to the bath Hamiltonians, Eq.~\eqref{eq:H-source-drain}, and
integrate the Schr\"odinger equation for the state $\ket{\Psi(t)}$ of the dot
and reservoirs in small time steps $\Delta t$. In Born and Markov
approximations, the increment of the wave function,
$\Delta \ket{\Psi(t)} = \ket{\Psi(t + \Delta t)} - \ket{\Psi(t)}$, is given
by~\cite{Gardiner2015}
\begin{equation}
  \label{eq:Delta-Psi}
  \Delta\ket{\Psi(t)} = \left[ -\imag \hat{H}_{\mathrm{eff}}^{} \Delta t +
    \sqrt{2} \hat{L}_-^{} \Delta \hat{B}_-^{\dagger}(t) +
    \sqrt{2} \hat{L}_+^{} \Delta \hat{B}_+^{}(t) \right]
  \ket{\Psi(t)}.
\end{equation}
The effective Hamiltonian reads
\begin{equation}
  \label{eq:H-eff}
  \hat{H}_{\mathrm{eff}}^{} =  \hat{H}_0 - \imag \sum_{\alpha = \pm} \hat{L}_{\alpha}^{\dagger}
  \hat{L}_{\alpha}^{},
\end{equation}
where $\hat{L}_- = \sqrt{\gamma_{\lind}/2} \, \hat{\psi}$ and
$\hat{L}_+ = \sqrt{\gamma_+/2} \, \hat{\psi}^{\dagger}$ are the jump operators
defined in Eq.~\eqref{eq:jump-operators-FC} for a single lattice site. Furthermore,
in Eq.~\eqref{eq:Delta-Psi}, we have introduced quantum It\^o increments,
\begin{equation}
  \Delta \hat{B}_{\alpha}(t) = \sqrt{\frac{2}{\gamma_{\alpha}}} \int_t^{t +
    \Delta t} \diff t'
  \sum_k T_{\alpha,k} \hat{b}_{\alpha,k} \e^{-\imag \omega_{\alpha,k} t'},
\end{equation}
which obey the anticommutation relations
\begin{equation}
  \label{eq:Ito-increment-anticommutation}
  \left\{ \Delta \hat{B}_{\alpha}^{\vphantom{\dagger}}(t), \Delta
    \hat{B}_{\alpha}^{\dagger}(t') \right\} = \Delta t \,
  \delta_{t,t'}^{\vphantom{\dagger}},
\end{equation}
where $\delta_{t,t'}$ is the Kronecker delta for discrete times $t$ and
$t'$. The operators $\Delta \hat{B}_{\alpha}(t)$ and
$\Delta \hat{B}_{\alpha}^{\dagger}(t)$ destroy and create fermions in the drain
and source at a given time $t$, as opposed to the operators
$\hat{b}_{\alpha, k}$ and $\hat{b}_{\alpha, k}$, which are associated with a
frequency $\omega_{\alpha, k}$. Finally, the loss and gain rates appearing in
Eq.~\eqref{eq:Delta-Psi} are defined as
$\gamma_{\alpha} = 2 \pi \rho_{\pm}(\omega_0) \abs{T_{\pm}(\omega_0)}$, where
$\rho_{\alpha}(\omega_0)$ and $T_{\alpha}(\omega_0)$ are the density of states
and tunneling amplitudes at the energy of the dot, which are introduced upon
taking the continuum limit such that
$\sum_k \to \int_0^{\infty} \diff \omega \, \rho_{\alpha}(\omega)$ and
$T_{\alpha, k} \to T_{\alpha}(\omega)$~\cite{Wiseman2010}.

The operators $\hat{L}_-^{} \Delta \hat{B}_-^{\dagger}(t)$ and
$\hat{L}_+^{} \Delta \hat{B}_+^{}(t)$ in Eq.~\eqref{eq:Delta-Psi} describe the
transfer of a fermion from the quantum dot to the drain and from the source to
the dot, respectively. Since the It\^o increments anticommute at different
times, a fermion that has been transferred from the dot to the drain at time $t$
cannot be transferred back at a later time $t'$. Formally, the state
$\Delta \hat{B}_-^{\dagger}(t) \ket{0_-}$ is annihilated by
$\Delta \hat{B}_-(t')$. An analogous argument applies to the transfer of
fermions from the source to the dot. Therefore, in Eq.~\eqref{eq:Delta-Psi}, the
operators $\Delta \hat{B}_-(t)$ and $\Delta \hat{B}_+^{\dagger}(t)$ do not
occur.

In analogy to photon counting in quantum optics, fermion counting is implemented
by measuring the number operators,
\begin{equation}
  \label{eq:Ito-increment-number-ops}
  \hat{N}_{\alpha}(t) = \frac{\Delta \hat{B}_{\alpha}^{\dagger}(t)}{\sqrt{\Delta t}} \frac{\Delta
    \hat{B}_{\alpha}(t)}{\sqrt{\Delta t}},
\end{equation}
at each time step, with possible measurement outcomes $N_{\alpha}(t) = 0, 1$. To
provide a physical interpretation of these measurements, we may again take
inspiration from mesoscopic electronics~\cite{Wiseman2010}. In this context, we
can imagine the reservoirs as being connected via ohmic contacts to an external
circuit, such that a fermion in the drain is quickly absorbed in the circuit,
and a fermion transferred from the source to the quantum dot is quickly
replenished. These processes lead to measurable current spikes in the external
circuit. The presence or absence of such spikes at time $t$ can thus be
interpreted as a projective measurements of the occupation numbers
$\hat{N}_{\alpha}(t)$.

To describe the statistics of the measurement outcomes and the change of the
state of the quantum dot for a given outcome, it is convenient to rewrite
Eq.~\eqref{eq:Delta-Psi} in the form
\begin{equation}
  \label{eq:app-Ito-SSE-FC}
  \ket{\Psi(t + \Delta t)} = \left[ \hat{M}_0^{} + \hat{M}_-^{} \frac{\Delta
      \hat{B}_-^{\dagger}(t)}{\sqrt{\Delta t}} + \hat{M}_+^{} \frac{\Delta
      \hat{B}_+^{}(t)}{\sqrt{\Delta t}} \right] \ket{\Psi(t)},
\end{equation}
where the measurement operators are defined as
\begin{equation}
  \hat{M}_0 = 1 - \imag \hat{H}_{\mathrm{eff}} \Delta t, \quad \hat{M}_- =
  \sqrt{2 \Delta t} \, \hat{L}_-, \quad \hat{M}_+ = \sqrt{2 \Delta t} \, \hat{L}_+.
\end{equation}
A measurement of the number operators, Eq.~\eqref{eq:Ito-increment-number-ops},
performed on the state $\ket{\Psi(t + \Delta t)}$,
Eq.~\eqref{eq:app-Ito-SSE-FC}, yields $N_-(t) = 0$ and $N_+(t) = 1$ with
probability
\begin{equation}
  p_0(t) = \braket{\psi(t) \middle| \hat{M}_0^{\dagger} \hat{M}_0 \middle|
    \psi(t)}.
\end{equation}
After a measurement with this outcome, the state of the quantum dot is given by
\begin{multline}
  \label{eq:M0-psi}
  \ket{\psi(t + \Delta t)} = \frac{\hat{M}_0 \ket{\psi(t)} }{\sqrt{p_0(t)}} =
  \left[ 1 - \imag \hat{H}_{\mathrm{eff}}^{} \Delta t + \sum_{\alpha = \pm}
    \tilde{p}_{\alpha}(t) \Delta t \right] \ket{\psi(t)},
\end{multline}
where we have expanded the denominator up to order $\Delta t$, and where, as in
Eq.~\eqref{eq:nonnormalized-probabilities-FC}, we have defined
\begin{equation}
  \tilde{p}_{\alpha}^{}(t) = \braket{\psi(t) \middle| \hat{L}_{\alpha}^{\dagger}
    \hat{L}_{\alpha}^{} \middle| \psi(t)}.
\end{equation}
The measurement outcome above and the change of the state in
Eq.~\eqref{eq:M0-psi} correspond to the case that no quantum jump has occurred
during the time step, and the states of drain and source remain unchanged. In
contrast, for $N_-(t) = 1$ and $N_+(t) = 1$, one fermion has been transferred
from the quantum dot to the drain. This happens with probability
$p_-(t) = 2 \tilde{p}_-(t) \Delta t$, and the state after the measurement is
\begin{equation}
  \label{eq:M1-psi}
  \ket{\psi(t+\Delta t)} = \frac{\hat{M}_- \ket{\psi(t)}}{\sqrt{p_-(t)}} =
  \frac{\hat{L}_- \ket{\psi(t)} }{\sqrt{\tilde{p}_-(t)}}.
\end{equation}
For $N_-(t) = 0$ and $N_+(t) = 0 $, a fermion has been transferred from the
source to the dot, which happens with probability
$p_+(t) = 2 \tilde{p}_+(t) \Delta t$. Then, the state after the measurement is
\begin{equation}
  \label{eq:M2-psi}
  \ket{\psi(t+\Delta t)} = \frac{\hat{M}_+ \ket{\psi(t)}
  }{\sqrt{p_+(t)}} = \frac{\hat{L}_+ \ket{\psi(t)}}{\sqrt{\tilde{p}_+(t)}}.
\end{equation}

Performing measurements of the number operators,
Eq.~\eqref{eq:Ito-increment-number-ops}, in each time step results in stochastic
dynamics of the wave function of the quantum dot, conditional to the measurement
outcomes. To formulate these dynamics as an evolution equation, we define
stochastic increments $\Delta N_{\alpha}(t) = 0, 1$ that count the number of
fermions that are transferred from the dot to the drain and from the source to
the dot during the step from $t$ to $t + \Delta t$. Neglecting terms of order
$\Delta N_{\alpha}(t) \Delta t$~\cite{Wiseman2010}, we can thus write the
increment of the wave function as
\begin{multline}
  \Delta \ket{\psi(t)} = \left\{ \left[ - \imag \hat{H}_{\mathrm{eff}}^{} +
      \sum_{\alpha = \pm} \tilde{p}_{\alpha}(t) \right] \Delta t \right. \\
  \left. + \sum_{\alpha = \pm} \left[
      \frac{\hat{L}_{\alpha}}{\sqrt{\tilde{p}_{\alpha}(t)}} - 1 \right] \Delta
    N_{\alpha}(t) \right\} \ket{\psi(t)}.
\end{multline}
In the limit of infinitesimal time steps, $\Delta t \to \diff t$, the stochastic
increments $\Delta N_{\alpha}(t) \to \diff N_{\alpha}(t)$ describe a Poisson
point process with the properties specified in
Sec.~\ref{sec:stochastic-schrodinger-equation}~\cite{Breuer2007}, and the
evolution of the quantum dot is then described by the stochastic Schr\"odinger
equation~\eqref{eq:stochastic-schroedinger-equation} with $L = 1$ and where we
omit the contribution from $\hat{H}_0$, Eq.~\eqref{eq:H-quantum-dot}, which
leads to a global phase factor for each trajectory.

\subsection{Generalized occupation measurements}
\label{sec:app-physical-models-OM}

Measurements of the occupation of a quantum dot can be implemented by modifying
the above setup as outlined in Sec.~\ref{sec:stochastic-schrodinger-equation}:
Instead of tunneling from the reservoirs to the dot as described by
Eq.~\eqref{eq:H-coup-FC}, we consider tunneling directly from the source to the
drain which, however, is possible only when the dot is
occupied~\cite{Wiseman2010},
\begin{equation}
  \hat{H}_{\mathrm{coup}} = \imag \sum_{k,q} \hat{\psi}^{\dagger} \hat{\psi} \left(
    \chi \hat{b}_{-,k}^{\dagger} \hat{b}_{+,q}^{} - \chi^\ast \hat{b}_{+,q}^{\dagger}
    \hat{b}_{-,k}^{} \right).
\end{equation}
This coupling Hamiltonian, with a mode-independent tunneling rate $\chi$, is a
simplified version of a setup that has been realized in mesoscopic devices, for
example, in Ref.~\cite{Gustavsson2009}.

As above, we integrate the Schr\"odinger equation for the dot and reservoirs in
small time steps, leading to
\begin{equation}
  \label{eq:app-Ito-SSE-occupation}
  \Delta\ket{\Psi (t)} = \left[ - \imag \hat{H}_{\mathrm{eff}} \Delta t +
    \sqrt{2} \hat{L} \Delta \hat{B}(t) \right] \ket{\Psi(t)},
\end{equation}
where $\hat{n} = \hat{\psi}^{\dagger} \hat{\psi}$ and
$\hat{L} = \sqrt{\gamma} \, \hat{n}$. The measurement rate is
$\gamma = 2 \pi \rho_-(\omega_0) \rho_+(\omega_0) \abs{\chi}^2 \left( \mu_+ -
  \mu_- \right)$,
where $\mu_{\alpha}$ are the chemical potentials of the reservoirs, and the
effective Hamiltonian is given by
\begin{equation}
  \hat{H}_{\mathrm{eff}} = \hat{H}_0 - \imag \hat{L}^{\dagger} \hat{L}.
\end{equation}
We have again introduced quantum It\^o increments,
\begin{equation}
  \label{eq:Ito-increment-occupation}
  \Delta \hat{B}(t) = \sqrt{\frac{\gamma}{2}} \chi^{*} \int_t^{t + \Delta t}
  \diff t' \sum_{k,q} \hat{b}_{-,k}^{\dagger} \hat{b}_{+,q}^{} \e^{-\imag \left(
      \omega_{+,q} - \omega_{-,k} \right) t'}.
\end{equation}
which obey canonical anticommutation relations,
\begin{equation}
  \left\{ \Delta \hat{B}(t), \Delta \hat{B}^{\dagger}(t') \right\} = \Delta t \,
  \delta_{t,t'}.
\end{equation}
From here on, the derivation of the stochastic Schr\"odinger equation describing
the dynamics of the quantum dot is analogous to the one for fermion counting
outlined above.

\section{Implementation of generalized measurements}
\label{sec:impl-gener-meas}

The generalized measurements described by Eqs.~\eqref{eq:measurement-ops-FC}
and~\eqref{eq:measurement-ops-OM} can be implemented through projective
measurements, performed on auxiliary degrees of freedom, as explained in the
following.

\subsection{Fermion counting}

As in Appendix~\ref{sec:app-physical-models}, we consider first a single
fermionic site or quantum dot, coupled to two reservoirs, designated as drain
and source. However, the reservoirs comprise a single state each, and not a
continuum of states as above. Our goal is to implement a generalized measurement
on the quantum dot with measurement operators $\hat{M}_- = \hat{\psi}$ and
$\hat{M}_+ = \hat{\psi}^{\dagger}$. Before the measurement, the drain is empty,
the source is filled, and the quantum dot is in an arbitrary superposition
state,
\begin{equation}
  \label{eq:dot-drain-source-initial-state-FC}
  \ket{\Psi} = \left( c_0 + c_1 \hat{\psi}^{\dagger} \right) \hat{b}_+^{\dagger}
  \ket{0},
\end{equation}
where $\ket{0}$ is the vacuum state of dot, drain, and source. To implement the
measurement, we first entangle the quantum dot with the reservoirs by applying
two gates sequentially. The first gate, $\hat{U}_-$, transfers a fermion from
the dot to the drain,
\begin{equation}
  \hat{U}_- = \e^{- \imag \hat{V}_-}, \qquad \hat{V}_- = \frac{\imag \pi}{2}
  \left( \hat{\psi}^{\dagger} \hat{b}_-^{} - \hat{b}_-^{\dagger}
    \hat{\psi} \right).
\end{equation}
The second gate, $\hat{U}_+$, fills the dot from the source. To avoid a fermion
being transferred from the source to the drain through the application of
$\hat{U}_+ \hat{U}_-$, the action of the second gate is conditioned on the drain
being empty,
\begin{equation}
  \hat{U}_+ = \e^{- \imag \hat{V}_+}, \qquad \hat{V}_+ = \frac{\imag \pi}{2}
  \left( 1 - \hat{b}_-^{\dagger} \hat{b}_-^{} \right) \left(
    \hat{\psi}^{\dagger} \hat{b}_+^{} - \hat{b}_+^{\dagger} \hat{\psi} \right).
\end{equation}
Applying these gates to the state in
Eq.~\eqref{eq:dot-drain-source-initial-state-FC} leads to
\begin{equation}
  \label{eq:dot-drain-source-entangled-state-FC}
  \hat{U}_+ \hat{U}_- \ket{\Psi} = \left( c_0 \hat{\psi}^{\dagger} - c_1
    \hat{b}_-^{\dagger} \hat{b}_+^{\dagger} \right) \ket{0}.
\end{equation}
Now we perform a projective measurement of the occupation of the reservoirs,
$\hat{N}_b = \frac{1}{2} \sum_{\alpha = \pm} \hat{n}_{\alpha}$, with possible
outcomes 0 and 1. For the outcome 0, the nonnormalized state after the
measurement is
\begin{equation}
  \label{eq:outcome-0}
  \left( 1 - \hat{N}_b \right) \hat{U}_+ \hat{U}_-
  \ket{\Psi} = c_0 \hat{\psi}^{\dagger} \ket{0} = \hat{\psi}^{\dagger} \hat{b}_+
  \ket{\Psi}.
\end{equation}
The norm of this state is the probability for this outcome to occur,
$p_0 = \abs{c_0}^2 = \braket{\Psi \middle| \hat{\psi}^{\dagger} \hat{\psi}
  \middle| \Psi}$.
For the outcome 1, the nonnormalized state after the measurement is
\begin{equation}
  \label{eq:outcome-1}
  \hat{N}_b \hat{U}_+ \hat{U}_- \ket{\Psi} = - c_1 \hat{b}_-^{\dagger}
  \hat{b}_+^{\dagger} \ket{0} = \hat{\psi} \hat{b}_-^{\dagger} \ket{\Psi},
\end{equation}
with probability
$p_1 = \abs{c_1}^2 = \braket{\Psi \middle| \hat{\psi} \hat{\psi}^{\dagger}
  \middle| \Psi}$.
In both cases, through the measurement, the entangled state in
Eq.~\eqref{eq:dot-drain-source-entangled-state-FC} collapses, and the dot and
reservoirs become disentangled. The reservoirs can then be reinitialized for the
next measurement, and we may consider the effect of the measurement process on
the dot alone. As desired, the states after the measurement in
Eqs.~\eqref{eq:outcome-0} and~\eqref{eq:outcome-1} as well as the corresponding
probabilities are equivalent to a generalized measurement with the measurement
operators $\hat{M}_{\pm}$ given above.

It remains to generalize this measurement process to an extended lattice
system. This is achieved straightforwardly by first selecting a lattice site
$l \in \{ 1, \dotsc, L \}$ with uniform probability $1/L$ and then carry out the
generalized measurement at this lattice site as described above. This is
equivalent to the generalized measurement with measurement operators given in
Eq.~\eqref{eq:measurement-ops-FC}, where the factors
$\left. 1 \middle/ \sqrt{L} \right.$ and the corresponding factors of $1/L$ in
the probabilities in Eq.~\eqref{eq:probabilities-FC} can be interpreted as
resulting from randomly picking a lattice site.

\subsection{Generalized occupation measurements}

The effect of a generalized measurement described by
Eq.~\eqref{eq:measurement-ops-OM} on a single quantum dot is trivial. Therefore,
we consider the general case of $N$ particles on $L$ lattice sites, coupled to
$L$ drain states with annihilation operators $\hat{b}_{-, l}$, but only a single
source mode with annihilation operator $\hat{b}_+$. Note that this is different
from the derivation of the stochastic Schr\"odinger equation discussed in
Appendix~\ref{sec:app-physical-models-OM} and the sketch in
Fig.~\ref{fig:1}. The initial state is a superposition of states with $N$
fermions on sites $l_1, \dotsc, l_N \in \{ 1, \dotsc, L \}$, with respective
amplitudes $\psi_{l_1, \dotsc, l_N}$,
\begin{equation}
  \label{eq:dot-drain-source-initial-state-OM}
  \ket{\Psi} = \sum_{l_1 < \dotsb < l_N} \psi_{l_1, \dotsc, l_N} \left(
    \prod_{n = 1}^N \hat{\psi}_{l_n}^{\dagger} \right) \hat{b}_+^{\dagger}
  \ket{0}.
\end{equation}
There is a single fermion in the source mode, while the drain states are
empty. As above, to implement the measurement, we first entangle the system with
the reservoirs. We apply a unitary gate that transfers the fermion from the
source to the drain state at the sites $l$, conditioned on the corresponding
sites in the system being occupied,
\begin{equation}
  \hat{U} = \e^{- \imag \hat{V}}, \quad \hat{V} = \frac{\imag \pi}{2 \sqrt{N}}
  \sum_{l = 1}^L \hat{n}_l^{\phantom{\dagger}} \left( \hat{b}_{-, l}^{\dagger}
    \hat{b}_+^{\phantom{\dagger}} - \hat{b}_+^{\dagger} \hat{b}_{-,
      l}^{\phantom{\dagger}} \right).
\end{equation}
Consequently, in each term in the sum in
Eq.~\eqref{eq:dot-drain-source-initial-state-OM}, the reservoir fermion ends up
in an equal superposition of the sites $l_n$ that are occupied in the system,
\begin{equation}
  \hat{U} \ket{\Psi} = \frac{1}{\sqrt{N}} \sum_{l_1 < \dotsb < l_N} \psi_{l_1,
    \dotsc, l_N} \left( \prod_{n = 1}^N \hat{\psi}_{l_n}^{\dagger} \right)
  \sum_{n = 1}^N \hat{b}_{-, l_n}^{\dagger} \ket{0}.
\end{equation}
Next, we perform a projective measurement of the position of the fermion in the
drain, described by the operator
$\hat{X}_- = \sum_{l = 1}^L l \hat{b}_-^{\dagger} \hat{b}_-^{}$, with possible
measurement outcomes $l \in \{ 1, \dotsc, L \}$. For the result $l$, the state
after the measurement is
\begin{equation}
  \begin{split}
    \hat{b}_{-, l}^{\dagger} \hat{b}_{-, l}^{\phantom{\dagger}} \hat{U}
    \ket{\Psi} & = \frac{1}{\sqrt{N}} \sum_{l_1 < \dotsb < l_N} \delta_{l \in \{
    l_1, \dotsc, l_N \}} \psi_{l_1,
      \dotsc, l_N} \left( \prod_{n = 1}^N \hat{\psi}_{l_n}^{\dagger} \right)
    \hat{b}_{-, l}^{\dagger} \ket{0} \\ & = \frac{1}{\sqrt{N}} \hat{n}_l^{}
    \hat{b}_{-, l}^{\dagger} \hat{b}_+^{} \ket{\Psi},
  \end{split}
\end{equation}
where the factor $\delta_{l \in \{ l_1, \dotsc, l_N \}}$ indicates that only
those terms should be retained in the sum over occupied sites $l_1, \dotsc,
l_N$, for which $l$ is occupied. The probability for the outcome $l$, given by
the norm of the state, evaluates to $p_l = \braket{\Psi | \hat{n}_l | \Psi} \!
/N$. Focusing again on the effect of the measurement on the system, we see that
it can equivalently be described as a generalized measurement with measurement
operators given in Eq.~\eqref{eq:measurement-ops-OM}.

\section{Telegraph-reduced autocorrelation function}
\label{sec:calc-telegr-reduc}

The telegraph-reduced autocorrelation function,
Eq.~\eqref{eq:z-string-z-correlation}, can be calculated analytically in the
limit $\gamma \gg J$. For simplicity of notation, we assume that the evolution
has already reached a stationary state at $t' = 0$. A closed equation of motion
can be obtained for the correlation function defined as
\begin{equation}
  \label{eq:Q-l-1-l-2-l-3}
  Q_{l_1, l_2, l_3}(t) = \overline{q_{l_1, l_2, l_3}(t) z_{l_3}(0)}, \qquad
  q_{l_1, l_2, l_3}(t) = z_{l_1, l_2}(t) s_{l_3}(t),
\end{equation}
where
\begin{equation}
  z_{l, l'}(t) = 2 D_{l, l'}(t) - 1, \quad z_l(t) = z_{l, l}(t), \quad s_l(t) =
  \left( - 1 \right)^{N_l(t)},
\end{equation}
with the single-particle density matrix defined in
Eq.~\eqref{eq:single-particle-density-matrix}, and
\begin{equation}
  N_l(t) = \int_0^t \diff N_l(t'), \qquad \diff N_l(t) = \sum_{\alpha = \pm}
  \diff N_{\alpha, l}(t).
\end{equation}
The telegraph-reduced autocorrelation function,
Eq.~\eqref{eq:z-string-z-correlation}, is obtained by setting $l_1 = l_2 = l_3$
in Eq.~\eqref{eq:Q-l-1-l-2-l-3}, $Q(t) = Q_{l, l, l}(t)$. To obtain the
evolution equation of $Q_{l_1, l_2, l_3}(t)$, we consider first the stochastic
increment in a single trajectory,
\begin{equation}
  \label{eq:d-q-l-1-l-2-l-3}
  \diff q_{l_1, l_2, l_3}(t) = \diff z_{l_1, l_2}(t) s_{l_3}(t) + z_{l_1,
    l_2}(t) \diff s_{l_3}(t) + \diff z_{l_1, l_2}(t) \diff s_{l_3}(t).
\end{equation}
The product of two increments in the last term yields a nonvanishing
contribution: For Poisson processes, $\diff t \diff N_{\alpha, l}(t) = 0$, but
$\diff N_{\alpha, l}(t) \diff N_{\alpha', l'}(t) = \delta_{\alpha, \alpha'}
\delta_{l, l'} \diff N_{\alpha, l}(t)$
is nonzero~\cite{Wiseman2010}. To obtain the increment $\diff z_{l, l'}(t)$, we
first rewrite the stochastic Schr\"odinger
equation~\eqref{eq:stochastic-schroedinger-equation} as a stochastic master
equation for the conditional state
$\hat{\rho}(t) = \ket{\psi(t)} \bra{\psi(t)}$. Then, using Wick's theorem, the
latter can be recast as an equation for the single-particle density matrix,
which leads to
\begin{multline}
  \label{eq:d-z}
  \diff z_{l, l'}(t) = - \imag \left[ H, z(t) \right]_{l, l'} \diff t - 2
  \sum_{m = 1}^L \left[ \frac{D_{l, m}(t) D_{m, l'}(t)}{D_{m, m}(t)} \diff N_{-,
      m}(t) \right. \\ \left. - \frac{\left( \delta_{l, m} -
        D_{l, m}(t) \right) \left( \delta_{m, l'} - D_{m, l'}(t) \right)}{1 -
      D_{m, m}(t)} \diff N_{+, m}(t) \right],
\end{multline}
with $H$ defined in Eq.~\eqref{eq:Hamiltonian-matrix}. We insert this result in
Eq.~\eqref{eq:d-q-l-1-l-2-l-3}, along with the expression for $\diff s_l(t)$
that follows from
\begin{equation}
  s_l(t + \diff t) = s_l(t) \left( 1 - 2 \diff N_l(t) \right) = s_l(t) + \diff
  s_l(t).
\end{equation}
Thus, with the averages of the Poisson increments $\diff N_{\alpha, l}(t)$ given
in Sec.~\ref{sec:models}, we obtain
\begin{equation}
  \label{eq:d-Q-diagonal}
  \frac{\diff Q_{l, l, l}(t)}{\diff t} = 2 J \Im \! \left[ Q_{l, l + 1, l}(t) +
    Q_{l, l - 1, 1}(t) \right].
\end{equation}
The equation of motion for $Q_{l_1, l_2, l_3}(t)$ with unequal indices has an
additional contribution proportional to $\gamma$; for example,
\begin{multline}
  \label{eq:d-Q-offdiagonal}
  \frac{\diff Q_{l, l \pm 1, l}}{\diff t} = \imag J \left[ Q_{l + 1, l \pm 1,
      l}(t) + Q_{l - 1, l \pm 1, l}(t) \right. \\ \left. - Q_{l, l \pm 1 - 1,
      l}(t) - Q_{l, l \pm 1 + 1, l}(t) \right] - 2 \gamma Q_{l, l \pm 1, l}.
\end{multline}
Our goal is to solve these coupled equations of motion perturbatively in
$J \ll \gamma$. To zeroth order, from Eq.~\eqref{eq:d-Q-diagonal}, we obtain
$Q(t) = Q_{l, l, l}(t) = \mathrm{const.}$---by construction, $Q(t)$ does not
decay in the absence of coherent hopping. The decay of $Q_{l, l, l}(t)$ is due
to the coupling to $Q_{l, l \pm 1, l}(t)$ at first order in $J$. More generally,
the commutator with $H$ in Eq.~\eqref{eq:d-z} couples $Q_{l, l, l}(t)$ to
$Q_{l_1, l_2, l_3}(t)$ at order $\abs{l_1 - l_3} + \abs{l_2 - l_3}$. Therefore,
the leading correction to the zeroth order result is obtained by setting the
right-hand side of Eq.~\eqref{eq:d-Q-offdiagonal} to zero, keeping only
contributions involving $Q_{l, l, l}(t)$ and $Q_{l, l \pm 1, l}(t)$, which leads
to
\begin{equation}
  Q_{l, l \pm 1, l}(t) = - \frac{\imag J}{2 \gamma} Q(t).
\end{equation}
Inserting this in Eq.~\eqref{eq:d-Q-diagonal}, and using $Q(0) = 1$, we obtain
$Q(t) = \e^{- 2 \nu t}$ with $\nu = J^2/\gamma$ as claimed in the main text.

\section{Regularization, causality structure, and normalization in Keldysh field
  theory}
\label{sec:caus-struct-regul}

Here, we address points (i)~and (ii)~listed below
Eq.~\eqref{eq:G-symmetric-limit}. Furthermore, we discuss the causality
structure and normalization of the Keldysh partition function in the replica
limit.

\subsection{Regularization of the Keldysh action}

(i)~In the construction of the Keldysh functional integral, in
Eq.~\eqref{eq:keldysh-partition}, time evolution from $t_0$ to $t$ is split into
$N$ time steps of duration $\Delta t = (t - t_0)/N$. At each discrete time
$t_n = t_0 + \Delta t n$, a resolution of the identity in terms of coherent
states,
$\ket{\pm \psi_{\pm, n}} = \e^{\pm \sum_{r = 1}^R \sum_{l = 1}^L \psi_{\pm, r,
    l, n} \hat{\psi}_{r, l}^{\dagger}} \ket{0}$,
where $\ket{0}$ is the vacuum state, is inserted. If a measurement is
performed between times $t_n$ and $t_{n + 1}$, we thus obtain a matrix element
of the measurement operator between the coherent states $\ket{\psi_{\pm, n}}$
and $\ket{\psi_{\pm, n + 1}}$. For generalized occupation measurements with
measurement operators given in Eq.~\eqref{eq:measurement-ops-OM},
the matrix elements on the forward and backward branches contain the factors
\begin{equation}
  \begin{split}
    \braket{\psi_{+, n + 1} | \hat{n}_{r, l} | \psi_{+, n}} & = \psi^*_{+,
      r, l, n + 1}
    \psi_{+, r, l, n} \braket{\psi_{+, n + 1} | \psi_{+, n}}, \\
    \braket{- \psi_{-, n} | \hat{n}_{r, l} | - \psi_{-, n + 1}} & =
    \psi^*_{-, r, l, n} \psi_{-, r, n + 1, l} \braket{- \psi_{-, n} | -
      \psi_{-, n + 1}},
  \end{split}
\end{equation}
where the fields that represent the operators $\hat{n}_{r, l}$ are one discrete
time step apart. However, as shown in Ref.~\cite{Yang2023a}, in the functional
integral, we can replace
\begin{equation}
  \begin{split}
    \psi^*_{+, r, l, n + 1} \psi_{+, r, l, n} & \to \psi^*_{+, r, l, n}
    \psi_{+, r, l, n} + 1, \\ \psi^*_{-, r, l, n} \psi_{-, r, l, n + 1} & \to
    \psi^*_{-, r, l, n} \psi_{-, r, l, n} + 1,
  \end{split}
\end{equation}
where now the fields are evaluated at equal times. This is a prerequisite for
introducing fermionic bilinears at equal times as new variables. The vertex in
Eq.~\eqref{eq:OM-vertex} is thus modified as
\begin{equation}
  \label{eq:OM-vertex-equal-time}
  V[\psi^{*}, \psi] = \frac{1}{n} \left( \psi^*_+ \psi_+ + 1
  \right) \left( \psi^*_- \psi_- + 1 \right).
\end{equation}
For fermion counting, with the linear measurement operators in
Eq.~\eqref{eq:measurement-ops-FC}, this issue does not arise.

(ii)~In the discrete-time formulation of the theory, the Green's function is
defined as
$\imag G_{n, n'} = \left\langle \psi_n \psi^{\dagger}_{n'}
\right\rangle$~\cite{Kamenev2023}.
Working in the basis introduced through the Larkin-Ovchinnikov rotation,
Eq.~\eqref{eq:Larkin-Ovchinnikov-rotation}, the continuous-time limit is given
by
\begin{equation}  
  \label{eq:continous-time-limit}
  G(t, t') = \lim_{n, n', N \to \infty} \left( G_{n, n'} + \frac{\imag}{2}
    \delta_{n, n'} \sigma_x \right),
\end{equation}
where the limit is taken such that $t_n \to t$ and $t_{n'} \to t'$. In the last
term, we leave the identity matrices in replica and lattice space implicit. The
shift $\imag \delta_{n, n'} \sigma_x/2$ on the right-hand side leads to the
continuous-time Green's function at equal times being given by the symmetrized
limit as in Eq.~\eqref{eq:G-symmetric-limit},
\begin{equation}
  \label{eq:symmetrized-equal-time-limit}
  G(t, t) = \frac{1}{2} \left[ G(t, t + 0^+) + G(t, t - 0^+) \right].
\end{equation}
In turn, this implies that the Keldysh Green's function is continuous at $t = t'$,
and that the anti-Keldysh component of the Green's function vanishes at all
times---which is not the case for the discrete-time Green's function $G_{n, n'}$
that has a nonvanishing anti-Keldysh component at $n = n'$~\cite{Kamenev2023}.

For most applications of Keldysh field theory, the shift on the right-hand side
of Eq.~\eqref{eq:continous-time-limit} is irrelevant and can be
omitted. However, here we have to take it into account to ensure normalization
of the Keldysh partition function as discussed below. The shift can be
incorporated in the action by switching to a regularization in which the free
discrete-time Green's function is
$G^{\mathrm{(reg)}}_{n, n'} = G_{n, n'} + \imag \delta_{n, n'}
\sigma_x/2$~\cite{Poboiko2023}.
Changing the regularization is achieved through the following relation:
\begin{equation}
  \label{eq:regularization-Z}
  \begin{split}
    Z_R(t) & = \int \frac{\Diff[\psi^*, \psi]}{\Det \! \left( - \imag
        G_0^{-1} \right)} \, \e^{\imag \left( \psi^{\dagger} G_0^{-1} \psi + \gamma
        \int \diff^2 \mathbf{x} \, \mathcal{L}_M[\psi^*_l, \psi_l] \right)} \\ &
    = \int \frac{\Diff[\psi^*, \psi]}{\Det \! \left( - \imag
        G_0^{\mathrm{(reg)} -1} \right)} \, \e^{\imag \left( \psi^{\dagger}
        G_0^{\mathrm{(reg)} -1} \psi + \gamma \int \diff^2 \mathbf{x} \,
        \mathcal{L}_M^{\mathrm{(reg)}}[\psi^*_l, \psi_l] \right)},
  \end{split}
\end{equation}
where we make the determinants that are usually absorbed in the integration
measure explicit~\cite{Kamenev2023}. The regularized measurement Lagrangian can
be constructed order by order in an expansion in $\gamma$. To first order,
$\mathcal{L}_M^{\mathrm{(reg)}}$ is obtained by adding the sum over all partial
contractions, where each contraction is
$\left\langle \psi_n \psi^{\dagger}_{n'} \right\rangle = \delta_{n, n'} \sigma_x/2$,
to the original measurement Lagrangian $\mathcal{L}_M$.  Since these
contractions are diagonal in replica space, they can be added to each factor of
the product over replicas in Eq.~\eqref{eq:measurement-Lagrangian}, that is, to
the vertices Eqs.~\eqref{eq:FC-vertices} and~\eqref{eq:OM-vertex-equal-time}. As
stated in the main text, the vertices for fermion counting do not change under
the regularization. The regularized vertex for occupation measurements is given
in Eq.~\eqref{eq:OM-vertex-rotated}.

\subsection{Causality structure and normalization of the Keldysh partition
  function}

For closed quantum systems undergoing unitary dynamics without measurements, the
Keldysh partition function is normalized to unity,
$Z(t) = \tr(\hat{\rho}(t)) = 1$, as follows from the normalization of the
density matrix $\hat{\rho}(t)$. However, the replica Keldysh partition function
$Z_R(t)$ is defined in Eq.~\eqref{eq:keldysh-partition} in terms of $R$ copies
of the nonnormalized density matrix $\hat{D}(t)$. In the replica limit
$R \to 1$, the completeness relation of measurement operators implies again that
$Z_R(t) \to 1$. For $R > 1$, on the other hand, the replica Keldysh partition
function is not normalized. This is because the products of measurement
operators over $r$ replicas, which occur in $\otimes_{r = 1}^R \hat{D}_r(t)$, do
not obey a completeness relation.

In Keldysh field theory for closed systems, the normalization of the partition
function is reflected in the causality structure of the action, in particular,
in the property that the action vanishes for $\psi_+ = \psi_-$ and
$\psi^*_+ = \psi^*_-$~\cite{Altland2010a, Kamenev2023, Sieberer2016a,
  Sieberer2023}. To see this, we consider a system with Hamiltonian
$\hat{H}$. The Keldysh partition function is
$Z(t) = \int \Diff[\psi^*, \psi] \, \e^{\imag S}$ where
$S = \int_{t_0}^t \diff t' \mathcal{L}[\psi^*, \psi]$, and with the
Lagrangian
\begin{equation}
  \label{eq:Lagrangian-closed-system}
  \mathcal{L}[\psi^*, \psi] = \sum_{\sigma = \pm} \sigma \left(
    \psi^{\dagger}_{\sigma} \partial_t \psi_{\sigma} - H[\psi^*_{\sigma},
    \psi_{\sigma}] \right).
\end{equation}
From $Z(t) = \tr(\hat{\rho}(t)) = 1$ it follows that
\begin{equation}
  \label{eq:causality-structure-Lagrangian}
  \partial_t Z = \imag \left\langle \mathcal{L}[\psi^*, \psi] \right\rangle =
  \imag \sum_{\sigma = \pm} \sigma \left\langle
    \psi^{\dagger}_{\sigma} \partial_t \psi_{\sigma} - H[\psi^*_{\sigma},
    \psi_{\sigma}] \right\rangle = 0.
\end{equation}
Upon taking the sum over $\sigma = \pm$, both terms in the average vanish
individually. Indeed, the cyclic property of the trace implies that the
expectation value of any operator $\hat{O}$ at time $t$ can be evaluated
equivalently on the forward and on the backward branch of the Keldysh contour,
and, therefore,
\begin{equation}
  \label{eq:causality-Lagrangian-expectation}
  \left\langle \mathcal{L}[\psi^*, \psi] \right\rangle = \left\langle
    \left. \mathcal{L}[\psi^*, \psi] \right.\rvert_{\psi_+ = \psi_-,
      \psi^*_+ = \psi^*_-} \right\rangle.
\end{equation}
Then, $\left\langle \mathcal{L}[\psi^*, \psi] \right\rangle = 0$ follows for
the Lagrangian of a closed system in Eq.~\eqref{eq:Lagrangian-closed-system}
from the causality structure,
\begin{equation}
  \label{eq:causality-Lagrangian}
  \left. \mathcal{L}[\psi^*, \psi] \right.\rvert_{\psi_+ = \psi_-,
    \psi^*_+ = \psi^*_-} = 0.
\end{equation}
This shows that the causality structure of the action guarantees the
normalization of the Keldysh partition function.

As mentioned in Sec.~\ref{sec:repl-keldysh-action}, the term
$\psi^*_2 \psi_1$ in the regularized vertex for occupation measurements,
Eq.~\eqref{eq:OM-vertex-rotated}, violates the usual causality
structure. Inverting the Larkin-Ovchinnikov rotation,
Eq.~\eqref{eq:Larkin-Ovchinnikov-rotation}, shows that this term couples the
forward and backward branches. The above argument based on the cyclic property
of the trace does, in general, not apply, and
Eqs.~\eqref{eq:causality-Lagrangian-expectation}
and~\eqref{eq:causality-Lagrangian} do not hold for such terms. To see that the
partition function is normalized even though the action does not obey the usual
causality structure, one should note that according to
Eq.~\eqref{eq:causality-structure-Lagrangian}, normalization is ensured when the
expectation value of the Lagrangian vanishes. We will now confirm that this
condition is met.

The free evolution described by the first term in the Keldysh action in
Eq.~\eqref{eq:S-fermionic} obeys the causality structure for any value of $R$,
and, therefore, we only need to consider the measurement Lagrangian
Eq.~\eqref{eq:measurement-Lagrangian}. For $R = 1$, we thus require
\begin{equation}
  \label{eq:L-M-causality}
  \left\langle \imag \mathcal{L}_M[\psi^*, \psi] \right\rangle = \sum_{n =
    \pm} \left\langle V_n[\psi^*, \psi] \right\rangle - 1 = 0.
\end{equation}
We first check this condition for fermion counting. Using the form of the
vertices given in Eq.~\eqref{eq:FC-vertices-rotated}, we obtain
\begin{equation}
  \label{eq:average-L-M-FC}
  \left\langle \imag \mathcal{L}_M[\psi^*_l, \psi_l] \right\rangle = \imag
  \left[ G_{l, l}^R(t, t) - G_{l, l}^A(t, t) \right] - 1.
\end{equation}
To evaluate the Green's functions at equal times, we express them through
fermionic field operators~\cite{Altland2010a}:
\begin{equation}
  \begin{split}    
    \imag G_{l, l'}^R(t, t') & = \left\langle \psi_{1, l}^{}(t) \psi_{1,
        l'}^{*}(t') \right\rangle = \theta(t - t') \left\langle \left\{
        \hat{\psi}_l^{}(t), \hat{\psi}_{l'}^{\dagger}(t') \right\} \right\rangle, \\
    \imag G^A_{l, l'}(t, t') & = \left\langle \psi_{2, l}^{}(t) \psi_{2,
        l'}^{*}(t') \right\rangle = - \theta(t' - t) \left\langle \left\{
        \hat{\psi}_l^{}(t), \hat{\psi}_{l'}^{\dagger}(t') \right\} \right\rangle, \\
    \imag G_{l, l'}^K(t, t') & = \left\langle \psi_{1, l}^{}(t) \psi_{2,
        l'}^{*}(t') \right\rangle = \left\langle \left[ \hat{\psi}_l^{}(t),
        \hat{\psi}_{l'}^{\dagger}(t') \right] \right\rangle,
  \end{split}
\end{equation}
where two-time averages are defined as in Eq.~\eqref{eq:quantum-regression}. The
values of the Green's functions in the symmetrized limit of equal times,
Eq.~\eqref{eq:symmetrized-equal-time-limit}, follow directly from fermionic
anticommutation relations and the expectation value
$\left\langle \hat{n}_l \right\rangle_{\mathrm{ss}}$ in the unconditional steady
state. According to Sec.~\ref{sec:stochastic-schrodinger-equation},
$\left\langle \hat{n}_l \right\rangle_{\mathrm{ss}} = n$ for occupation
measurements, with $n = 1/2$ for fermion counting with equal rates of loss and
gain. We thus obtain
\begin{equation}
  \label{eq:equal-time-GFs}  
  \imag G_{l, l'}^{R/A}(t, t) = \pm \frac{1}{2} \delta_{l, l'}, \quad
  \imag G_{l, l'}^K(t, t) = \left( 1 - 2 n \right) \delta_{l, l'}.
\end{equation}
Inserting these values in Eq.~\eqref{eq:average-L-M-FC} yields
$\left\langle \imag \mathcal{L}_M[\psi^*_l, \psi_l] \right\rangle = 0$,
confirming that the Keldysh partition function is normalized.

Finally, we check Eq.~\eqref{eq:L-M-causality} for occupation
measurements. Since the regularized vertex
Eq.~\eqref{eq:OM-vertex-rotated} is quartic, the expectation value
in Eq.~\eqref{eq:L-M-causality} cannot be evaluated exactly. At lowest order in
$\gamma$, we can use Wick's theorem to obtain
\begin{equation}
  \left\langle \imag \mathcal{L}_M[\psi^*_l, \psi_l] \right\rangle =
  \frac{1}{n} \left[ \frac{1}{4} - \frac{\imag}{2} G^K_{l, l}(t, t) + G^R_{l,
      l}(t, t) G^A_{l, l}(t, t) \right] - 1,
\end{equation}
where now the Green's functions are those of the system without
measurements. However, Eq.~\eqref{eq:equal-time-GFs} still applies, and we again
find $\left\langle \imag \mathcal{L}_M[\psi^*_l, \psi_l] \right\rangle = 0$.

\section{The symmetry underlying the NLSM}
\label{sec:app-replicon-symmetry}

In the following, we first show that rotations of the form
$\mathcal{R}_{\Phi} = \e^{\imag \Phi \sigma_x/2}$, with a traceless Hermitian
matrix $\Phi$, are a symmetry of the measurement Lagrangian for fermion
counting, Eq.~\eqref{eq:lagrangian-FC}. Then, we derive this symmetry of the
Lagrangian from a strong symmetry of the superoperator that describes the time
evolution of the $R$-replica nonnormalized density matrix~\cite{Buca2012,
  Albert2014}. This derivation leads to conditions (i)~and (ii)~for the symmetry
to occur given in Sec.~\ref{sec:symmetry-keldysh-action}. Finally, we discuss
modifications of the theory for inefficient detection, which lead to a breaking
of the symmetry and thus show that also condition (iii)~is necessary.

\subsection{Symmetry of the measurement Lagrangian under $\mathcal{R}_{\Phi}$}

To show that rotations of the form $\mathcal{R}_{\Phi}$ are a symmetry of the
measurement Lagrangian for fermion counting, Eq.~\eqref{eq:lagrangian-FC}, we
first perform two auxiliary calculations. We define the adjoint action of a
matrix $A$ on a matrix $B$ as $\mathcal{A}_A B = \left[ A, B \right]$. Then,
\begin{equation}
  \begin{split}
    \mathcal{R}_{\Phi}^{-1} \tau_{\pm} \mathcal{R}_{\Phi} & = \sum_{n =
      0}^{\infty} \frac{\left( - \imag \right)^n}{n!} \mathcal{A}_{\Phi
      \sigma_x/2}^n \tau_{\pm} = \sum_{n = 0}^{\infty} \frac{\left( \mp \imag
        \Phi \right)^n}{n!}  \tau_{\pm} = \tau_{\pm} \e^{\mp \imag \Phi},
  \end{split}
\end{equation}
where we have used that $\tau_{\pm}$ is invariant---up to a change of
sign---under multiplication by $\sigma_x$ from both the left and the right,
$\sigma_x \tau_{\pm} = \pm \tau_{\pm}$ and
$\tau_{\pm} \sigma_x = \mp \tau_{\pm}$, which can be combined to
$\mathcal{A}_{\sigma_x} \tau_{\pm} = \pm 2 \tau_{\pm}$. Next, for a matrix $A$
in Keldysh and replica space, we consider
\begin{multline}
  \trK \! \left( \mathcal{R}_{\Phi} \tau_{\pm} A
    \mathcal{R}_{\Phi}^{-1} \right) \\
  \begin{aligned}
    & = \sum_{n = 0}^{\infty} \frac{\imag^n}{n!} \trK \! \left(
      \mathcal{A}_{\Phi \sigma_x/2}^n \tau_{\pm} A \right) = \sum_{n =
      0}^{\infty} \frac{\imag^n}{n!}  \mathcal{A}_{\Phi/2}^n \trK \! \left(
      \sigma_x^n \tau_{\pm} A \right) \\ & = \sum_{n = 0}^{\infty}
    \frac{\left( \pm \imag \right)^n}{n!}  \mathcal{A}_{\Phi/2}^n \trK \! \left(
      \tau_{\pm} A \right) = \e^{\pm \imag \Phi/2} \trK(\tau_{\pm}
    A) \e^{\mp \imag \Phi/2},
  \end{aligned}
\end{multline}
These two results and
$\detR \! \left( \e^{\mp \imag \Phi} \right) = \e^{\mp \imag \trR(\Phi)} = 1$
lead to
\begin{multline}
  \detR \! \left[ \trK \! \left( \tau_{\pm} \mathcal{R}_{\Phi} \mathcal{G}
      \mathcal{R}_{\Phi}^{-1} \right) \right] = \detR \! \left[ \trK \! \left(
      \mathcal{R}_{\Phi} \tau_{\pm} \e^{\mp
        \imag \Phi} \mathcal{G} \mathcal{R}_{\Phi}^{-1} \right) \right] \\
  \begin{aligned}
    & = \detR \! \left[ \e^{\pm \imag \Phi/2} \trK \! \left( \tau_{\pm} \e^{\mp
          \imag \Phi} \mathcal{G} \right) \e^{\mp \imag \Phi/2} \right] \\ & =
    \detR \! \left( \e^{\mp \imag \Phi} \right) \detR \! \left[ \trK \! \left(
        \tau_{\pm} \mathcal{G} \right) \right] = \detR \! \left[ \trK \!  \left(
        \tau_{\pm} \mathcal{G} \right) \right],
  \end{aligned}
\end{multline}
showing that the transformation $\mathcal{G} \mapsto \mathcal{R}_{\Phi}
\mathcal{G} \mathcal{R}_{\Phi}^{-1}$ is indeed a symmetry of
Eq.~\eqref{eq:lagrangian-FC} for $R > 1$.

The above derivation relies on $\tau_{\pm}$ being invariant under multiplication
by $\sigma_x$. This invariance does not apply to multiplication by $\sigma_y$ or
$\sigma_z$. Therefore, rotations with $\sigma_x$ replaced by $\sigma_y$ or
$\sigma_z$ are not symmetries of the measurement Lagrangian.

\subsection{Derivation from a superoperator symmetry}

We next derive the symmetry of the Keldysh action under
$\mathcal{R}_{\Phi} = e^{\imag \Phi \sigma_x/2}$ from the operator formalism. We
proceed in two steps: First, we discuss the different types of symmetries
associated with the transformation $\hat{G}_{\Phi}$ introduced in
Eq.~\eqref{eq:G-Phi}. In particular, we distinguish weak and strong symmetries
of the superoperator that describes the time evolution of the $R$-replica
nonnormalized density matrix, and explain how they are related to particle
number conservation~\cite{Buca2012, Albert2014, Sieberer2023}. Second, we show
that the symmetry of the Keldysh action under $\mathcal{R}_{\Phi}$ results the
strong symmetry of the superoperator under $\hat{G}_{\Phi}$.

\subsubsection{Symmetries of the time evolution superoperator}

We first want to understand under which conditions the unitary operator
$\hat{G}_{\Phi}$ defined in Eq.~\eqref{eq:G-Phi} is a symmetry of the time
evolution operator for $R$ replicas, $\prod_{r = 1}^R \hat{V}_r(t)$. Each factor
$\hat{V}_r(t)$ is of the form given in Eq.~\eqref{eq:V}, and comprises both
coherent dynamics and measurements.

For generality, let us consider coherent evolution generated by a generic
quadratic Hamiltonian. Then, $\prod_{r = 1}^R \hat{V}_r(t)$ contains an
$R$-replica Hamiltonian of the form
\begin{equation}
  \hat{H}_R = \sum_{r = 1}^R \sum_{l, l' = 1}^L \left( - J_{l, l'}^{} \hat{\psi}_{r,
      l}^{\dagger} \hat{\psi}_{r, l'}^{} + \Delta_{l, l'}^{} \hat{\psi}_{r, l}
    \hat{\psi}_{r, l'} + \mathrm{H.c.} \right).
\end{equation}
The action of $\hat{G}_{\Phi}$ on fermionic field operators is given by
\begin{equation}
  \label{eq:G-Phi-psi-r-l-G-Phi-dagger}
  \hat{G}_{\Phi}^{\dagger} \hat{\psi}_{r, l}^{} \hat{G}_{\Phi}^{} = \sum_{r' = 1}^R
  G_{r, r'}^{} \hat{\psi}_{r', l}^{}, \qquad G = \e^{\imag \Phi/2}.
\end{equation}
It is thus straightforward to see that particle-number conserving terms in the
Hamiltonian are symmetric under $\hat{G}_{\Phi}$. In contrast, pairing terms are
not symmetric. Such pairing terms are contained in the Majorana model of
Ref.~\cite{Fava2023}.

We next consider measurements. 
As above, for the sake of generality, we allow more generic types of measurement
operators than those considered in the main text. Our only requirement is that
each replica is at all times in a pure Gaussian state with a fixed number of
particles---these properties are also preserved by coherent evolution generated
by a particle-number conserving quadratic Hamiltonian. Such states are Slater
determinants as given for $R$ replicas in
Eq.~\eqref{eq:R-replica-N-particle-Slater-determinant}.
The above requirement is thus satisfied if each measurement operator maps an
$N$-particle Slater determinant to an $N + \Delta N$-particle Slater
determinant, where the integer $\Delta N$ may depend on the measurement
operator. For the operators in Eq.~\eqref{eq:measurement-ops-OM} that describe
generalized measurements of occupation measurements, $\Delta N = 0$, whereas for
fermion counting with measurement operators given in
Eq.~\eqref{eq:measurement-ops-FC}, $\Delta N = - 1$ and $\Delta N = + 1$ for the
measurement operators $\hat{M}_{-, l}$ and $\hat{M}_{+, l}$, respectively.

Consider now a generic measurement operator $\hat{M}$ that satisfies this
requirement. We want to infer how $\prod_{r = 1}^R \hat{M}_r$ transforms under
$\hat{G}_{\Phi}$. In fact, for our purposes it is sufficient to determine the
transformation of the restriction of $\prod_{r = 1}^R \hat{M}_r$ to the space
spanned by $R$-replica Slater determinants that are symmetric under permutations
of replicas as given in
Eq.~\eqref{eq:R-replica-N-particle-Slater-determinant}. This is because by
construction of the replica formalism in Sec.~\ref{sec:replica-trick}, the
initial $R$-replica state is symmetric under permutations of replicas, and this
property is conserved due to the product structure of the time evolution
operator, $\prod_{r = 1}^R \hat{V}_r(t)$.  Combined with the above assumptions
on preservation of Gaussianity and a fixed number of particles in each replica,
we see that $\prod_{r = 1}^R \hat{M}_r$ maps permutation-symmetric $R$-replica
$N$-particle Slater determinants to permutation-symmetric $R$-replica
$N + \Delta N$-particle Slater determinants.

As a preliminary step, we consider the transformation of the product of field
operators over the replica index,
\begin{equation}
  \hat{G}_{\Phi}^{\dagger} \left( \prod_{r = 1}^R \hat{\psi}_{r, l} \right)
  \hat{G}_{\Phi} = \prod_{r = 1}^R \sum_{r' = 1}^R G_{r, r'} \hat{\psi}_{r', l}
  = \sum_{r_1', \dotsc, r_R' = 1}^R \prod_{r = 1}^R G_{r, r_r'}
  \hat{\psi}_{r_r', l}.
\end{equation}
The product in the last equality can be split into two independent products. A
further simplification is due to fermionic statistics: The product
$\prod_{r = 1}^R \hat{\psi}_{r_r', l}$ is nonzero only if all replica indices
$r_r'$ are different, that is, if $r_r' = \sigma_r$ where
$\sigma \in \mathrm{S}_R$ is an element of the symmetric group of permutations
of $R$ elements. Then, upon rearranging the product we obtain
$\prod_{r = 1}^R \hat{\psi}_{\sigma_r, l} = \sgn(\sigma) \prod_{r = 1}^R
\hat{\psi}_{r, l}$. We thus find
\begin{equation}
  \label{eq:G-Phi-prod-r-psi-r-l-G-Phi-dagger}
  \begin{split}
    \hat{G}_{\Phi}^{\dagger} \left( \prod_{r = 1}^R \hat{\psi}_{r, l} \right)
    \hat{G}_{\Phi} & = \sum_{\sigma \in \mathrm{S}_R} \sgn(\sigma) \left(
      \prod_{r = 1}^R G_{r, \sigma_r} \right) \left( \prod_{r = 1}^R
      \hat{\psi}_{r, l} \right) \\ & = \det(G) \prod_{r = 1}^R \hat{\psi}_{r, l}
    = \e^{\imag \trR(\Phi)/2} \prod_{r = 1}^R \hat{\psi}_{r, l},
  \end{split}
\end{equation}
which shows that the product of field operators is invariant under the
transformation $\hat{G}_{\Phi}$ if $\trR(\Phi) = 0$. Otherwise, the product
acquires a phase factor.

We next consider the action of $\hat{G}_{\Phi}$ on a permutation-symmetric
$R$-replica $N$-particle Slater determinant as given in
Eq.~\eqref{eq:R-replica-N-particle-Slater-determinant}. With the above results,
we can straightforwardly deduce the action of $\hat{G}_{\Phi}^{\dagger}$ on this
state:
\begin{equation}
  \label{eq:G-Phi-dagger-R-N-Slater-determinant}
  \begin{split}
    \hat{G}_{\Phi}^{\dagger} \ket{\psi_{R, N}} & = s_{N, R} \prod_{n = 1}^N
    \prod_{r = 1}^R \sum_{l = 1}^L \psi_{l, n} \hat{G}_{\Phi}^{\dagger}
    \hat{\psi}_{r, l}^{\dagger} \hat{G}_{\Phi} \ket{0} \\ & = s_{N, R} \prod_{n
      = 1}^N \prod_{r = 1}^R \sum_{r' = 1}^R G_{r, r'}^{*} \sum_{l = 1}^L
    \psi_{l, n} \hat{\psi}_{r, l}^{\dagger} \ket{0} \\ & = \e^{- \imag N
      \trR(\Phi)/2} \ket{\psi_{R, N}}.
  \end{split}
\end{equation}
In the first equality, we have used that
$\hat{G}_{\Phi}^{\dagger} \ket{0} = \ket{0}$, and that exchanging the order of
products over $r$ and $n$ in
Eq.~\eqref{eq:R-replica-N-particle-Slater-determinant} leads merely to a sign
factor
$s_{N, R} = \left( - 1 \right)^{N \left( N - 1 \right) R \left( R - 1 \right) \!
  /4}$;
in the second equality, we have employed
Eq.~\eqref{eq:G-Phi-psi-r-l-G-Phi-dagger}; finally, proceeding as in
Eq.~\eqref{eq:G-Phi-prod-r-psi-r-l-G-Phi-dagger} establishes the third equality.

As explained above, we may focus on the restriction of
$\prod_{r = 1}^R \hat{M}_r$ to a map from $R$-replica $N$-particle Slater
determinants $\ket{\psi_{R, N}}$,
Eq.~\eqref{eq:R-replica-N-particle-Slater-determinant}, to the same type of
state but with $N + \Delta N$ particles. Under this restriction and with
Eq.~\eqref{eq:G-Phi-dagger-R-N-Slater-determinant}, we thus find
\begin{equation}
  \label{eq:G-Phi-dagger-prod-r-M-r-G-Phi}
  \hat{G}_{\Phi}^{\dagger} \left( \prod_{r = 1}^R \hat{M}_r \right)
  \hat{G}_{\Phi} = \e^{- \imag \Delta N \trR(\Phi)/2} \prod_{r = 1}^R
  \hat{M}_r.
\end{equation}
This result clarifies under which conditions $\hat{G}_{\Phi}$ is a symmetry of
$\prod_{r = 1}^R \hat{M}_r$: It is always a symmetry if $\Delta N = 0$, that is,
if measurements preserve the number of particles. In contrast, if measurements
change the number of particles and $\Delta N \neq 0$, $\hat{G}_{\Phi}$ is a
symmetry of $\prod_{r = 1}^R \hat{M}_r$ only for $\trR(\Phi) = 0$.

Returning now to the full time evolution described by
$\prod_{r = 1}^R \hat{V}(t)$ that includes both measurements and coherent
dynamics, we see that $\hat{G}_{\Phi}$ is a symmetry of
$\prod_{r = 1}^R \hat{V}(t)$ if both the Hamiltonian and measurements conserve
the number of particles. If the Hamiltonian is particle-number conserving but
measurements change the number of particles by an integer $\Delta N$,
$\hat{G}_{\Phi}$ is a symmetry of the dynamics only for $\trR(\Phi) = 0$.
Finally, $\hat{G}_{\Phi}$ is not a symmetry if particle-number conservation is
broken by the Hamiltonian.

So far, we have focused on $\prod_{r = 1}^R \hat{V}_r(t)$. However, in
Eq.~\eqref{eq:keldysh-partition}, both $\prod_{r = 1}^R \hat{V}_r(t)$ and
$\prod_{r = 1}^R \hat{V}_r(t)^{\dagger}$ appear, acting on the initial density
matrix $\hat{\rho}_{R, 0} = \otimes_{r = 1}^R \hat{\rho}_{0, r}$ from the left-
and right-hand-side, respectively. Therefore, we next study symmetries of the
superoperator $\mathcal{V}(t)$ defined by
\begin{equation}
  \label{eq:mathcal-V}
  \mathcal{V}(t) \hat{\rho} = \left( \prod_{r = 1}^R \hat{V}_r(t) \right)
  \hat{\rho} \left( \prod_{r = 1}^R \hat{V}_r(t)^{\dagger} \right),
\end{equation}
such that Eq.~\eqref{eq:keldysh-partition} can be written as
\begin{equation}
  \label{eq:keldysh-partition-mathcal-V}
  Z_R(t) = \sum_{\{ \alpha_m, l_m, t_m \}} \tr \! \left[ \mathcal{V}(t)
    \hat{\rho}_{R, 0} \right].
\end{equation}
For a superoperator $\mathcal{V}(t)$, one distinguishes between weak and strong
symmetries~\cite{Buca2012, Albert2014}: The unitary operator $\hat{G}_{\Phi}$ is
a weak symmetry if $\mathcal{V}(t)$ is invariant under the simultaneous
transformation with $\hat{G}_{\Phi}$ and $\hat{G}_{\Phi}^{\dagger}$ from the
left- and right-hand side, respectively. That is, defining the superoperator
$\mathcal{G}_{\Phi}$ as this simultaneous transformation,
$\mathcal{G}_{\Phi}^{} \hat{\rho} = \hat{G}_{\Phi}^{} \hat{\rho}
\hat{G}_{\Phi}^{\dagger}$,
the weak symmetry condition reads
$\left[ \mathcal{V}(t), \mathcal{G}_{\Phi} \right] = 0$. In contrast,
$\hat{G}_{\Phi}$ is a strong symmetry if $\mathcal{V}(t)$ is invariant under
independent transformations to on the left- and right-hand-sides. Defining
$\mathcal{G}_{+, \Phi} \hat{\rho} = \hat{G}_{\Phi} \hat{\rho}$ and
$\mathcal{G}_{-, \Phi} \hat{\rho} = \hat{\rho} \hat{G}_{\Phi}^{\dagger}$, the
strong symmetry condition can be stated equivalently as
$\left[ \mathcal{V}(t), \mathcal{G}_{+, \Phi} \right] = 0$ or
$\left[ \mathcal{V}(t), \mathcal{G}_{-, \Phi} \right] = 0$. Evidently, the
strong symmetry condition implies the weak symmetry condition. To check for both
possibilities at the same time, we consider
\begin{multline}
  \label{eq:mathcal-V-transformed}
  \mathcal{G}_{-, \Phi_-}^{\dagger} \mathcal{G}_{+, \Phi_+}^{\dagger}
  \mathcal{V}(t) \mathcal{G}_{+, \Phi_+} \mathcal{G}_{-, \Phi_-} \hat{\rho} \\
  \begin{aligned}
    & = \hat{G}_{\Phi_+}^{\dagger} \left( \prod_{r = 1}^R \hat{V}_r(t) \right)
    \hat{G}_{\Phi_+}^{} \hat{\rho} \hat{G}_{\Phi_-}^{\dagger} \left( \prod_{r =
        1}^R \hat{V}_r(t)^{\dagger} \right) \hat{G}_{\Phi_-}^{} \\ & = \e^{-
      \frac{\imag \Delta N}{2} \left[ \trR \! \left( \Phi_+ \right) - \trR \!
        \left( \Phi_- \right) \right]} \mathcal{V}(t) \hat{\rho}.
  \end{aligned}
\end{multline}
In the last equality, we have used
Eq.~\eqref{eq:G-Phi-dagger-prod-r-M-r-G-Phi}. However, here we denote by
$\Delta N$ the total change of the number of particles in a single replica due
to the measurement operators included in $\hat{V}(t)$. The transformation of
$\mathcal{V}(t)$ in the first line in Eq.~\eqref{eq:mathcal-V-transformed} is a
symmetry if the phase factor in the last equality reduces to unity. This is the
case for $\Phi_+ = \Phi_-$, which shows that $\hat{G}_{\Phi}$ is a weak symmetry
for any value of $\trR(\Phi)$. To check whether $\hat{G}_{\Phi}$ is also a
strong symmetry, we may set $\Phi = \Phi_+$ and $\Phi_- = 0$. Then, the phase
factor evaluates to unity and $\hat{G}_{\Phi}$ is a strong symmetry if
measurements conserve the number of particles, $\Delta N = 0$. However, even if
$\Delta N \neq 0$, there is still a strong $\mathrm{SU}(R)$ symmetry described
by $\hat{G}_{\Phi}$ with $\trR(\Phi) = 0$. As we show next, this last symmetry
underlies the NLSM.

\subsubsection{Connection between operator and Keldysh formulations}

The weak and strong symmetries of $\mathcal{V}(t)$ are directly related to
symmetries of Keldysh action. To establish this connection, we start from the
Keldysh partition function given in
Eq.~\eqref{eq:keldysh-partition-mathcal-V}. We assume that $\mathcal{V}(t)$
obeys the symmetry condition Eq.~\eqref{eq:mathcal-V-transformed} with the phase
factor on the right-hand side evaluating to unity. Then, the Keldysh partition
function can be expressed in terms of transformed operators as
\begin{multline}
  \label{eq:Z-R-Phi-+-Phi--}
  Z_R(t) = \sum_{\{ \alpha_m, l_m, t_m \}} \tr \! \left[ \left( \prod_{r = 1}^R
      \hat{G}_{\Phi_+}^{\dagger} \hat{V}_r(t) \hat{G}_{\Phi_+}^{} \right)
  \right. \\ \left. \times \hat{\rho}_{R, 0} \left( \prod_{r = 1}^R
      \hat{G}_{\Phi_-}^{\dagger} \hat{V}_r(t)^{\dagger} \hat{G}_{\Phi_-}^{}
    \right) \right].
\end{multline}
The operators $G_{\Phi_+}^{\dagger}$ and $G_{\Phi_-}^{\dagger}$ can be pulled
through $\hat{V}_r(t)$ and $\hat{V}_r(t)^{\dagger}$, respectively, by
transforming all field operators that appear in $\hat{V}_r(t)$ and
$\hat{V}_r(t)^{\dagger}$ according to
Eq.~\eqref{eq:G-Phi-psi-r-l-G-Phi-dagger}. Using the resulting expression as the
starting point for the construction of the Keldysh functional integral
representation as in Eq.~\eqref{eq:Z-R-M-M-dagger}, we find the same action as
in Eq.~\eqref{eq:S-fermionic}, but with transformed fields. Leaving matrix
multiplication in replica space and the lattice site index implicit, the
transformation of fields is
\begin{equation}
  \label{eq:psi-pm-G-Phi-pm}
  \psi_{\pm} \mapsto G_{\Phi_{\pm}} \psi_{\pm}, \qquad \psi^{\dagger}_{\pm} \mapsto
  \psi^{\dagger}_{\pm} G_{\Phi_{\pm}}^{\dagger}.
\end{equation}
We thus see that the weak symmetry of $\mathcal{V}(t)$ corresponds to the
symmetry of the Keldysh action under identical transformations of fields on the
forward and backward branches. In contrast, the strong symmetry of
$\mathcal{V}(t)$ for $\Delta N = 0$ or $\trR(\Phi) = 0$ is reflected in the
symmetry of the Keldysh action under independent transformations on the forward
and backward branches~~\cite{Sieberer2023}.

Finally, we perform a Larkin-Ovchinnikov transformation of the fields,
Eq.~\eqref{eq:Larkin-Ovchinnikov-rotation}. At the same time, we transform the
matrices $\Phi_{\pm}$ as
\begin{equation}
  \bar{\Phi} = \frac{1}{2} \left( \Phi_+ + \Phi_- \right), \qquad \Phi =
  \frac{1}{2} \left( \Phi_+ - \Phi_- \right).
\end{equation}
Then, for $\psi = \left( \psi_1, \psi_2 \right)^{\transpose}$,
Eq.~\eqref{eq:psi-pm-G-Phi-pm} becomes
\begin{equation}
  \psi \mapsto \e^{\frac{\imag}{2} \left( \bar{\Phi} + \Phi \sigma_x \right)} \psi.
\end{equation}
In this formulation, we see that the weak symmetry of $\mathcal{V}(t)$ under
$\hat{G}_{\Phi}$ implies symmetry with $\bar{\Phi} = \Phi_+ = \Phi_-$ and
$\Phi = 0$; the strong symmetry implies the additional symmetry of the Keldysh
action under transformations with $\Phi = \Phi_+ = - \Phi_-$. For
$\bar{\Phi} = 0$, these are just the desired rotations $\mathcal{R}_{\Phi} =
\e^{\imag \Phi \sigma_x/2}$.

The above analysis shows that the $\mathrm{SU}(R)$ symmetry of the Keldysh
action under rotations of the form
$\mathcal{R}_{\Phi} = \e^{\imag \Phi \sigma_x/2}$ is a consequence of the strong
$\mathrm{SU}(R)$ symmetry of the time evolution superoperator $\mathcal{V}(t)$
under the unitary transformation $\hat{G}_{\Phi}$, Eq.~\eqref{eq:G-Phi}, with
$\trR(\Phi) = 0$. They key property of $\mathcal{V}(t)$ which underlies this
symmetry is that the dynamics described by $\mathcal{V}(t)$ are restricted to
the space of permutation-symmetric $R$-replica $N$-particle Slater determinants,
Eq.~\eqref{eq:G-Phi-dagger-R-N-Slater-determinant}. As detailed above, this is
the case for particle-number conserving quadratic Hamiltonians and measurements
that map $N$-particle Slater determinants to $N + \Delta N$-particle Slater
determinants. These requirements are formulated as condition (i)~in
Sec.~\ref{sec:symmetry-keldysh-action}.

\subsection{Breaking of the symmetry due to inefficient detection}

Finally, we elaborate on condition (ii)~in
Sec.~\ref{sec:symmetry-keldysh-action}, that the outcomes of all measurements
have to be recorded. To prove the necessity of imposing this condition, we
demonstrate the breaking of the symmetry due to inefficient detection. As
discussed, for example, in Ref.~\cite{Wiseman2010}, inefficient detection can be
modeled by averaging over a finite fraction of the measurement outcomes. Here,
we focus on fully inefficient detection, as described by the unconditional
dynamics of $R$ replicas.

But for the moment, let us disregard measurements and focus on the Hamiltonian
dynamics of an isolated system. The total system is subdivided into two parts,
the system of interest and a reservoir. Conservation of the total number of
particles is reflected, via the Noether theorem, in a global continuous
$\mathrm{U}(1)$ symmetry of the Hamiltonian. When we integrate out the reservoir
to obtain a description of the system alone, the $\mathrm{U}(1)$ symmetry of the
total system becomes a strong or a weak $\mathrm{U}(1)$ symmetry of the system
dynamics if particles cannot or can be exchanged between system and reservoir,
respectively~\cite{Sieberer2016a, Sieberer2023}. The former case is realized in
the unconditional dynamics obtained by averaging over the outcomes of occupation
measurements; the latter case applies to the unconditional dynamics under
fermion counting. For a single replica, the relevant phase rotation symmetries
are $\psi_{\pm} \mapsto \e^{\imag \phi_{\pm}} \psi_{\pm}$. The strong symmetry
is equivalent to invariance of the Keldysh action under this transformation for
arbitrary choices of $\phi_+$ and $\phi_-$. For the weak symmetry, invariance of
the action is required only for $\phi_+ = \phi_-$. Let us now consider the
unconditional dynamics of $R$ replicas under fermion counting. This is described
by a contribution to the Keldysh action of the form given in
Eq.~\eqref{eq:measurement-Lagrangian}, but with a sum instead of a product over
replicas, that is,
$\sum_{r = 1}^R \left( \psi_{+, r} \psi^*_{-, r} + \psi^*_{+, r}
  \psi_{-, r} \right) = \psi^*_+ \psi_- - \psi^*_- \psi_+$.
Evidently, this term is symmetric under the transformation in
Eq.~\eqref{eq:psi-pm-G-Phi-pm} with $\Phi_+ = \Phi_-$, which generalizes the
weak $\mathrm{U}(1)$ symmetry associated with conservation of the number of
particles in the total system to $R$ replicas. However, the strong
$\mathrm{SU}(R)$ symmetry underlying the NLSM, given by
Eq.~\eqref{eq:psi-pm-G-Phi-pm} with $\Phi_+ = - \Phi_-$, is broken. The breaking
of the latter symmetry can also be understood on the level of quantum states: In
the unconditional dynamics, an initially pure state with $N$ particles evolves
into an incoherent mixture of states with different numbers of
particles~\cite{Buca2012, Albert2014}. However, as shown above, the strong
$\mathrm{SU}(R)$ symmetry requires the system to be in a pure state with a fixed
number of particles at all times.

These considerations show that the conservation of the total number of particles
in the system and reservoirs is not sufficient to guarantee the existence of a
strong $\mathrm{SU}(R)$ symmetry. An additional necessary requirement is that
each exchange of particles between the system and reservoirs has to be detected,
such that the system remains in a pure state.

\section{Fluctuation expansion of the Keldysh action}
\label{sec:fluctuation-expansion}

To obtain the variation of the Keldysh action with respect to $\mathcal{G}$ and
thereby find the saddle-point manifolds described by
Eqs.~\eqref{eq:saddle-points-FC}
and~\eqref{eq:saddle-points-OM}, and to derive the action
that governs the Gaussian theory discussed in Sec.~\ref{sec:gaussian-theory}, we
perform an expansion of the Keldysh action around its saddle points.

\subsection{Fluctuation expansion of the measurement Lagrangian for fermion
  counting}

The existence of a manifold of saddle points rather than just a single isolated
saddle point is a direct consequence of the symmetries of the action. Therefore,
to establish the full manifold, it is sufficient to show that a single point on
the manifold is indeed a saddle point. For the manifold described by
Eqs.~\eqref{eq:G-saddle-point} and~\eqref{eq:saddle-points-FC}, a
convenient choice is given by $Q = \sigma_z$, where we omit the identity $1_R$
in replica space for brevity. We parametrize fluctuations around this
configuration as
$\mathcal{G} = - \imag \left( \sigma_z + \dQG \right) \! \big/2$. Inserting this
expression in Eq.~\eqref{eq:lagrangian-FC}, we obtain
\begin{equation}
  \imag \mathcal{L}_M = \frac{1}{2^R} \sum_{\alpha = \pm} \detR\! \left[ 1 - \imag
    \trK \! \left(\tau_{\alpha} \dQG \right) \right] - 1.
\end{equation}
By expanding the determinant using 
\begin{equation}
  \label{eq:det-expansion}
  \detKR(1 + \epsilon X) = 1 + \epsilon \trKR(X) - \frac{\epsilon^2}{2} \left[ \trKR \!
    \left( X^2 \right) - \trKR(X)^2 \right] + O \! \left( \epsilon^3 \right),
\end{equation}
we find, to zeroth and first order in $\dQG$,
\begin{equation}
  \label{eq:L-M-expansion-FC-0-1}
  \imag \mathcal{L}_M^{(0)} =  \frac{1}{2^{R-1}} - 1, \qquad
  \imag \mathcal{L}_M^{(1)} = \frac{1}{2^R}
  \trKR\!\left( \sigma_z \dQG{} \right).
\end{equation}
There are two contributions of second order. The first one is
\begin{equation}
  \label{eq:L-M-2-1-definition}
  \begin{split}
    \imag \mathcal{L}_M^{(2,1)} & = \frac{1}{2^{R + 1}}
    \sum_{\alpha = \pm} \trR\!\left[ \trK\!\left(\tau_{\alpha}
        \dQG\right)^2\right] \\ & = - \frac{1}{2^{R + 2}} \sum_{r, r' = 1}^R
    \left[ \trK \! \left( \sigma_z \delta Q_{\mathcal{G}, r, r'} \right) \trK \!
      \left( \sigma_z \delta Q_{\mathcal{G}, r', r} \right) \vphantom{\sigma_y} \right. \\ & \peq
    \left. - \trK \! \left( \sigma_y \delta Q_{\mathcal{G}, r, r'} \right) \trK
      \! \left( \sigma_y \delta Q_{\mathcal{G}, r', r} \right) \right].
  \end{split}
\end{equation}
For fixed replica indices $r$ and $r'$, $\delta Q_{\mathcal{G}, r, r'}$ is a
$2 \times 2$ matrix in Keldysh space. To simplify
Eq.~\eqref{eq:L-M-2-1-definition} we can, therefore, use the following relation
between determinants and traces of $2 \times 2$ matrices $A$ and $B$:
\begin{equation}
  \label{eq:det-sum-2x2}
  \det(A + B) = \det(A) + \det(B) + \tr(A) \tr(B) - \tr(A B).
\end{equation}
This allows us to express the products of traces in
Eq.~\eqref{eq:L-M-2-1-definition} as
\begin{multline}  
  \trK \! \left( \sigma_z \delta Q_{\mathcal{G}, r, r'} \right) \trK \! \left(
    \sigma_z \delta Q_{\mathcal{G}, r', r} \right) = \trK \! \left( \sigma_z
    \delta Q_{\mathcal{G}, r, r'} \sigma_z \delta Q_{\mathcal{G}, r', r} \right)
  \\ - \detK \! \left( \delta Q_{\mathcal{G}, r, r'} + \delta Q_{\mathcal{G},
      r', r} \right) + \detK \! \left( \delta Q_{\mathcal{G}, r, r'} \right) +
  \detK \! \left( \delta Q_{\mathcal{G}, r', r} \right),
\end{multline}
and analogously for the terms involving $\sigma_y$. The terms that contain
$\detK$ cancel in the difference between terms involving $\sigma_z$ and
$\sigma_y$ in Eq.~\eqref{eq:L-M-2-1-definition}. We thus obtain
\begin{equation}
  \label{eq:L-M-expansion-FC-2-1}
  \imag \mathcal{L}_M^{(2, 1)} = -\frac{1}{2^{R + 2}}
  \trKR\!\left[ \left( \sigma_z \dQG \right)^2 - \left( \sigma_y \dQG \right)^2
  \right].
\end{equation}
The second contribution of second order reads
\begin{equation}
  \label{eq:L-M-expansion-FC-2-2}
  \begin{split}
    \imag \mathcal{L}_M^{(2,2)} & = - \frac{1}{2^{R + 1}}
    \sum_{\alpha = \pm} \trR\!\left[ \trK\!\left(\tau_{\alpha}
        \dQG\right)\right]^2 \\
    & = \frac{1}{2^{R + 2}} \left[ \trKR\!\left(\sigma_z \dQG \right)^2 -
      \trKR\!\left( \sigma_y \dQG \right)^2 \right].
  \end{split}
\end{equation}

Using the first-order contribution in Eq.~\eqref{eq:L-M-expansion-FC-0-1},
we can derive the saddle-point equation
$\delta S/\delta \mathcal{G} = \imag 2 \left( \delta S/\dQG{} \right) = 0$,
leading to $\Sigma = \gamma Q/2^{R - 1}$ with $Q = \sigma_z$ as expected.

\subsection{Fluctuation expansion of the measurement Lagrangian for occupation
  measurements}

To obtain the saddle-point manifold and derive the Gaussian theory of
Sec.~\ref{sec:gaussian-theory}, it is sufficient to expand the measurement
Lagrangian in fluctuations around a particular saddle point, which can be chosen
as $Q = \Lambda$ for occupation measurements. However, the derivation of the
NLSM for occupation measurements requires an expansion around $Q_0$ given in
Eq.~\eqref{eq:Q0-parametrized}, which reduces to $Q_0 = \Lambda$ for
$\mathcal{R}_{\phi} = \mathcal{R}_{\theta} = 1$. From $\Lambda$ given in
Eq.~\eqref{eq:lambda-OM}, $Q_0$ inherits the properties
$Q_0^2 = 1$, $\trK(Q_0) = 0$, and $\detK(Q_0) = -1$. Inserting
$\mathcal{G} = - \imag \left( Q_0 + \dQG \right) \! \big/2$ in
Eq.~\eqref{eq:lagrangian-occupation}, we obtain
\begin{multline}
  \imag \mathcal{L}_M = \frac{1}{\left( 4 n \right)^R} \left\{
    \detKR \!  \left[ 1 - \sigma_x \left( Q_0 + \dQG \right) \right] \right. \\
  \left. + \trK \! \left( \sigma_x \dQG \right) + 2 R \left( 1 - 2 \rho_0
    \right) \right\} - 1,
\end{multline}
where $\rho_0$ is defined in Eq.~\eqref{eq:rho-0}. A more convenient form for
the expansion of the determinant in $\dQG$ is given by
\begin{multline}
  \detKR \! \left[ 1 - \sigma_x \left( Q_0 + \dQG \right) \right] \\ = \left[ -
    \detK(\sigma_x - Q_0) \right]^R \detKR \! \left[ 1 - \left( \sigma_x - Q_0
    \right)^{-1} \dQG \right].
\end{multline}
To calculate the inverse $\left( \sigma_x - Q_0 \right)^{-1}$, we use the
following relation, which is valid for any invertible $2 \times 2$ matrix $A$:
\begin{equation}
  \label{eq:inverse-2x2}
  A^{-1} = \frac{1}{\det(A)} \left[ \tr(A) - A \right].
\end{equation}
Furthermore, using Eq.~\eqref{eq:det-sum-2x2}, we obtain
$\detK(\sigma_x - Q_0) = - 4 \rho_0$.
We thus find
\begin{multline}
  \frac{1}{\left( 4 n \right)^R} \detKR \! \left[ 1 - \sigma_x \left( Q_0 + \dQG
    \right) \right] \\ = \frac{\rho_0^R}{n^R} \detKR \! \left( 1 + \frac{Q_0 -
      \sigma_x}{4 \rho_0} \dQG \right).
\end{multline}
Then, employing Eq.~\eqref{eq:det-expansion} to expand
$\imag \mathcal{L}_M$ to second order in $\dQG$, we obtain
\begin{equation}
  \label{eq:L-M-expansion-OM}
  \begin{split}
    \imag \mathcal{L}_M^{(0)} & = \frac{\rho_0^R}{n^R} + \frac{2
      R}{\left( 4 n \right)^R} \left( 1 - 2 \rho_0 \right) - 1, \\
    \imag \mathcal{L}_M^{(1)} & = \frac{\rho_0^{R - 1}}{4 n^R}
    \trKR \! \left[ \left( Q_0 - \sigma_x \right) \dQG \right] + \frac{1}{\left(
        4 n \right)^R} \trKR \! \left( \sigma_x \dQG \right), \\ \imag
    \mathcal{L}_M^{(2, 1)} & = - \frac{\rho_0^{R - 2}}{32 n^R}
    \trKR \! \left\{ \left[ \left( Q_0 - \sigma_x \right) \dQG \right]^2 \right\},
    \\ \imag \mathcal{L}_M^{(2, 2)} & = \frac{\rho_0^{R - 2}}{32
      n^R} \trKR \! \left[ \left( Q_0 - \sigma_x \right) \dQG \right]^2.
  \end{split}
\end{equation}

To show that $Q = \Lambda$ is a saddle point, we can replace $Q_0$ by $\Lambda$
in the expansion above, leading to $\rho_0 = n$. Then, the solution to
$\delta S/\delta \mathcal{G} = 0$ is given by
\begin{equation}
  \Sigma = \frac{\gamma}{2 n} \left\{ \Lambda - \left[ 1 - \frac{1}{\left( 4 n
        \right)^{R - 1}} \right] \sigma_x \right\}.
\end{equation}
The Pauli matrix $\sigma_x$ in Keldysh space corresponds to a nonvanishing
anti-Keldysh component of the self-energy and thus violates the usual causality
structure. However, in the replica limit $R \to 1$, the coefficient of
$\sigma_x$ vanishes, and we recover the expected causal result
$\Sigma = \gamma Q/(2 n)$ with $Q = \Lambda$.

\subsection{Expansion of the full action}

Next, we expand the action Eq.~\eqref{eq:S-0-G-Sigma} in fluctuations around the
particular saddle points given by Eq.~\eqref{eq:saddle-points-FC}
and~\eqref{eq:saddle-points-OM} with
$\mathcal{R}_{\Phi} = \mathcal{R}_{\phi} = 1$. For simplicity, we set $R = 1$ in
all numerical factors. Then, fluctuations around the saddle points can be
parametrized as
$\Sigma = \gamma \left( \Lambda + \dQS{} \right) \! \big/ (2 n)$, where
$n = 1/2$ for fermion counting or occupation measurements at half filling.

The saddle-point self-energy $\Sigma = \gamma \Lambda/(2 n)$ obeys the familiar
causality structure of propagators in Keldysh field theory. Therefore, at the
saddle point, $\Tr \! \left[ \ln \! \left( G^{-1} \right) \right] = 0$ and
$\Tr \! \left( \mathcal{G} \Sigma \right) = 0$~\cite{Kamenev2023}. For fermion
counting with $A = 0$ in Eq.~\eqref{eq:S-0-G-Sigma}, this implies that
$\imag S_0^{(0)} = 0$. Furthermore, according to
Eq.~\eqref{eq:L-M-expansion-FC-0-1}, also $\imag \mathcal{L}_M^{(0)} = 0$, so
that the Keldysh action vanishes at the saddle point, in agreement with the
expected normalization $Z_R(t) = 1$ of the Keldysh partition function in the
replica limit $R \to 1$. For occupation measurements, the matrix $A$ in
Eq.~\eqref{eq:anti-Keldysh-self-energy} involves an anti-Keldysh component,
leading to a nonvanishing contribution
\begin{equation}
  \label{eq:Tr-G-A-saddle-point}
  \imag S_0^{(0)} = \imag \Tr \! \left( G A \right) = - \frac{\gamma L T}{2 n}
  \left( 1 - 2 n \right),
\end{equation}
where we have set $G = \Lambda$, $L$ is the system size and $T = t - t_0$ is the
total evolution time. However, this contribution is cancelled by the measurement
action: from Eq.~\eqref{eq:L-M-expansion-OM}, with
$\rho_0 = n$ at the saddle point $Q_0 = \Lambda$, we obtain
\begin{equation}
  \imag \gamma S_M^{(0)} = \imag \gamma \int \diff^2 \mathbf{x} \,
  \mathcal{L}_M^{(0)} = \frac{\gamma L T}{2 n} \left( 1 - 2 n \right),
\end{equation}
such that $\imag \left( S_0^{(0)} + \imag \gamma S_M^{(0)}
\right) = 0$.

In an expansion around a saddle point, there is no contribution of first
order. To second order in $\dQS$, we find
\begin{equation}
  \label{eq:S-0-2}
  \imag S_0^{(2)} = \frac{1}{4 \tau_0} \Tr\!\left(\frac{1}{2 \tau_0} G
    \dQS{}G\dQS{} - \dQS{}\dQG{} \right),
\end{equation}
where $\tau_0 = n/\gamma$. The dressed Green's function in momentum space is
defined as
\begin{equation}
  \begin{split}
    G_q(\omega) & = \left[ \omega - \xi_q + \imag \Lambda/(2 \tau_0) \right]^{-1} \\
    & = \frac{1}{2} G^R_q(\omega) \left(1 + \Lambda\right) + \frac{1}{2}
    G^A_q(\omega) \left(1 - \Lambda\right),
  \end{split}
\end{equation}
where $\xi_q = -2J \cos(q)$ and retarded $G^R$ and advanced $G^A$ Green's
functions are given by
\begin{equation}
  G^{R/A}_q(\omega) = \frac{1}{\omega - \xi_q \pm \imag/(2 \tau_0)}.
\end{equation}
Indeed, as stated in the main text in Eq.~\eqref{eq:S-0-2-B}, only the diffusion
block $\mathcal{B}_l(t) = G^R_l(t) G^A_{-l}(- t)$ contributes to
Eq.~\eqref{eq:S-0-2}.

\section{Derivation of the NLSM}
\label{sec:derivation-nlsm}

As explained in Sec.~\ref{sec:effective-field-theory}, the parametrization in
Eqs.~\eqref{eq:Q-parametrized} and~\eqref{eq:Q0-parametrized} is chosen such
that the symmetries of the action that are broken spontaneously by the saddle
point $Q = \Lambda$ and thus give rise to Goldstone modes appear explicitly. For
fermion counting, these are rotations of the form $\mathcal{R}_{\Phi}$,
Eq.~\eqref{eq:R-Phi-R-Theta}; for generalized occupation measurements, also
rotations of the form $\mathcal{R}_{\phi}$, Eq.~\eqref{eq:R-phi-R-theta}, are a
symmetry of the action. The NLSM describes fluctuations of the corresponding
massless Goldstone modes, and is obtained by integrating out the massive
fluctuations described by $\Theta$, $\theta$, and, for fermion counting, $\phi$.

To derive the NLSM, we first take the spatial continuum limit, in which the
matrix $H$ defined in Eq.~\eqref{eq:Hamiltonian-matrix} is replaced by a
differential operator containing derivatives with respect to a continuous
spatial coordinate $x$. Inserting Eq.~\eqref{eq:G-Sigma-nonlinear-fluctuations}
in the action in Eq.~\eqref{eq:S-0-G-Sigma} and performing an expansion in
spatial and temporal derivatives yields a Lagrangian of the
form~\cite{Kamenev2023}
\begin{equation}
  \label{eq:L-0-NLSM}
  \imag \mathcal{L}_0[Q] = \Tr \! \left[ \frac{1}{2} \Lambda
    \mathcal{R}^{-1} \partial_t \mathcal{R} - \frac{\nu}{8} \left( \partial_x Q
    \right)^2 \right] - \frac{\gamma}{2 n} \left( 1 - 2 n \right),
\end{equation}
where $\nu = 2 n J^2/\gamma$. In the term $\imag \Tr(G A)$ in
Eq.~\eqref{eq:S-0-G-Sigma}, we have neglected derivatives of $\mathcal{R}$,
which would be multiplied by an additional small parameter $A \sim \gamma$.
Rotations of the form $\mathcal{R}_{\Phi}$, Eq.~\eqref{eq:R-Phi-R-Theta}, being
a symmetry of the action and, in particular, commuting with $A$, drop out of
$\imag \Tr(G A)$. Then, expanding $\imag \Tr(G A)$ in the massive mode $\Theta$
leads to terms of second order $\gamma$, which should be dropped on our order of
approximation. The same reasoning can be repeated for the rotations given in
Eq.~\eqref{eq:R-phi-R-theta}. Consequently, in Eq.~\eqref{eq:L-0-NLSM} we have
replaced the term $\imag \Tr(G A)$ by its value on the saddle point given in
\eqref{eq:Tr-G-A-saddle-point}.

The expansion of the derivative terms in Eq.~\eqref{eq:L-0-NLSM} in $\Theta$ is
detailed in Ref.~\cite{Poboiko2023} and yields
\begin{multline}
  \label{eq:L-0-NLSM-expanded}
  \imag \mathcal{L}_0[Q] = \Tr \! \left[ \frac{1}{2} \Lambda
    \mathcal{R}_0^{-1} \partial_t \mathcal{R}_0 - \frac{\nu}{8}
    \left( \partial_x Q_0 \right)^2 \right] \\ + \frac{1}{4} \trK(Q_0 \sigma_z)
  \trR \! \left[ \Theta U^{-1/2} \left( \partial_t U \right) U^{-1/2} \right] \\
  - \nu \rho_0 \left( 1 - \rho_0 \right) \trR \! \left[ \left( \partial_x U^{-1}
    \right) \left( \partial_x U \right) \right] - \frac{\gamma}{2 n} \left( 1 -
    2 n \right),
\end{multline}
where $\mathcal{R}_0 = \mathcal{R}_{\phi} \mathcal{R}_{\theta}$,
$U = \e^{\imag \Phi}$, and $\rho_0$ is defined in Eq.~\eqref{eq:rho-0}.

We focus now for the moment on the replica-symmetric sector of the theory,
setting $R = 1$. Since $\theta$ is a massive mode for both fermion counting and
occupation measurements, we expand $\mathcal{L}_0$ to second order in $\theta$.
Furthermore, we drop terms that contain $\partial_t \theta$ and $\partial_t \phi$
and no other fields. Such terms contribute to the action only at the initial and
final time. However, these contributions vanish. We thus find, with
$f = 1 - 2 n$,
\begin{multline}
  \label{eq:L-0-replica-symmetric}
  \imag \mathcal{L}_0[Q_0] = - \frac{\imag}{4} \left( 2 + f \theta \right)
  \theta
  \partial_t \phi - \frac{\nu}{4} \left\{ \left[ 1 - f^2 + 2 f \theta
    \right. \right. \\ \left. \left. - \left( 1 - f^2 \right) \theta^2 \right]
    \left( \partial_x \phi \right)^2 - \imag 2 f \left( f - \theta \right)
    \left( \partial_x \phi \right) \left( \partial_x \theta \right) \right. \\
  \left. + \left( 1 + f^2 \right) \left( \partial_x \theta \right)^2 \right\} -
  \frac{ f \gamma}{2 n}.
\end{multline}

Returning to $R > 1$, we next consider the contribution from the measurement
Lagrangian, given in Eqs.~\eqref{eq:lagrangian-FC}
and~\eqref{eq:lagrangian-occupation} for fermion counting and occupation
measurements, respectively, where we insert
Eq.~\eqref{eq:G-Sigma-nonlinear-fluctuations}. The parametrization of rotations
in Eq.~\eqref{eq:Q-parametrized} is chosen such that $\mathcal{R}_{\Phi}$, being
a symmetry of the Lagrangian for both models, drops out. For fermion counting,
$\mathcal{R}_{\Phi}$ is the only symmetry. Therefore, we can expand $Q$ to
second order in the massive modes $\Theta$, $\phi$, and $\theta$,
\begin{equation}
  \label{eq:Q-Theta-FC}
  \mathcal{R}_{\Theta} Q_0 \mathcal{R}_{\Theta}^{-1} = \sigma_z + \delta Q,
\end{equation}
where
\begin{equation}
  \label{eq:delta-Q-FC}
  \delta Q = \phi \sigma_y - \theta \sigma_x - \frac{\theta^2 }{2} \sigma_z -
  \frac{\phi^2}{2}  \sigma_z - \frac{\Theta^2}{2} \sigma_z - \Theta \left(
    \sigma_x + \sigma_z \theta \right).
\end{equation}
Then, using Eqs.~\eqref{eq:L-M-expansion-FC-0-1},
\eqref{eq:L-M-expansion-FC-2-1}, and~\eqref{eq:L-M-expansion-FC-2-2}, we obtain
\begin{equation}
  \label{eq:delta-L-M-FC}
  \imag \mathcal{L}_M[Q] = - \frac{1}{2} \left[ \trR \!
    \left( \Theta^2 \right) + \theta^2 + \phi^2 \right],
\end{equation}
where we have set $R = 1$ in all numerical factors, such that
$\mathcal{L}_M^{(0)}$, given in Eq.~\eqref{eq:L-M-expansion-FC-0-1}, vanishes.

For occupation measurements, also $\mathcal{R}_{\phi}$ is a symmetry. Therefore,
we cannot expand $Q$ in $\phi$ as we did for fermion counting, and
Eqs.~\eqref{eq:Q-Theta-FC} and~\eqref{eq:delta-Q-FC} have to be replaced by
\begin{equation}
  \label{eq:Q-Theta-OM}
  \mathcal{R}_\Theta Q_0 \mathcal{R}_\Theta^{-1} = Q_0 + \delta Q,
\end{equation}
where
\begin{equation}
  \delta Q = \frac{\imag}{2} \Theta \left[ \sigma_y, Q_0 \right] -
  \frac{\Theta^2 }{4} \left( Q_0 - \sigma_y Q_0 \sigma_y \right).
\end{equation}
Inserting Eq.~\eqref{eq:Q-Theta-OM} in Eq.~\eqref{eq:L-M-expansion-OM} we obtain
an expansion of the measurement Lagrangian in $\Theta$,
\begin{equation}
  \label{eq:delta-L-M-OM}
  \imag \mathcal{L}_M = \frac{f}{2 n}   -
  \frac{1}{32 n \rho_0 } \trK \! \left(\sigma_z Q_0\right)^2
  \trR\!\left(\Theta^2\right).
\end{equation}

To find the replica-symmetric Lagrangian for fermion counting, we can simplify
Eq.~\eqref{eq:L-0-replica-symmetric} further by setting $f = 0$, performing an
expansion to second order in both $\phi$ and $\theta$, and omitting the last
term that stems from $\imag \tr(G A)$. Then, combining
Eq.~\eqref{eq:L-0-replica-symmetric} with Eq.~\eqref{eq:delta-L-M-FC} yields
Eq.~\eqref{eq:L-replica-symmetric-FC}. For occupation measurements, combining
Eq.~\eqref{eq:L-0-replica-symmetric} with Eq.~\eqref{eq:delta-L-M-OM} we obtain
Eq.~\eqref{eq:L-replica-symmetric-OM}. In the replicon sector described by the
modes $\Phi$ and $\Theta$, integration over $\Theta$ leads for both models to
Eq.~\eqref{eq:L-replicon}.

\section{Modified NLSM for particle-hole symmetric Hamiltonians}
\label{sec:NLSM-with-PHS}

Through the gauge transformation $\hat{\psi}_l \mapsto \imag^l \hat{\psi}_l$,
the hopping matrix $H$ in Eq.~\eqref{eq:Hamiltonian-matrix} becomes purely
imaginary, and obeys the PHS condition $H = - H^{\transpose}$. As we discuss
here, PHS modifies the target manifold of the NLSM~\cite{Jian2022, Fava2024,
  Poboiko2025}. We summarize only the key steps in the derivation of the
modified NLSM. Details of the derivation will be presented elsewhere.

We begin by explicitly symmetrizing the action with respect to PHS. Consider
first the Hamiltonian contribution to the action in Eq.~\eqref{eq:G-0}. After
the above gauge transformation and the Larkin-Ovchinnikov rotation,
Eq.~\eqref{eq:Larkin-Ovchinnikov-rotation}, this contribution can be written as
\begin{equation}
  \psi^{\dagger} H \psi = \frac{1}{2} \psi^{\dagger} \left( H - H^{\transpose}
  \right) \psi = \frac{1}{2} \left( \psi^{\dagger} H \psi + \psi^{\transpose} H
    \psi^{*} \right) = \bar{\Psi} H \Psi,
\end{equation}
where we have introduced the vectors of fields,
\begin{equation}
  \label{eq:doubled-spinors}
  \Psi = \frac{1}{\sqrt{2}}
  \begin{pmatrix}
    \psi \\ \imag \sigma_y \psi^{*}
  \end{pmatrix},
  \quad
  \bar{\Psi} = \frac{1}{\sqrt{2}} \left( \psi^{\dagger}, \psi^{\transpose}
    \left( - \imag \sigma_y \right) \right) = \Psi^{\transpose} \mathcal{C}.
\end{equation}
For fixed lattice site index $l$, the vectors $\Psi_l(t)$ and $\bar{\Psi}_l(t)$
have $4 R$ components in the combined Keldysh, charge-conjugation, and replica
space. The charge-conjugation matrix is defined as
$\mathcal{C} = \sigma_y \otimes \sigma_y$, where the first and second factors of
$\sigma_y$ act in Keldysh and charge-conjugation space, respectively, and the
identity in replica space is left implicit.

The matrices $\pm \imag \sigma_y$ in Eq.~\eqref{eq:doubled-spinors} are not
required to symmetrize the Hamiltonian contribution to the action. However, for
this definition of $\Psi$ and $\bar{\Psi}$, also the measurement vertices in
Eqs.~\eqref{eq:FC-vertices-rotated} and Eq.~\eqref{eq:OM-vertex-rotated} take
forms that do not have any structure in charge-conjugation space and are thus
manifestly particle-hole symmetric: For fermion counting, we obtain
\begin{equation}
  V_{\pm}[\bar{\Psi}, \Psi] = \imag \bar{\Psi} \tau_{\pm} \Psi,
\end{equation}
with matrices $\tau_{\pm}$, which are defined below
Eq.~\eqref{eq:lagrangian-FC}, acting in Keldysh space; for generalized
occupation measurements, we can write the measurement vertex in the
exponentiated form
\begin{equation}
  V[\bar{\Psi}, \Psi] = \frac{1}{4 n} \e^{2 \bar{\Psi} \sigma_x \Psi},
\end{equation}
again with the matrix $\sigma_x$ acting in Keldysh space.

As for broken PHS, the next step is to perform a generalized
Hubbard-Stratonovich transformation. However, Eq.~\eqref{eq:HS-identity} has to
be modified to an integration over Hermitian $4 R \times 4 R$ matrices
$\mathcal{G}$ and $\Sigma$ that obey the condition
$\mathcal{G} = - \mathcal{C} \mathcal{G}^{\transpose} \mathcal{C}$, which
follows from $\mathcal{G} = - \imag \Psi \bar{\Psi}$ and
Eq.~\eqref{eq:doubled-spinors}. The measurement Lagrangian for fermion counting
in Eq.~\eqref{eq:lagrangian-FC} is thus replaced by
\begin{equation}
  \imag \mathcal{L}_M = \sum_{\alpha = \pm} \e^{\beta \trCR
    \left\{ \ln \left[ \trK \left( \frac{1}{\beta} \tau_{\alpha}
          \mathcal{G} \right) \right] \right\}} - 1, 
\end{equation}
where $\beta = 1/2$. Here and in the following, the corresponding expressions
for broken PHS are recovered by setting $\beta = 1$. The measurement Lagrangian
for occupation measurements, Eq.~\eqref{eq:lagrangian-occupation}, becomes
\begin{equation}
  \imag \mathcal{L}_M = \frac{1}{n^R} \e^{
    \beta \tr \left[ \ln \left( \frac{1}{2} - \frac{\imag}{\beta} \sigma_x
        \mathcal{G} \right) \right] } + \frac{\imag}{2^{2 R - 1} n^R} \tr \! \left( \sigma_x
    \mathcal{G} \right) - 1.
\end{equation}
Finally, also the action in Eq.~\eqref{eq:S-0-G-Sigma} acquires a factor of
$\beta = 1/2$,
\begin{equation}
  \imag S_0[\mathcal{G}, \Sigma] = \Tr \! \left\{ \beta \left[ \ln \! \left(
        G^{-1} \right) + \imag G A \right]- \imag \mathcal{G} \Sigma \right\},
\end{equation}
where $A = 0$ for fermion counting; for occupation measurements, $A$ is defined
in Eq.~\eqref{eq:anti-Keldysh-self-energy}.

Next, we consider the saddle points of the particle-hole symmetrized
action. Equation~\eqref{eq:G-saddle-point} becomes
$\mathcal{G} = - \imag \beta Q/2$; in contrast, the self-energy can still be
expressed through $Q$ as in Eqs.~\eqref{eq:saddle-points-FC}
and~\eqref{eq:saddle-points-OM}. Note, however, that the replica-symmetric
saddle point, $Q = Q_0 \otimes 1_R$, is now determined by a $4 \times 4$ matrix
$Q_0$ in Keldysh and charge-conjugation space:
\begin{equation}
  Q_0 =
  \begin{pmatrix}
    \Lambda & 0 \\ 0 & - \sigma_y \Lambda^{\transpose} \sigma_y
  \end{pmatrix},
\end{equation}
with $\Lambda$ given in Eq.~\eqref{eq:lambda-OM}.

The target manifold of the NLSM can now be determined in three steps:
(i)~Identify the group $G$ of transformations
$Q \mapsto \mathcal{R} Q \mathcal{R}^{-1}$ that preserve Hermiticity as well as
the generalized skew-symmetry condition
$Q = - \mathcal{C} Q^{\transpose} \mathcal{C}$, and are symmetries of the
action. In Keldysh space, these transformations are
\begin{equation}
  \mathcal{R} = M
  \begin{pmatrix}
    \mathcal{V} & 0 \\ 0 & \sigma_y \mathcal{V}^{*} \sigma_y
  \end{pmatrix} M,
\end{equation}
where
$M = \frac{1}{\sqrt{2}} \left( \begin{smallmatrix} 1 & 1 \\ 1 & -
    1 \end{smallmatrix} \right)$
is the matrix form of the Larkin-Ovchinnikov rotation,
Eq.~\eqref{eq:Larkin-Ovchinnikov-rotation}. For fermion counting, the matrices
$\mathcal{V}$ are in $G = \mathrm{SU}(2 R)$; for occupation measurements,
conservation of the number of particles results in an enlarged symmetry group
$G = \mathrm{U}(2 R) = \mathrm{U}(1) \rtimes \mathrm{SU}(R)$. (ii)~Identify the
subgroup $H \subset G$ of transformations that leave the saddle point $Q_0$
invariant. This is the case for matrices $\mathcal{V}$ that obey the condition
$\mathcal{V}^{\transpose} \sigma_y \mathcal{V} = \sigma_y$ and thus form the
compact symplectic group $H = \mathrm{Sp}(R)$. (iii)~The target manifold of the
NLSM is $G / H$. Focusing on the replicon sector of the theory, the target
manifold of the NLSM for both fermion counting and generalized occupation
measurements is thus $\mathrm{SU}(2 R) / \mathrm{Sp}(R)$. Crucially, also in the
presence of PHS, the target manifold of the replicon NLSM is not modified for
fermion counting as compared to models with a conserved number of
particles~\cite{Fava2024, Poboiko2025}. Particle-number conservation affects
only the replica-symmetric sector: The $\mathrm{U}(1)$ NLSM becomes massive when
particle-number conservation is broken by fermion counting.

\bibliography{bibliography}

\end{document}